\documentclass[12pt]{article}
\usepackage{amssymb,amsmath,amsthm,mathrsfs,latexsym,amsxtra,graphicx,appendix,hyperref,setspace,subfig,fix-cm,color,cite,dsfont,longtable,multirow,tikz,longtable,array,soul,fancybox,comment}
\numberwithin{equation}{section}
\setul{}{1pt}

\newcommand{\bea}{\begin{eqnarray}}
\newcommand{\eea}{\end{eqnarray}}
\newcommand{\be}{\begin{equation}}
\newcommand{\ee}{\end{equation}}

\newcommand{\nn}{\nonumber}

\newcommand{\cA}{\mathcal A}
\newcommand{\cD}{\mathcal D}
\newcommand{\cE}{\mathcal E}
\newcommand{\cF}{\mathcal F}
\newcommand{\cN}{\mathcal N}
\newcommand{\cG}{\mathcal G}
\newcommand{\cz}{\mathcal Z}
\newcommand{\bz}{\mathbb Z}
\newcommand{\fm}{\mathfrak m}

\newcommand{\bt}{\mathbb T}
\newcommand{\ch}{\mathcal H}
\newcommand{\uno}{{\mathds 1}}
\newcommand{\m}{\imath}

\newcommand{\wh}[1]{\widehat{#1}}
\newcommand{\bq}{\bar q}

\newcommand{\g}{\Gamma}

\newcommand{\op}{\hspace{1pt}}
\newcommand{\opm}{\hspace{-3pt}}

\newcommand{\vs}[1]{\vspace{#1pt}}
\newcommand{\hs}[1]{\hspace{#1pt}}

\newcommand{\msc}[1]{\mbox{\scriptsize$#1$}}
\newcommand{\msm}[1]{\mbox{\small$#1$}}

\newcommand{\tsm}[1]{\mbox{\small#1}}
\newcommand{\tsc}[1]{\mbox{\scriptsize#1}}

\newcommand{\vtgen}[2]{\vartheta\!
\Big[\!
	\hspace*{-1mm}\begin{array}{c}
		\msm{{#1}}\\[-1mm]
		\msm{{#2}}
	\end{array}
	\hspace*{-1mm}\!\Big]}

\newcommand{\hvtgen}[2]{\widehat\vartheta
\Big[\!
	\hspace*{-1mm}\begin{array}{c}
		\msm{{#1}}\\
		\msm{{#2}}
	\end{array}
	\hspace*{-1mm}\!\Big]}
\DeclareMathOperator{\spanop}{span}
\DeclareMathOperator{\Aut}{Aut}

\begin{document} 

\begin{titlepage}
\begin{center}

\hfill \\
\hfill \\
\vskip 0.75in

{\Large 
	\bf Asymmetric Orbifolds and\\[10pt] 4-dimensional $\mathcal N=4$ String Islands }\\
\vskip 0.4in

{\large Gerardo Aldazabal${}^{a,b,c}$, Eduardo C.~Andr\'es${}^{a,c}$, Anamar\'ia Font${}^{d,e}$,\\[2mm]
Elisa I.~Marieni${}^{f}$, Kumar Narain${}^{g}$, and Ida G.~Zadeh${}^{f}$
}\\
\vskip 4mm

${}^{a}$
{\it G. F\'isica CAB-CNEA, Centro At\'omico Bariloche, Av. Bustillo 9500, Bariloche, Argentina} \vskip 1mm
${}^{b}$
{\it Consejo Nacional de Investigaciones Cient\'ificas y T\'ecnicas {\rm(}CONICET{\rm)}} \vskip 1mm
${}^{c}$
{\it Instituto Balseiro, Universidad Nacional de Cuyo {\rm(}UNCUYO{\rm)}, Av. Bustillo 9500, R8402AGP,  Bariloche, Argentina} \vskip 1mm
${}^{d}$
{\it Fac. de Ciencias, Universidad Central de Venezuela, A.P.20513, Caracas 1020-A, Venezuela} \vskip 1mm
${}^{e}$
{\it Max-Planck-Institut f\"ur Gravitationsphysik, Albert-Einstein-Institut, 14476 Golm, Germany} \vskip 1mm
${}^{f}$
{\it Mathematical Sciences \& STAG Research Centre, University of Southampton, Southampton SO17 1BJ, UK} \vskip 1mm
${}^{g}$
{\it International Centre for Theoretical Physics, Strada Costiera 11, 34151 Trieste, Italy} \vskip 1mm

\end{center}

\vskip 0.35in

\begin{center} {\bf ABSTRACT } \end{center}

We formulate heterotic and type II $\bt^6/\bz_n$ asymmetric orbifolds 
preserving all 16 right-moving supersymmetries and leading
to 4-dimensional islands without massless moduli other than the dilaton. 
We establish the consistency of the orbifold conformal field theory, including
the validity of operator interpretation. In the heterotic case we classify all islands based 
on the $\Gamma(22,6)$ lattice constructed via the Leech lattice. 
In the type II case we also provide a classification of allowed $\Gamma(6,6)$
lattices and orbifold generators. In both cases we further examine whether a particular
orbifold admits an archipelago of different islands arising from inequivalent
realizations of the orbifold action. 
\vfill

\noindent November 18, 2026 %\today

\end{titlepage}
\pagenumbering{roman}
\setcounter{tocdepth}{2}

\tableofcontents

\newpage
\pagenumbering{arabic}
\setcounter{page}{1}

\section{Introduction and overview}\label{section_intro}

String islands are exotic string vacua which have no continuous moduli except for the dilaton \cite{Dabholkar:1998kv}.
What is the distribution of world-sheet conformal field theories describing string islands in the moduli space of world-sheet theories? 
This is a difficult question, the answer to which is in general not known.
On the other hand, the formalism of asymmetric orbifolds \cite{Narain:1986qm,Narain:1990mw} has 
proven to be an efficient approach to access exotic points in the world-sheet moduli space.
Its power lies in the non-geometric orbifold action on bosons and fermions, which includes a
subgroup of the automorphism group of the indefinite even self-dual lattice encoding the momenta of the compact bosons in the 
un-orbifolded world-sheet theory  \cite{Narain:1985jj,Narain:1986am}. Depending on how the orbifold generator acts on left and 
right-moving bosonic and fermionic degrees of freedom, various amounts of supersymmetry may be preserved. 
In this paper we will use the toroidal asymmetric orbifold framework developed in our earlier work \cite{Aldazabal:2025zht}, to carry 
out a systematic study of string islands with 16 supersymmetries in four spacetime dimensions.
Such islands do not have moduli belonging to massless vector multiplets and therefore the gauge group, excluding Abelian factors from 
graviphotons, has rank zero.

Asymmetric orbifolds have been extensively applied to study string vacua with 16 supersymmetries 
and reduced rank of the gauge group, including the case of islands 
\cite{Chaudhuri:1995bf,Chaudhuri:1995dj,Chaudhuri:1995ee,Bianchi:1997rf,Dabholkar:1998kv,
Mikhailov:1998si, deBoer:2001wca,Mizoguchi:2001cp,Aharony:2007du,Persson:2015jka,Bossard:2017wum,
Font:2021uyw, Fraiman:2021soq,Fraiman:2021hma,Fraiman:2022aik,Baykara:2024vss,Aldazabal:2025zht,Baykara:2025lhl}. 
The first heterotic vacua with maximal supersymmetry and reduced rank, including a 4-dimensional island, were 
found in \cite{Chaudhuri:1995fk} in the fermionic formulation. Type II islands with 16 supersymmetries in four dimensions have 
as well been obtained as asymmetric Gepner models \cite{Blumenhagen:2016rof}. 
A myriad of string vacua with a lower amount of supersymmetry, inclusive non-supersymmetric ones, have also been constructed as
asymmetric orbifolds \cite{Ibanez:1987pj,Lust:1987ts,
Harvey:1987da,Aoki:2004sm,Beye:2013moa,Tan:2015nja,Satoh:2015nlc,GrootNibbelink:2017usl,Harvey:2017rko,
Harvey:2017xdt,GrootNibbelink:2020dib,Baykara:2021ger,Bianchi:2022tbr,Acharya:2022shu,Cheng:2022nso,Montero:2022vva,
Gkountoumis:2023fym,Nakajima:2023zsh, Baykara:2023plc,Israel:2023tjw,Gkountoumis:2024dwc,Gkountoumis:2024boj,Baykara:2024vss,
Baykara:2024tjr, Angelantonj:2024jtu,Ishige:2026ihp,Larotonda:2026hxy,Ishige:2026ylw, Fraiman:2026irf}. 
Other interesting applications of asymmetric orbifolds in world-sheet string theory include \cite{Gaberdiel:2002jr,Plauschinn:2018wbo,Faraggi:2023mkn,Israel:2025ouq,Faraggi:2026bhv} and references therein.

In \cite{Aldazabal:2025zht}, a general world-sheet conformal field theory formalism employing toroidal 
asymmetric orbifolds was developed to obtain heterotic vacua with 16 supersymmetries and diverse ranks in various 
spacetime dimensions. One motivation behind the work was to verify a conjecture in \cite{Fraiman:2022aik} on the existence of a
6-dimensional heterotic island with 16 supersymmetries realized as a $\bt^4/\bz_{10}$ asymmetric orbifold. 
As a proof of concept, $\bt^D/\bz_{2m}$ heterotic asymmetric orbifolds, where $m$ is an odd prime number, 
were constructed and in particular shown to consistently describe two $\bt^4/\bz_{10}$ islands, a populated archipelago of $\bt^6/\bz_{22}$ islands, 
and several models with reduced rank in six dimensions. The classification of 6-dimensional string islands with 16 supersymmetries
was originally proposed in \cite{Fraiman:2022aik}, and recently established in \cite{Baykara:2025lhl}.

In the context of strong/weak string duality, supersymmetry preserving symmetries of type IIA theory on K3 are identified with 
supersymmetry preserving symmetries of the $(20,4)$ even self-dual lattice of the heterotic string on $\bt^4$. The former are 
symmetries of the internal conformal field theory of type IIA string theory, which is a non-linear sigma-model on K3, and preserve 
its small $\cN=(4,4)$ supersymmetry, while the latter are symmetries that act trivially on the four right-moving directions of the 
heterotic theory. It has been shown that all supersymmetry preserving symmetries of the K3 sigma-model are elements of the 
Conway group $\text{Co}_0$, which is the automorphism group of the Leech lattice \cite{Gaberdiel:2011fg}. Hence, asymmetric orbifolds 
of the heterotic theory on $\bt^4/\bz_n$ may be exhaustively constructed using cyclic automorphisms of the Leech lattice classified 
in \cite{HL}. In four spacetime dimensions, however, an analogous classification theorem for supersymmetry preserving symmetries 
of the type IIA theory on K3$~\times \bt^2$ or that of the heterotic string on $\bt^6$ is not currently known. Consequently, the full 
classification of asymmetric orbifolds describing maximally supersymmetric heterotic theories with reduced rank in four dimensions is 
a challenging problem. In particular, only sporadic instances of 4-dimensional islands are known to date 
\cite{Chaudhuri:1995fk,Chaudhuri:1995ee,Dabholkar:1998kv,Mizoguchi:2001cp,Blumenhagen:2016rof,Aldazabal:2025zht}. 
It is therefore important to make progress along this direction.

The art of constructing string islands via asymmetric orbifolds lies in crafting the even self-dual momentum lattice of internal bosons of 
string theory compactified on $\bt^D$. In four spacetime dimensions, the signature of this lattice, which we call $\Gamma$, is $(22,6)$ 
for heterotic theory and $(6,6)$ for type II theory. Nikulin's gluing technique \cite{Nikulin80int} enables one to adjoin two definite even 
lattices to build $\Gamma$. The signatures of the two definite lattices in the case of heterotic theory are $(22,0)$ and $(0,6)$, and in the 
case of type II theory are $(6,0)$ and $(0,6)$. Then, by construction, the lattice $\Gamma$ has an automorphism $\Theta$ which fixes 
the $(0,6)$ lattice component pointwise, and rotates all directions orthogonal to that. This is the automorphism we use to construct our 
islands: it preserves precisely 16 right-moving supercharges in both heterotic and type II theories. This is because the automorphism acts 
trivially on the right-moving sector and hence, preserves all right-moving supersymmetries. Furthermore, it acts nontrivially along all 
left-moving directions and breaks all left-moving supersymmetries in type II theory. 
In the case of $\bt^6/\bz_n$ asymmetric orbifolds, $\Theta$ is chosen as a cyclic subgroup of the automorphism group of $\Gamma$. 
The fixed-point sublattice under the action of $\Theta$ is denoted the invariant lattice $I$. The orthogonal complement of $I$ in 
$\Gamma$ is denoted the normal lattice $N$. As we will see in more detail in section \ref{section_asym}, the automorphism $\Theta$ of 
the glued $\Gamma$ must leave the conjugacy classes in $N^*\!/N$ invariant.

In this paper we focus on heterotic and type II $\bt^6/\bz_n$ asymmetric orbifolds, aiming at several objectives.
Our first goal is to apply the method of \cite{Aldazabal:2025zht} to catalog
4-dimensional heterotic islands with 16 supersymmetries obtained using the Leech lattice to build $\Gamma$ of signature $(22,6)$. 
Our second objective is to fully classify 4-dimensional type II islands with 16 supersymmetries in which $\Gamma$ of signature 
$(6,6)$ is formed gluing ADE root lattices, thereby completing earlier attempts \cite{Dabholkar:1998kv,Mizoguchi:2001cp}. 
In both settings we provide the explicit orbifold partition function and analyze the consistency of the orbifold conformal field theory.
To achieve the latter we extend the general formalism of \cite{Aldazabal:2025zht} for orders $n=2m$, with $m$ prime greater 
than 2, to $n=3m$, with $m$ prime greater than 3, while other values of $n$ are treated case by case.

String islands with 16 supersymmetries realized as asymmetric orbifolds do not contain any massless states in their twisted sectors. 
The only massless states come from their untwisted sector and belong to the supergravity multiplet. In the case of heterotic theories, 
the absence of massless vector multiplets in the untwisted sector demands that the normal lattice $N$ does not contain 
any vectors of length squared 2, namely root vectors. This motivated us in \cite{Aldazabal:2025zht} to choose $N$ as a sublattice of 
the Leech lattice, which does not have root vectors. The task of enumerating $T^6/\bz_n$ asymmetric orbifolds which are built upon 
the Leech lattice then translates into enumerating cyclic subgroups of the automorphism group of the Leech lattice. The orbits of 
fixed-point sublattices of the Leech lattice under cyclic subgroups of its automorphism group were classified in \cite{HL}. There are 
72 such orbits. To construct 4-dimensional heterotic islands, we are interested in fixed-point sublattices of rank 2 because their 
orthogonal complements have rank 22. In fact, every one of these rank-22 normal lattices possess at least one $\bz_n$ generator which leaves 
the conjugacy classes in $N^*\!/N$ intact. We therefore classify the Leech lattice-based heterotic islands by orders of these 
$\bz_n$ automorphisms, which are $n=6,8,9,12,14,15,16,18,20,21,22,23,24,28,30$.
Some orders allow multiple fixed-point sublattice orbits. 
In total, there are 28 4-dimensional heterotic island \emph{candidates} collected in Table \ref{HMisl}.
Five of them have already been found \cite{Chaudhuri:1995ee,Aldazabal:2025zht}.

Why do we call the entries of Table \ref{HMisl} island \emph{candidates}? To address this relevant question we first note that 
the generator of the orbifold group, which we denote by $g$, acts on a generic state with momentum $P\!\in\!\Gamma$ as 
$g|P\rangle\!=\!e^{2\pi i P\cdot v}|\Theta P\rangle$, where $v$ is a constant vector, called the \emph{shift} vector.
This vector plays an important role by shifting the momenta and providing a mechanism to eliminate massless states in twisted sectors. 
As we will show in sections \ref{section_asym} and \ref{section_het_story}, modular invariance of the partition function of the 
asymmetric orbifold and absence of massless states in all twisted sectors result in a set of constraints that $v$ must satisfy as necessary, 
but not sufficient, conditions for the theory to describe an island. It turns out that each entry in Table \ref{HMisl} admits shift 
vectors which satisfy these necessary conditions. To prove that the theory is an island, one has to show in addition that the orbifold 
conformal field theory is fully consistent. This is achieved by proving the validity of operator interpretation. In particular, consistency 
of the group action on degeneracy factors and phases dressing the states must be established. It is only then that the asymmetric 
orbifold realizes an island. In section \ref{section_het} we prove the full consistency of the world-sheet theory for several models in 
Table \ref{HMisl}. Applying our methods to other cases is feasible but beyond the scope of this work. Thus, since
we do not yet have a general proof of full consistency for all values of the order $n$, other models in  
Table \ref{HMisl} remain as island candidates. We hope to make progress on the general proof in the future.

In the setting of type II theories, to realize islands as $T^6/\bz_n$ asymmetric orbifolds, we build the even self-dual lattice 
$\Gamma$ with signature $(6,6)$ in terms of root lattices $\Lambda_{\cG}$ of rank-6 semisimple simply-laced Lie algebras $\cG$, 
i.e. ADE root lattices \cite{Narain:1985jj, Ginsparg:1986bx}.
The idea behind is to glue $\Lambda_{\cG}$ with itself, but with opposite signature.  
The $\bz_n$ automorphism is then chosen to be an element in the Weyl group of 
the Lie algebra $\cG$ which acts only on the left-moving sector and, by construction, preserves the conjugacy classes in $N^*\!/N$. 
This class of asymmetric orbifolds has been used to construct type II theories with reduced rank \cite{Dixon:1987yp}, and also
examples of islands \cite{Dabholkar:1998kv, Mizoguchi:2001cp, Baykara:2025lhl}, 
though without proving the full consistency of the orbifold.

As we will discuss in section \ref{section_typeii_story}, the requirement that all right-moving 
supersymmetries are preserved under the orbifold action, all left-moving ones are broken, and that the $\bz_n$ automorphism 
acts crystallographically on the normal lattice, puts stringent constraints on the order and eigenvalues. Table \ref{tab_triple} lists all 
possibilities. For 4-dimensional islands we impose the extra condition that the automorphism does not leave invariant directions,
i.e. does not have eigenvalues equal to one. It then follows that of all entries in Table  \ref{tab_triple} satisfying this condition,  only
21 correspond to $\bz_n$ automorphisms that can be realized as elements of the Weyl group of a rank-6 semisimple simply-laced Lie algebra. 
They comprise 11 different orders $n=4,6,8,12,14,15,18,20,24,30$, and are tabulated in Table \ref{tab_t}. The ADE root lattices 
and their respective automorphisms are also presented in the Table. In fact, there is a unique choice for each distinct $\bz_n$,
except for the $\bz_4$ and a particular $\bz_6$. Furthermore, each orbifold admits shift vectors which satisfy modular invariance 
conditions and remove massless states in all the twisted sectors. We therefore obtain in total 23 asymmetric orbifolds and 
complete the ADE root lattice classification of 4-dimensional type II island \emph{candidates} with 16 right-moving 
supersymmetries which, to the best of our knowledge, had not been accomplished before. 
Entries 3, 7, 12, 13, and 15 of Table \ref{tab_t} are the islands studied in \cite{Dabholkar:1998kv, Mizoguchi:2001cp}. 
In section \ref{section_typeii_isl} we construct specific type II islands from Table \ref{tab_t} and verify the full consistency of the 
asymmetric orbifold.

For each $\bt^6/\bz_n$ asymmetric orbifold, there are in general several shift vectors $v$ which satisfy the necessary conditions. 
Some vectors, however, may lead to a theory with the same spectrum. We call these vectors \emph{equivalent}. For all string islands 
we construct in this work, we classify families of non-equivalent shift vectors giving rise to theories with different massive spectra. 
These families determine the population of the \emph{archipelago} of islands for each $\bt^6/\bz_n$ orbifold.

In summary, our main results are : ($a$) a Leech lattice classification of heterotic island candidates and a full ADE 
root lattice classification of type II island candidates, all with 16 supersymmetries in four dimensions; ($b$) explicit construction of 
several string islands in each case, showing consistency of the orbifold conformal field theories, 
and determining the population of archipelagos; (c) a general analysis of $\bt^D/\bz_{3m}$ heterotic asymmetric
orbifolds, with $m$ prime greater than 3.

The paper is organized as follows. In section \ref{section_asym}, we review salient features of asymmetric orbifold world-sheet theories 
and analyze the consistency conditions that the conformal field theory must satisfy. Section \ref{section_het_story} presents the Leech 
lattice classification of heterotic island candidates. The level matching conditions imposed by modular invariance are derived in a systematic 
way in subsection \ref{subsection_het_Z}. In section \ref{section_het} we construct several heterotic islands and show that, except 
for the $\bt^6/\bz_8$ theory, they form archipelagos of isolated models. Section \ref{section_typeii_story} presents the ADE 
root lattice classification of type II island candidates. In section \ref{section_typeii_isl} we construct three specific asymmetric 
orbifolds $\bt^6/\bz_9$, $\bt^6/\bz_6$, and $\bt^6/\bz_4$, which realize type II islands for entries 1, 4, and 7 in Table \ref{tab_t}. 
The former construction revisits the original Dabholkar-Harvey island \cite{Dabholkar:1998kv} and, by examining the equivalence 
classes of the shift vectors, we show that it is indeed a unique theory. We end by making some final comments in section 
\ref{section_disc}. Details of proofs and computations are deferred to a collection of appendices. Appendices \ref{app_Z8}-\ref{app_Z3m} 
contain the computation of the partition function and proof of the consistency of operator interpretation in heterotic islands. 
In app. \ref{appv} we show how to obtain non-equivalent families of shift vectors in asymmetric orbifolds. 
Finally, in app. \ref{app_triple} we list all possible orders and eigenvalues of $\bz_n$ automorphisms acting crystallographically 
and breaking the 16 left-moving supersymmetries in the type II setting.

\section{Basics of asymmetric orbifolds}\label{section_asym}

In this section we outline the basic ingredients needed to construct asymmetric orbifolds $\bt^D/\bz_n$.
We will mostly adopt the notation and conventions of \cite{Aldazabal:2025zht}, where further details can be found.
In sections \ref{section_het_story} and \ref{section_typeii_story} we will discuss more specific features of the heterotic and type II
asymmetric orbifolds giving rise to 4-dimensional islands.

The starting point is a 10-dimensional closed superstring theory, either heterotic or type IIB for concreteness.
In both theories, the light-cone, right-moving world-sheet degrees of freedom consist of 8 bosons and 8 fermions.
On the other hand, the left-movers are also 8 bosons plus 8 fermions in type II,
whereas in the heterotic theory they comprise 24 bosons, 16 of which live on an even self-dual
lattice $\Gamma_0$. Recall that the 16-dimensional lattice $\Gamma_0$ can be either the $E_8+E_8$ root lattice,
or the $SO(32)$ root lattice with a spinor weight added.
The world-sheet fermions will be bosonized, i.e. described by bosons in the $SO(8)$ dual lattice, with weights restricted 
to the vector ($V$) and the spinor ($Sp$) classes by the GSO projection. 

The next stage is compactification on a $D$-dimensional torus $\bt^D$, including generic backgrounds for the metric,
the Kalb-Ramond field, as well as the Wilson lines in the heterotic case. The resulting momenta of the left- and right-moving
internal bosons can be combined to form an even self-dual lattice that will be denoted $\Gamma$ unless otherwise stated.
In type II and heterotic theories $\Gamma$ has signature $(D,D)$ and $(16+D,D)$ respectively. Notice that the
lattice $\Gamma$ depends on moduli, namely the background values mentioned above. 

The asymmetric orbifold is obtained by modding the $\bt^D$ conformal field theory by the $\bz_n$ 
with generator denoted $g$. We will next define the action of $g$ on the world-sheet degrees of freedom.
Generalities about the orbifold partition function will be discussed after introducing additional properties of 
$\Gamma$ and invariant sublattices.

\subsection{Orbifold action}\label{subsection_Zn}

The $\bz_n$ action on the underlying even self-dual lattice $\Gamma$ must be an order $n$ automorphism $\Theta$ 
which acts separately on left and right movers, i.e. \mbox{$\Theta=(\Theta_L, \Theta_R)$}. 
According to the formalism developed in\cite{Narain:1986qm, Narain:1990mw},
the action of $g$ on $\Gamma$ is specified in terms of the sublattice $I$ left invariant by $\Theta$, and the corresponding 
normal sublattice $N$, equal to the orthogonal complement of $I$ in $\Gamma$. These sublattices are defined by
\vs{-5}\be\label{INlattices}
I=\big\{P \in \Gamma \, ||\,  \Theta P = P \big\}\, , \quad
N:=I^\perp = \big\{P \in \Gamma \, || \, P\cdot X = 0,  \, \forall\op X \in I \big\}\ .\vs{-5}
\ee
The dual lattices of $I$ and $N$ are denoted $I^*$ and $N^*$. 
Vectors $P\!\in\!\Gamma$ can be decomposed as $P\!=\!(P_N, P_I)$, with $P_N \!\in\! N^*$ and $P_I \!\in\! I^*$ \cite{Narain:1986qm}. 
The great advantage of this decomposition is that $\Theta P=\Theta(P_N, P_I)= (\Theta P_N, P_I)$. The action of $g$ on $P \in \Gamma$
is then defined as
\vs{-5}
\be\label{gactlatt}
g|P_N, P_I \rangle = e^{2\pi i P \cdot v} |\Theta P_N, P_I \rangle\ ,\vs{-5}
\ee 
where $v$ is a constant shift vector that can be taken along the invariant directions without loss of generality.

We are mainly interested in asymmetric $\bt^D/\bz_n$ orbifolds preserving 16 supercharges, and we will specialize to even 
values of $D\le6$. For type II we will only examine the 
class of models, originally studied in \cite{Dixon:1987yp}, in which all right-moving supersymmetries are unbroken. Thus, in both 
type II and heterotic setups, we require the $D$ right-movers to be inert under the $\bz_n$ action, which in turn
implies $\Theta_R=\uno$ and $\Theta_L=\Theta$. Since all $D$ right directions are invariant, in both theories the normal lattice $N$ has
signature $(s,0)$ while the invariant lattice $I$ has signature $(16+D-s,D)$, with $0 \leq s \leq 16+D,$ in the heterotic theory, and 
$(D-s,D)$, with $0 \leq s \leq D$, in type II. The eigenvalues of $\Theta$ can be written as 
$e^{\pm2\pi i t_a}$, with $n t_a\in\bz$, and $1\le a\le\frac s2$.

The action of $g$ on the left-moving internal bosons depends on the eigenvalues of $\Theta$. Concretely,
\vs{-5}
\be\label{gXa}
g X_a = e^{2\pi i  t_a} X_a   \qquad (\text{no sum in}\  a) \ ,\vs{-0}
\ee
where the $X_a$ are complex. 
To maintain world-sheet supersymmetry in type II, the action on left-moving fermions is also determined by $\Theta$
as specified by
\vs{-5}\be\label{defvf}
g:|p \rangle \to e^{2\pi i p \cdot v_f} |p \rangle\ , \qquad\qquad v_f=(0,t_1,t_2,t_3) \ .\vs{-5}
\ee
Here $p$ is an $SO(8)$ weight belonging to vector ($V$) or spinor ($Sp$) classes. Since the action is order $n$ it must
be that $p\cdot n v_f \in \bz$ for all $p \in V$ and $p \in Sp$. Thus, in the type II setup there is an additional constraint 
\mbox{$n(t_1+t_2+t_3) =0\; \text{mod}\; 2$}.

\subsection{\texorpdfstring{More properties of $\Gamma$ and invariant sublattices}{TEXT}}\label{subsection_Gclasses}

Since $\Gamma$ is even self-dual, it can further be shown that $N^*/N \cong I^*/I$, and there exists
a bijection $\varsigma$ between conjugacy classes in the cosets.
More precisely, given a coset representative $\mathrm{w} \in N^*\!/N$,  the image $\varsigma(\mathrm{w})$ belongs to
$I^*\!/I$ while the discriminant forms verify \mbox{$q_I(\varsigma(\mathrm{w}))=-q_N(\mathrm{w})$}.
It follows \cite{Nikulin80int} that the lattices $N$ and $I$ can be glued to construct $\Gamma$ by adding correlated classes, i.e.  
\vs{-5}\be\label{GNI}
\Gamma = (N, I) + \coprod_{\mathrm{w} \in N^*\!\!/N} \hspace*{-2mm} (\mathrm{w}, \varsigma(\mathrm{w}))\ ,\vs{-5}
\ee
where $(\mathrm{w}, \varsigma(\mathrm{w}))$ are glue vectors that have even norm and integer scalar product with each other. 
Notice that to be an automorphism of $\Gamma$, $\Theta$ must preserve the conjugacy classes in $N^*/N$, since those in $I^*/I$, are 
invariant. 

When $n=2m$, with $m$ prime greater than 2, a thorough analysis of the decomposition of $\Gamma$ into the direct sum of 
$N$ and $I$ plus correlated classes was presented in \cite{Aldazabal:2025zht}. In this work we extend the results to the case
$n=3m$, with $m$ prime greater than 3, to obtain the general expression in eq. \eqref{Gamma3m}. For other $n$, a case by 
case characterization will be presented.

We also need to introduce the sublattices $I_\ell$ left invariant by $\Theta^\ell$, with $1\le\ell<n$.
Their orthogonal complements in $\Gamma$ are denoted $I_\ell^\perp$.
As before for $\ell=1$,
any vector $P\!\in\!\Gamma$ may be projected along $I_\ell$ and $I_\ell^\perp$  directions as
$(P_{I_\ell^\perp},P_{I_\ell})$, where $P_{I_\ell}\in I_\ell^*$ and $P_{I_\ell^\perp}\in I_\ell^{\perp*}$ \cite{Narain:1986qm}.
We have
\be\label{theta}
\Big(1+\Theta^\ell+\Theta^{2\ell}+\ldots+\Theta^{(\frac{n}{(\ell,n)}-1)\ell}\Big) (P_{I_\ell^\perp},P_{I_\ell})=
\Big(0,\frac{n}{(\ell,n)}P_{I_\ell}\Big)\in I_\ell\ ,
\ee
where $(\ell,n):=\text{gcd}(\ell,n)$.
Thus, $\frac{n}{(\ell,n)}P_{I_\ell}\!\in\! I_\ell$ for all $P_{I_\ell}\!\in\! I_\ell^*$, which in turn implies that $I_\ell^*/I_\ell$ 
generically contains $\bz_{q_i}$ conjugacy classes with $1\!\le\! i\!\le\!\xi$, where $\xi$ is the total 
number of divisors of $\frac{n}{(\ell,n)}$. 
Hence,\vs{-5}
\be
I_\ell^*/I_\ell\cong \bz_{q_1}^{\ell_{q_1}}\times\bz_{q_2}^{\ell_{q_2}}\times\ldots\times\bz_{q_\xi}^{\ell_{q_\xi}}
\ ,\qquad\ell_{q_i}\ge0\ ,\qquad q_i\Big|\frac{n}{(\ell,n)}\ .\vs{-5}
\ee
For instance, for $n=8$, $I_2$ generically contains $\bz_2$ and $\bz_4$ classes and $I_4$ contains only $\bz_2$ classes.
Notice also that for $\ell$ and $n$ coprime, i.e. $(\ell,n)=1$, $I_\ell=I_1=I$.

As will become clear, the number of invariant states in the orbifold spectrum ultimately depends
on degeneracy factors $\cF_\ell$ given by
\vs{-5}\be\label{fixpts}
\cF_\ell :=\sqrt{\frac{{\rm det}^{'}(1-\Theta^\ell)}{|I_\ell^*/I_\ell|}}\ ,\vs{-5}
\ee
where $\det'$ excludes eigenvalues equal to one. It has been shown that these 
degeneracy factors are integers \cite{Lepowsky, Narain:1986qm}. 

We are interested in theories with rank reduction, specially string islands without vector multiplets. In the heterotic setup 
a necessary condition for rank reduction is that the normal lattice $N$ does not have roots, i.e. vectors of length squared equal to 2. 
Following \cite{Aldazabal:2025zht}, to guarantee this condition we will choose $\Theta$ to be an element of the automorphism
group $\text{Co}_0$ of the Leech lattice $\Lambda$. 
Now, H\"ohn and Mason have shown that there are 290 orbits of invariant sublattices of $\Lambda$ under
$\text{Co}_0$, and they have also made available the corresponding normal sublattices \cite{HM}. 
As explained in \cite{Aldazabal:2025zht}, see also \cite{Baykara:2021ger}, even self-dual  
lattices of signature $(16+D, D)$, admitting $\bz_n$ automorphisms, and having normal sublattices without roots, can then be
constructed using the data provided in \cite{HM}. 

In the type II setup we will rely on a well-known class of even self-dual $(D,D)$ lattices that
admit purely left (or right) automorphisms \cite{Narain:1985jj, Ginsparg:1986bx}. 
These lattices have been considered in type II asymmetric orbifolds
with 16 supercharges by many authors, see e.g. \cite{Dixon:1987yp, Dabholkar:1998kv, Mizoguchi:2001cp, Baykara:2025lhl}.
In section \ref{section_typeii_story} we will review how the 
lattices depend on the choice of an ADE Lie algebra $\cG$, and have a group of left automorphisms equal to
the Weyl group of $\cG$. For $D=6$ there are only a few possibilities, further restricted because we
require all left-moving supersymmetries to be broken in order to keep only 16 supercharges.
Thus, it must be that $\Theta_L=\Theta$ lies in $SO(6)$ but not in $SU(3)$.
In practice we will first determine the possible eigenvalues of $\Theta$ that satisfy this latter condition, and then
identify which $\cG$'s admit Weyl group automorphisms having such eigenvalues.
To obtain islands all eigenvalues must be different from 1, i.e.  the invariant and normal lattices
must have signature $(0,6)$ and $(6,0)$ respectively.

\subsection{Partition function, modular invariance, and operator interpretation}\label{subsub_pf_mi_oi}

To determine the spectrum of invariant states and analyze the consistency of the orbifold conformal field theory
we will explicitly construct the partition function $\cz(\tau, \bar\tau)$. This is a fundamental object of the theory, entering in the
1-loop vacuum amplitude as
$Z_{\text{1-loop}}=\int_{\mathfrak F}\,  \frac{d^2\tau} {\tau_2^2} \, \cz(\tau, \bar\tau)$, where $\tau$ is the torus modular 
parameter, and $\mathfrak F$ is the fundamental region of $SL(2,\bz)$.
It follows that $\cz(\tau, \bar\tau)$ must be invariant under modular transformations with generators $S: \tau \to -1/\tau$
and $T: \tau \to \tau +1$. 

In the orbifolds preserving all right-moving supersymmetries the partition function can be written as
\be\label{pfgen}
\cz (\tau,\bar \tau) = \cz_X (\tau,\bar \tau)  \ \cz_{\bar\psi}(\bar\tau)  \ \cz_g(\tau, \bar\tau) \, .
\ee
Here $\cz_X$ and $\cz_{\bar\psi}$ arise respectively from the uncompactified light-cone world-sheet bosons and the 
right-moving world-sheet fermions, which are both completely inert under $g$. These standard pieces, see e.g. 
\cite{Blumenhagen:2013fgp}, are given by
\vs{-5}\be
\cz_X(\tau,\bar \tau) =\frac1{\left(\sqrt{\tau_2} \eta \bar\eta\right)^{8-D}}\ , \quad
\cz_{\bar\psi}(\bar\tau) = \frac1{2\op \bar\eta^4} (\bar\vartheta_3^4 - \bar\vartheta_4^4 - \bar\vartheta_2^4 + 
\bar\vartheta_1^4 ) =\frac{1}{\bar\eta^4} \left(\sum_{r \in V}-\sum_{r \in Sp}\right)\opm \bar q^{\frac12 r^2} \ ,
\label{zxzpsidef}
\ee
where $V$ and $Sp$ stand for the vector and spinor classes of $SO(8)$.
In contrast, $\cz_g(\tau, \bar\tau)$ is due to all world-sheet degrees of freedom that transform under $g$,
namely the internal bosons with momenta living on $\Gamma$, and in type II the left-moving fermions.
This $g$-dependent piece can be expanded as the double sum 
 \vs{-5}\be\label{fullzg}
 \cz_g(\tau,\bar \tau) = \sum_{\ell=0}^{n-1} \left[ \frac1{n} \sum_{k=0}^{n-1} Z(g^\ell, g^k) \right]\ .\vs{-5}
 \ee
Here $Z(g^\ell, g^k) = \text{Tr}_{\ch_\ell}\!\! \left(g^k \, q^{L_0} \, \bar q^{\bar L_0}\right)$, where $\ch_\ell$ is the Hilbert 
space of the $g^\ell$-twisted sector in which strings propagating on the torus close up to the action of $g^\ell$.
The untwisted sector is labelled by $\ell=0$.
The insertion of $g^k$ in the trace corresponds to twisted boundary conditions along the propagation direction
so that the sum over $k$ implements the projection on invariant states.
The double sum is necessary for modular invariance because modular transformations mix the 
boundary conditions along the two directions of strings propagating on the torus.

The purely untwisted term $Z(\uno,\uno)$ just comes from compactification on $\bt^D$. In the heterotic string
it takes the form
\vs{-5}\be\label{z00het}
Z(\uno,\uno)= 
\frac{1}{\eta^{16+D} \bar\eta^D}
\sum_{P\in \Gamma} q^{\frac{1}{2}P_L^2}\bar{q}^{\frac{1}{2}P_R^2}\, .\vs{-5}
\ee
In type IIB we instead have
\vs{-5}\be\label{z00typeii}
Z(\uno,\uno)= \Bigg\{\frac{1}{\eta^4}
\left(\sum_{p \in V}-\sum_{p \in Sp}\right)\opm  q^{\frac12 p^2} \Bigg\}
\Bigg\{ \frac{1}{\eta^D \bar\eta^D}
\sum_{P\in \Gamma} q^{\frac{1}{2}P_L^2}\bar{q}^{\frac{1}{2}P_R^2}\Bigg\}\, ,\vs{-5}
\ee
where the first factor is due to left-moving fermions.
The GSO projection yielding type IIA would be given by replacing the spinor class with the conjugate spinor in the sum.
Other untwisted terms $Z(\uno, g^k)$ can be computed since the action of $g$ on the known $\ch_0$ has been defined.
The explicit form will be presented shortly. For now we remark the implication, taking eq. \eqref{gactlatt} into account,
that the insertion of $g$ in the trace forces the lattice sum in $Z(\uno,g)$ to be over $P \in I$, 
and introduces a phase $e^{2 i \pi P\cdot v}$.

A hallmark of asymmetric orbifolds is that the $g$ action in the untwisted sector 
cannot be realized geometrically on the bosonic coordinates because it distinguishes between
left- and right-movers. This lack of geometric realization hinders a direct determination of the twisted Hilbert spaces
$\ch_\ell$. The alternative proposed in \cite{Narain:1986qm} is to extract $\ch_\ell$ from $Z(g^\ell, \uno)$,
which can be obtained by applying an $S$ transformation to $Z(\uno, g^\ell)$. Since this transformation involves Poisson 
resummation, it transpires that states e.g. in $\ch_1$ have momentum in $I^* + v$. We can then already 
draw the important consequence that a necessary condition to avoid massless states in the $g$-twisted sector is that
$v \notin I^*$. Similarly, we will show that a necessary condition for absence of massless states in $g^\ell$-twisted sectors
is that $\ell v \notin I_\ell^*$.

For a $\bz_n$ generator all terms $Z(g^\ell, g^k)$ with $\ell>0$ can be derived from untwisted sector terms $Z(\uno,g^k)$
by applying a chain of $S$ and $T$ modular transformations. The double sum in eq. \eqref{fullzg} then formally 
leads to a modular invariant partition function. However, there are still
consistency conditions to be verified. In particular, since we want $g^n$ to act as the identity, it must be that $T^n$ transforms
$Z(g^\ell,\uno)$ into itself. As we will see, the resulting level matching conditions (LMC) put restrictions on $g$, in particular
on the shift $v$.

A second consistency requirement is validity of the operator interpretation. This means that the $Z(g^\ell, g^k)$ obtained
by modular transformations must be understood as $\text{Tr}_{\ch_\ell}\!\! \left(g^k \, q^{L_0} \, \bar q^{\bar L_0}\right)$ in 
operator language. An essential requisite is that all $g^k$ act on the same $\ch_\ell$ determined from $Z(g^\ell, \uno)$. 
Therefore, for all $k$, the lattice sum in $Z(g^\ell, g^k)$ must be over momenta that are contained in $I_\ell^*$,
since this is the lattice appearing in $Z(g^\ell, \uno)$.
For coprime $g^\ell$-twisted sectors, i.e. with $(\ell,n)=1$, this condition is satisfied automatically because all $Z(g^\ell, g^k)$ 
are obtained from $Z(g^\ell, \uno)$ by $T$-transformations that do not change the momentum lattice.
When $(\ell, n)\not=1$ the condition is non-trivial and has to be verified for the different $T$-orbits of
$Z(g^\ell, g^k)$ terms connected by $T$-transformations.

A valid operator interpretation further demands a consistent group action. This is a twofold requirement on phases and degeneracy factors.
The phases arising from the $g^k$ action on states in $\ch_\ell$, which are read from $Z(g^\ell, g^k)$, must be the $k^{\rm{th}}$ 
power of the phases occurring when $k=1$. As for the degeneracy factors, the action of $g^k$ on $\cF_{\ell}$ in $Z(g^\ell,\uno)$ 
must yield the corresponding degeneracy factor in $Z(g^\ell,g^{k})$, where $1\!\le\! k\!<\!(\ell,n)$. 
In coprime twisted sectors, in which there is a single $T$-orbit, both conditions are fulfilled because there is a common degeneracy
factor $\cF_1$ and $T$-transformations give rise to proper phases.
Instead, consistency of the $g$-action in twisted sectors with multiple $T$-orbits may impose further constraints.

After establishing the consistency of the orbifold conformal field theory, we still need to determine the shifts $v$ satisfying
modular invariance. If we are interested in theories with rank reduction, or supersymmetry breaking, we also have
to ensure that the allowed shifts are such that massless states are absent in twisted sectors. In principle there might be
several $v$'s fulfilling all requirements but we want to determine the set of inequivalent ones, i.e. those that
imply different spectra of massive states. 
To tackle this problem, we will first determine the full partition function in our class of models. Based on the results, 
we will then further discuss the equivalence relation among shifts in appendix \ref{appv}.

\section{The heterotic story}\label{section_het_story}

In heterotic models the even self-dual lattice $\Gamma$ has signature ${(16+D,D)}$. We refer the reader to \cite[\S 2]{Aldazabal:2025zht}
for details about the general construction of $\Gamma$, and the choice of automorphism, based on the Leech lattice $\Lambda$.
Here we focus on $D=6$.  Below we will first recall the main facts relevant for realizing islands without vector 
multiplets in four dimensions. We will then present the classification of 4-dimensional heterotic islands using $\Lambda$. 
In subsection \ref{subsection_het_Z} we will lay out the structure of the partition function of the asymmetric orbifolds and 
derive the level-matching conditions (LMCs) imposed by modular invariance. 
Explicit expressions, constraints, and consistency analysis will be presented in section \ref{section_het} for specific classes 
of models.

\subsection{Lattices, automorphisms and classification}\label{subsection_het_class}

The starting point is a $\bz_n$ automorphism of $\Lambda$ with invariant and normal sublattices $\tilde I$ and $N$ 
of signature $(2,0)$ and $(22,0)$ respectively.  The basis for these lattices, as well as the possible orders $n$, are given in \cite{HM}.
Next, as explained in \cite{Baykara:2021ger}, 
the invariant lattice $I$ of signature $(0,6)$, which is to be correlated with $N$ to form $\Gamma$, can be obtained by first embedding
$\tilde I$ primitively in $E_8$, then finding its orthogonal complement in $E_8$, and finally reversing the signature.
In practice we use the function \texttt{E8SublatticeOrth} defined in the \texttt{Mathematica} package of \cite{Baykara:2021ger} to 
obtain $I$, and then verify that indeed it can be glued with $N$ into $\Gamma$ of signature $(22,6)$.
The genus of this $I$ may contain several non-isometric lattices that can be glued to the same $N$, giving rise to distinct 
$\Gamma$'s  admitting the same automorphism.

There exist 72 orbits of non-trivial fixed-point sublattices of $\Lambda$ under cyclic subgroups of  $\text{Co}_0$.
They have been studied in \cite{HL}, and are a subset of the more general orbits classified in \cite{HM}. 
In Table \ref{HMisl} we collect all the possibilities in which the invariant sublattice $\tilde I$ has rank 2. 
The HM label refers to the entry number in \cite[Table 1]{HM}. The HL label in parentheses identifies the conjugacy class
of the $\bz_n$ generator, as given in \cite[Table 1]{HL}.
In all these cases an even self-dual lattice with signature $(22,6)$ can be assembled. Moreover, the $\bz_n$
automorphism can be identified. Concretely, since $N$ is known, using Magma \cite{magma} we can look for automorphisms of the 
given order that rotate all the 22 directions in $N$ and preserve the conjugacy classes of $N^*/N$. 
The generator $\Theta$ has eigenvalues $e^{\pm 2\pi i t_a}$, $a=1, \ldots, 11$,  with $0 < t_a \leq \frac12$.
In Table \ref{HMisl} the 11 exponents $t_a$ are displayed as components of a vector $t=(t_1^{n_1},t_2^{n_2},\ldots )$, 
where exponents indicate multiplicity. The eigenvalues can also be determined from the frame shape, listed in \cite[Table1]{HL}, 
associated to the conjugacy class. Notice that some HM labels are repeated. This means that the corresponding
$\Gamma$ admits different $\bz_n$ automorphisms. For instance, the $\Gamma$ obtained from HM222 allows for 
$\bz_6$, $\bz_{18}$, and two $\bz_{12}$, distinct automorphisms.
Four of these models appeared in \cite{Chaudhuri:1995ee}, concretely those labeled HM271, HM257, HM240, and HM227. The model HM251 was found and analyzed in \cite{Aldazabal:2025zht}.
For many cases in Table \ref{HMisl}, the genus of the $I$ lattice contains more than one isometry class. The exceptions are
the HM labels 222-224, 226, 227, 230-232.

Summarizing, for each entry in Table \ref{HMisl} we have one, or several, (22,6) lattices $\Gamma$ admitting the 
$\bz_n$ automorphism with the given non-trivial eigenvalues.
By construction, the corresponding $I$ sublattice is $(0,6)$ whereas the $N$ sublattice is
$(22,0)$ and has no roots. Thus, in each case we can build a heterotic $\bt^6/\bz_n$ asymmetric orbifold
to obtain an island {\it candidate} in four dimensions. In all cases we find that there exist shift vectors $v$ 
that satisfy modular invariance and eliminate all massless states in the twisted sectors.  This is a necessary but not sufficient 
condition for the model to be an island. The full consistency of the world-sheet conformal field theory, including validity of the operator interpretation, must be established to prove that the orbifold does give rise to an island.
For $n$ prime, our arguments in section \ref{subsub_pf_mi_oi} imply that this is always the case. For $n=2m$, with $m$ prime 
greater than 2,  a set of general constraints required for proper operator interpretation was derived in \cite{Aldazabal:2025zht}.  
In this work we obtain similar results for $n=3m$, with $m$ prime greater than 3. All models in Table \ref{HMisl} that fall into 
one of these three classes, as well as the outliers with $n=8$ and $n=9$, will be discussed further in section 
\ref{section_het}. In the current absence of general proof of consistency we leave the remaining models in
Table \ref{HMisl} as conjectured members of a complete catalog of Leech type 4-dimensional heterotic islands.

\begin{table}[h!]\begin{center}
\setlength{\extrarowheight}{-10pt}
\renewcommand{\arraystretch}{.8}
\begin{tabular}{|c|c|c|c|c|c|c|c|c|c|c|c|c|c|c|c|c|c|c|c|c|c|c|c|c|c|c|c|c|c|c|}
\hline
\multirow{2}{*}{\raisebox{-1ex}{$\bz_n$}} &\multicolumn{30}{c|}{ \multirow{2}{*}{\tsm{HM, HL labels}}}\\*[3mm]
{} & \multicolumn{30}{c|}{\msm{\Theta} \tsm{eigenvalues}}\\[1mm]\hline\hline
\multirow{2}{*}{\raisebox{-1ex}{*$\bz_6$}} &\multicolumn{30}{c|}{ \multirow{2}{*}{\tsc{HM222, HL(\msc{-6_F})}}}\\*[3mm]
{} & \multicolumn{30}{c|}{\msc{\frac16(1^6,2^3,3^2)}}\\[1mm]\hline
\multirow{2}{*}{\raisebox{-1ex}{*$\bz_8$}} &\multicolumn{30}{c|}{ \multirow{2}{*}{\tsc{HM224, HL(\msc{8_D})}}}\\*[3mm]
{} & \multicolumn{30}{c|}{\msc{\frac18(1^4,2^2,3^4,4)}}\\[1mm]\hline
\multirow{2}{*}{\raisebox{-1ex}{*$\bz_9$}} &\multicolumn{30}{c|}{ \multirow{2}{*}{\tsc{HM230, HL(\msc{9_B})}}}\\*[3mm]
{} & \multicolumn{30}{c|}{\msc{\frac19(1^3,2^3,3^2,4^3)}}\\[1mm]\hline
\multirow{2}{*}{\raisebox{-1ex}{$\bz_{12}$}} 
&\multicolumn{6}{c|}{\multirow{2}{*}{\tsc{HM222, HL(\msc{-12_D})}} }  
&\multicolumn{6}{c|}{\multirow{2}{*}{\tsc{HM231, HL(\msc{12_G})}} }
&\multicolumn{6}{c|}{\multirow{2}{*}{\tsc{HM228, HL(\msc{12_H})}}}  
&\multicolumn{6}{c|}{\multirow{2}{*}{\tsc{HM222, HL(\msc{-12_K})}} } 
&\multicolumn{6}{c|}{\multirow{2}{*}{\tsc{HM271, HL(\msc{12_M})}} }
 \\*[3mm]
{} 
& \multicolumn{6}{c|}{\msc{\frac1{12}(1^3,3^2,4^3,5^3)} }
& \multicolumn{6}{c|}{ \msc{\frac1{12}(1^2,2,3^4,4,5^2,6)}  }
& \multicolumn{6}{c|}{ \msc{\frac1{12}(1^2,2^3,4^2,5^2,6^2)} }
&\multicolumn{6}{c|}{ \msc{\frac1{12}(1^3,2^2,3^2,4,5^3)} }
& \multicolumn{6}{c|}{\msc{\frac1{12}(1^2,2^2,3^2,4^2,5^2,6)} }
\\[1mm]\hline
\multirow{2}{*}{\raisebox{-1ex}{*$\bz_{14}$}} &\multicolumn{30}{c|}{ \multirow{2}{*}{\tsc{HM232, HL(\msc{-14_B})}}}\\*[3mm]
{} & \multicolumn{30}{c|}{\msc{\frac1{14}(1^2,2,3^2,4,5^2,6,7^2)}}\\[1mm]\hline
\multirow{2}{*}{\raisebox{-1ex}{*$\bz_{15}$}} &\multicolumn{30}{c|}{ \multirow{2}{*}{\tsc{HM223, HL(\msc{15_E})}}}\\*[3mm]
{} & \multicolumn{30}{c|}{\msc{\frac1{15}(1^2,2^2,3,4^2,5,6,7^2)}}\\[1mm]\hline
\multirow{2}{*}{\raisebox{-1ex}{$\bz_{16}$}} 
&\multicolumn{15}{c|}{\multirow{2}{*}{\hs{50}\tsc{HM226, HL(\msc{16_A})}\hs{50}} }  
&\multicolumn{15}{c|}{\multirow{2}{*}{\tsc{HM224, HL(\msc{-16_B})}} }
 \\*[3mm]
{} 
& \multicolumn{15}{c|}{\msc{\hs{50}\frac1{16}(1^2,2,3^2,5^2,6,7^2,8)}\hs{50} }
& \multicolumn{15}{c|}{ \msc{\frac1{16}(1^2,2,3^2,4,5^2,6,7^2)}  }
\\[1mm]\hline
\multirow{2}{*}{\raisebox{-1ex}{$\bz_{18}$}} 
&\multicolumn{10}{c|}{\multirow{2}{*}{\tsc{HM230, HL(\msc{-18_B})}} }  
&\multicolumn{10}{c|}{\multirow{2}{*}{\tsc{HM222, HL(\msc{18_C})}} }
&\multicolumn{10}{c|}{\multirow{2}{*}{\tsc{HM245, HL(\msc{-18_C})}}}  
 \\*[3mm]
{} 
& \multicolumn{10}{c|}{\msc{\frac1{18}(1,2^2,3,4^2,5,6,7,8^2)} }
& \multicolumn{10}{c|}{ \msc{\frac1{18}(1^2,2,3,4,5^2,7^2,8,9)}  }
& \multicolumn{10}{c|}{ \msc{\frac1{18}(1,2^2,4^2,5,6,7,8^2,9)} }
\\[1mm]\hline
\multirow{2}{*}{\raisebox{-1ex}{$\bz_{20}$}} 
&\multicolumn{10}{c|}{\multirow{2}{*}{\tsc{HM257, HL(\msc{20_B})}} }  
&\multicolumn{10}{c|}{\multirow{2}{*}{\tsc{HM225, HL(\msc{20_C})}} }
&\multicolumn{10}{c|}{\multirow{2}{*}{\tsc{HM262, HL(\msc{-20_C})}}}  
 \\*[3mm]
{} 
& \multicolumn{10}{c|}{\msc{\frac1{20}(1,2,3,4,5^2,6,7,8,9,10)} }
& \multicolumn{10}{c|}{ \msc{\frac1{20}(1,2^2,3,4,6^2,7,8,9,10)}  }
& \multicolumn{10}{c|}{ \msc{\frac1{20}(1,2,3,4^2,6,7,8^2,9,10)} }
\\[1mm]\hline
\multirow{2}{*}{\raisebox{-1ex}{*$\bz_{21}$}} 
&\multicolumn{30}{c|}{ \multirow{2}{*}{\tsc{HM240, HL(\msc{21_C})}}}\\*[3mm]
{} & \multicolumn{30}{c|}{\msc{\frac1{21}(1,2,3,4,5,6,7^2,8,9,10)}}\\[1mm]\hline
\multirow{2}{*}{\raisebox{-1ex}{*$\bz_{22}$}} &\multicolumn{30}{c|}{ \multirow{2}{*}{\tsc{HM251, HL(\msc{22_A})}}}\\*[3mm]
{} & \multicolumn{30}{c|}{\msc{\frac1{22}(1,2,3,4,5,6,7,8,9,10,11)}}\\[1mm]\hline
\multirow{2}{*}{\raisebox{-1ex}{*$\bz_{23}$}} &\multicolumn{30}{c|}{ \multirow{2}{*}{\tsc{HM227, HL(\msc{23_A})}}}\\*[3mm]
{} & \multicolumn{30}{c|}{\msc{\frac1{23}(1,2,3,4,5,6,7,8,9,10,11)}}\\[1mm]\hline
\multirow{2}{*}{\raisebox{-1ex}{$\bz_{24}$}} 
&\multicolumn{10}{c|}{\multirow{2}{*}{\tsc{HM234, HL(\msc{24_E})}} }  
&\multicolumn{10}{c|}{\multirow{2}{*}{\tsc{HM229, HL(\msc{24_F})}} }
&\multicolumn{10}{c|}{\multirow{2}{*}{\tsc{HM253, HL(\msc{-24_F})}}}  
 \\*[3mm]
{} 
& \multicolumn{10}{c|}{\msc{\frac1{24}(1,3^2,4,5,7,8,9^2,11,12)} }
& \multicolumn{10}{c|}{ \msc{\frac1{24}(1,2,4^2,5,6,7,8,10,11,12)}  }
& \multicolumn{10}{c|}{ \msc{\frac1{24}(1,2,4,5,6,7,8^2,10,11,12)} }
\\[1mm]\hline
\multirow{2}{*}{\raisebox{-1ex}{$\bz_{28}$}} &\multicolumn{30}{c|}{ \multirow{2}{*}{\tsc{HM232, HL(\msc{-28_A})}}}\\*[3mm]
{} & \multicolumn{30}{c|}{\msc{\frac1{28}(1,3,4,5,7^2,8,9,11,12,13)}}\\[1mm]\hline
\multirow{2}{*}{\raisebox{-1ex}{$\bz_{30}$}} 
&\multicolumn{10}{c|}{\multirow{2}{*}{\tsc{HM237, HL(\msc{-30_D})}} }  
&\multicolumn{10}{c|}{\multirow{2}{*}{\tsc{HM246, HL(\msc{30_D})}} }
&\multicolumn{10}{c|}{\multirow{2}{*}{\tsc{HM223, HL(\msc{-30_E})}}}  
 \\*[3mm]
{} 
& \multicolumn{10}{c|}{\msc{\frac1{30}(1,3,5,6,7,9,10,11,12,13,15)} }
& \multicolumn{10}{c|}{ \msc{\frac1{30}(2,3,4,5,6,9,10,12,14,15)}  }
& \multicolumn{10}{c|}{ \msc{\frac1{30}(1,2,4,6,7,8,10,11,12,13,14)} }
\\[1mm]\hline
\end{tabular}
\caption{Heterotic island candidates $\bt^6/\bz_n$ with 16 supercharges constructed using sublattices of the Leech lattice. 
Models indicated with asterisks are those for which explicit construction of the orbifolds is provided in section \ref{section_het}, showing that the resulting theories are islands.
}\label{HMisl}
\end{center}\end{table}

\subsection{Partition function and LMC}\label{subsection_het_Z}

In the untwisted sector we start with eq. \eqref{z00het}, with $D=6$, to obtain the contributions\vs{-5}
\be\vs{-5}\label{hz1gk}
Z(\uno,g^\ell)\!=\! \frac1{\eta^{22} \bar\eta^{6}}\!\! \prod_{\substack{a=1 \\ \ell t_a \notin\bz}}^{11}
\frac{2 \sin (\pi  \ell t_a)\,  \eta^3}{\vtgen{\frac12}{\frac12 - \ell t_a} }\!
\sum_{P\, \in \, I_\ell} q^{\frac12 P_L^2} \bar q^{\frac12 P_R^2} \, e^{2\pi i P \cdot \ell v} \ ,
\ee
where $1\le\ell<n$. 
Throughout the text, scalar products are with Lorentzian metric, unless the subscripts $L$ or $R$ are explicitly 
written. For example, $P^2=P_L^2-P_R^2$, whereas \mbox{$P\cdot v =P_L \cdot v_L - P_R\cdot v_R$}.

In eq. \eqref{hz1gk} we have used that the action of $g$ on momentum states is given by eq. \eqref{gactlatt}.
The action of $g^{2j}$, for $j > 0$,  may have to be modified for some models with even orders $n$, to guarantee modular invariance and consistency of the even twisted sectors. For instance, in the $\bt^D/\bz_{2m}$ models with $m$ odd prime, the action takes the form\vs{-5}
\be\label{gactlatt2}
g^{2j}|P_N, P_I \rangle = e^{2\pi i  P \cdot j(2v+w)} |\Theta P_N, P_I \rangle\ ,\vs{-5}
\ee 
where $w$ is a vector in a $\bz_2$ class of $I^*/I$ such that $e^{2\pi iP^2}=e^{2\pi i P\cdot w}$ for all $P\in I^*$ -- see \cite[\S 3.1]{Aldazabal:2025zht}. For now, we schematically keep the vector $\ell v$ in the general expression \eqref{hz1gk}, and will analyze $w$ type vectors and potential modifications to the $g^{2j}$-action in each even order model we construct in section \ref{section_het}.

The partition function $Z(g^\ell,g^{k\ell})$ can be obtained from $T^kS$ modular transformation of $Z(\uno,g^\ell)$. In this
way we find
\vs{-5}\be\vs{-5}\begin{split}\label{hzglgkl}
\hs{-5}Z(g^\ell,g^{k\ell})& =\frac{\cF_\ell \op e^{i\pi \mu}}{\eta^{22} \bar\eta^{6}}\, 
\prod_{\substack{a=1\\\ell t_a\notin\bz}}^{11} \frac{ e^{i\pi\big(\frac12 -\ell  t_a+k(\ell t_a)^2\big)}\, \eta^3}
{\vtgen{\, \frac12-\ell t_a}{\frac12-k\ell t_a} }\sum_{P\, \in \, I_\ell^*} 
q^{\frac12 (P+\ell v)_L^2}\bar q^{\frac12 (P+\ell v)_R^2} e^{i\pi k (P+\ell v)^2}\\[-2pt]
&= \frac{\cF_\ell \op e^{i \pi 2kE_\ell}}{\eta^{22} \bar\eta^{6}}
\prod_{\substack{a=1\\\ell t_a\notin\bz}}^{11}
\frac{ \eta^3}{\hvtgen{ \frac12-\widetilde{\ell t_a}}{\frac12 -k\,\widetilde{\ell t_a}}}
\sum_{P\, \in \, I_\ell^*}q^{\frac12 (P+\ell v)_L^2}\bar q^{\frac12 (P+\ell v)_R^2}\op e^{i\pi k(P+\ell v)^2} \ ,\quad1\le\ell<n\ ,
\end{split}\ee
where in the second line we used 
\vs{-5}\be\label{that}
\hvtgen{\,\alpha}{\beta} := e^{-2 i\pi  \alpha \beta} \vtgen{\,\alpha}{\beta}\ ,\vs{-5}
\ee
which has a $q$-series expansion starting with 1. We also introduced the zero point energies $E_\ell$ defined as
\vs{-5}
\be\label{Fs}
E_\ell:=\frac12 \sum\limits_{\substack{a=1 \\ \ell t_a \notin\bz}}^{D/2} \widetilde{\ell t_a} (1 - \widetilde{\ell t_a}) \ ,\qquad
\widetilde{\ell t_a} := \ell t_a - \lfloor \ell t_a \rfloor \, .\vs{-5}
\ee
Notice that $0<\widetilde{\ell t_a} < 1$.
The degeneracy factor $\cF_\ell$ is given in eq. \eqref{fixpts}. The phase $e^{i\pi \mu}$ arises from
\vs{-5}\be\label{mu}
\prod_{a=1}^{D/2}  2\sin (\pi \ell t_a) = e^{i\pi \mu} \, 
\bigg|\prod_{\substack{a=1 \\ \ell t_a \notin\bz}}^{D/2}  2\sin (\pi\ell t_a) \bigg| =
e^{i\pi \mu}\, \sqrt{{\det}^{'}(1-\Theta^\ell)}\ ,\qquad
e^{i\pi \mu}:=\prod_{\substack{a=1 \\ \ell t_a \notin\bz}}^{D/2} e^{i\pi \lfloor\ell t_a\rfloor} \, . \vs{-10}
\ee
Clearly $e^{2\pi i\mu}=1$. 

If $(\ell,n)\ne1$, the twisted sector has more than one orbit and the partition functions $Z(g^\ell,g^{k\ell-j})$, $0<j<\ell$, may be obtained from a chain of $S$ and $T$ transformations starting form the untwisted sector. These orbits will be computed for each model individually in section \ref{section_het}.

To compute the LMC, and determine the conditions for absence of twisted massless states,
it is convenient to categorize the theories according to the order $n$.
We will next examine each category separately.

\subsubsection{$n$ prime}\label{N1}
The partition function in all twisted sectors is given by eq. \eqref{hzglgkl}, where $n v\!\in\! I$, $\cF_\ell\!=\!\cF_1$, $E_\ell=E_1$, and $I_\ell=I$ for all $1\le\ell<n$. In the $g$-twisted sector, the condition $Z(g,g^n)\!=\!Z(g,\uno)$ gives\vs{-5}
\be\label{lmc_g}
nP^2+2P\cdot nv+nv^2+2nE_1\in2\bz\ ,\qquad\forall\, P\in I^*\ .\vs{-5}
\ee
Suppose $j$ is an odd number such that $jP\in I$ for all $P$ in $\bz_j$ conjugacy classes in $I^*/I$. Then, $j^2P^2\!\in\!2\bz$ and $jP^2\!\in\!\bz$. Since $j$ is odd, $jP^2\!\in\!2\bz$. Thus, in the present case, $nP^2\!\in\!2\bz$. Furthermore, $P\cdot nv\!\in\!\bz$ since $nv\in I$. Hence, we obtain the LMC\vs{-5}
\be\label{lmc_g_i}
nv \in I\ ,\qquad\qquad nv^2+2nE_1\in2\bz\ .\vs{-5}
\ee
To avoid massless states we require $v\notin I^*$. The only prime order island in four dimensions is the $\bt^6/\bz_{23}$ asymmetric orbifold -- see section \ref{hm227}.

\subsubsection{$n=2m$, $m > 2$ prime}\label{N2}
The full partition function of these models is computed in \cite[\S 3]{Aldazabal:2025zht}. The LMCs in all twisted are derived in that reference and are of the form
\vs{-5}\be\label{lmc_g2mp}
2mv + w \in I\ ,\qquad\qquad 2mv^2+4mE_1\in2\bz\ ,\qquad\qquad m(2v+w)^2+2mE_2\in2\bz\ ,\vs{-5}
\ee
where $w\in\bz_2$ class of $I^*\!/I$. Removing massless states in the even twisted sectors requires $2v\notin I_2^*$.

There are three island candidates of this type: HM222 $(\bz_6)$, HM232 $(\bz_{14})$, and HM251 $(\bz_{22})$. The HM251 model was studied in \cite[\S5]{Aldazabal:2025zht} and is in fact a populated archipelago of islands. For the other two models HM222 and HM232, $2P^2\in2\bz$ for all $P\in\bz_2$ classes of $I^*$. As such, $w$ may be set to zero and so $2mv\in I$. The LMCs \eqref{lmc_g2mp} then simplify to
\vs{-5}\be\label{lmc_g2mp_ii}
2mv\in I\ ,\qquad\qquad2mv^2+4mE_1\in2\bz\ ,\qquad\qquad 2m(E_2-4E_1)\in2\bz\ .\vs{-5}
\ee
In sections \ref{hm222} and \ref{hm232} we will show that both models are indeed islands.

\subsubsection{$n=3m$, $m > 3$ prime}\label{N3}
In the coprime sectors we find $3mv^2+6mE_1\in2\bz$, where we used $3mP^2\in2\bz$, $\forall P\in I^*$, and $3mv\in I$. 
The LMC in the $g^3$ and $g^m$ sectors is respectively  $2(mE_3-3NE_1)\!\in\!2\bz$ and $2(3E_m-mNE_1)\!\in\!2\bz$. 
Here we used $mP^2\!\in\!2\bz$, $\forall P\!\in\! I_3^*$, and $3P^2\!\in\!2\bz$, $\forall P\!\in\! I_m^*$. 
This is because, as we will show in subsection \ref{order3m}, $I_3$ has uncorrelated $\bz_m$ classes and $I_m$ 
has uncorrelated $\bz_3$ classes. All in all, we obtain the conditions
\vs{-5}\be\label{lmc_3m}
3mv\in I\ ,\qquad Nv^2+2NE_1\in2\bz\ ,\qquad(mE_3-3NE_1)\in2\bz\ ,\qquad(3E_m-mNE_1)\in2\bz \ .\vs{-5}
\ee
To eliminate massless states we further require $v\notin I^*$, $3v\notin I_3^*$, and $mv\notin I_m^*$. 

As observed in Table \ref{HMisl}, there are two models of this type, namely HM223 ($\bz_{15}$) and HM240 ($\bz_{21}$), and they will be analyzed in sections \ref{hm223} and \ref{hm240}.

\subsubsection{$n=p^2$, $p$ prime}\label{N4}
The only island candidate is HM230 which has order $n=9$.  The LMC in the coprime and the $g^{3j}$ sectors is
\vs{-5}\be\label{lmc_9}
9v\in I\ ,\qquad\qquad9v^2+18E_1\in2\bz\ ,\qquad\qquad6(E_3-9E_1)\in2\bz\ .\vs{-5}
\ee
Requiring $v\notin I^*$ and $3v\notin I_3^*$ eliminates massless states in the twisted sectors. 
In section \ref{hm230} we will show that HM230 is an archipelago of 3 islands.

\subsubsection{$n=2m$, $m$ nonprime}\label{N5}
As observed in Table \ref{HMisl}, these models have orders $n\!=\!\{8,12,16,18,20,24,28,30\}$. In coprime sectors, which contain a single orbit, if $n P^2\!\in\!2\bz$, $\forall P\!\in\! I^*$, then $w$ may be set to zero, $nv\in I$, and the LMC \ref{lmc_g_i} holds. On the other hand, if $n P^2$ is odd for some $P\in I^*$, then one has to examine whether there exists a vector $w$ ($2w\!\in\! I$) that maintains modular invariance. Furthermore, the LMC in non-coprime twisted sectors has to be examined for more potential constrains. In section \ref{hm224} we will   show that the HM224 model is a $\bt^6/\bz_8$ island. The $\bt^6/\bz_{16}$ models HM224 and HM226 are quite similar to the $\bz_8$ model, and so are very likely to be islands as well. It would be very interesting to provide a general proof that all HM models of this type are islands. We leave this for future work.

\section{Classes of heterotic islands}\label{section_het}

In this section we analyze heterotic $\bt^6/\bz_n$ asymmetric orbifolds for some entries in Table \ref{HMisl}. We show that there exist 
shift vectors that satisfy the LMCs and remove all massless states in twisted sectors. Furthermore, we show that operator 
interpretation in all twisted sectors is consistent. Hence, each model in this section constitutes an island or an archipelago of islands. 
The latter occurs when there are several inequivalent shifts, according to the equivalence relation discussed in app. \ref{appv}.

\subsection{Prime order}\label{hm227}

The $\bz_{23}$ HM227 is the only prime-order model leading to a heterotic island candidate. It has the following properties \vs{-5}
\be\label{hm227dat}
t=\frac1{23}(1,2,3,4,5,6,7,8,9,10,11)\ ,\qquad E_1=\frac{22}{23}\ ,\qquad
I^*/I\cong \bz_{23} \ ,\qquad\cF_1=1\ .
\ee
The LMC \eqref{lmc_g_i} and the requirement of eliminating twisted sector moduli give\vs{-5}
\be\label{lmc227g}
23v^2\in2\bz\ ,\qquad\qquad 23v\in I\ ,\qquad v\notin I^*\ ,\vs{-10}
\ee
which have 84 equivalence classes of solutions. For instance, $\frac{1}{23}(2,1,1,1,3,2)$ is a representative of one class. 
Here and throughout this section, we express the vectors in the basis of the $I$ lattice. Since all the twisted sectors are coprime, 
the operator interpretation is automatically satisfied. HM227 is therefore an archipelago of many islands.

\subsection{Order $n=2m$}\label{order2m}

The general framework in \cite{Aldazabal:2025zht} was adapted for the $\bz_{2m}$ models with $m\!>\!2$ prime, and particular HM models were constructed: two $\bt^4/\bz_{10}$ islands in six dimensions (HM149) and an archipelago of $\bt^6/\bz_{22}$ islands in four dimensions (HM251). A consistent group action on the degeneracy factors was also defined for these models and a set of constraints for a consistent operator interpretation was derived. There are two more HM island candidates of this type in four dimensions, namely the $\bt^6/\bz_6$ (HM222) and the $\bt^6/\bz_{14}$ (HM232) models. The group action defined in \cite{Aldazabal:2025zht} applies to the HM232 model. For the HM222 model, as we will presently see, a different group action on the degeneracy factors must be defined  to obtain a consistent conformal field theory.

\subsubsection{\texorpdfstring{$\bz_6$}{TEXT} HM222}\label{hm222}

The model has the following properties:
\vs{-5}\be\vs{-5}\begin{split}\label{hm222dat}
&t=\frac1{6}(1,1,1,1,1,1,2,2,2,3,3)\ ,\qquad\qquad\; E_1=E_2=E_3=1\ ,\\
&I^*/I \cong \bz_2\times\bz_{6} \cong \bz_2^2\times\bz_{3}\ ,
\qquad\qquad I_2^*/I_2\cong\bz_3^5\ ,\qquad\qquad I_3^*/I_3\cong\bz_2^{8}\ ,\\
&\cF_1=6\ ,\qquad \cF_2=9\ ,\qquad \cF_3=16\ .
\end{split}\ee
Since $2P^2\in2\bz$ for all $P\in\bz_2$ classes of $I^*/I$, the vector $w$ may be set to zero. The LMC \eqref{lmc_g2mp_ii}, together with the absence of massless states in the twisted sectors, require
\vs{-8}\be\label{lmc222}
6v^2\in2\bz\ ,\qquad\qquad 6v\in I\ ,\qquad v\notin I^*\ ,\qquad
2v\notin I_2^*\ ,\qquad 3v\notin I_3^*\ .\vs{-7}
\ee
There are two equivalence classes of shift vectors which satisfy these conditions. Representatives of each class are $\frac{1}{6}( 1 , 0 , 1 , 0 , 1 , 1)$ and $\frac{1}{6}( 1 , 1 , 1 , 0 , 1 , 0) $.

\noindent\underline{\bf{Operator interpretation}}\\[2pt]
It is proven in \cite{Aldazabal:2025zht} that the lattice sums in the orbits $(g^{2j},g)$ and $(g^3,g^k)$, where $j,k=1,2$, are subsets of respectively $I_2^*$ and $I_3^*$. We next verify the group action on the degeneracy factors and phases. The degeneracy factors in the orbits $(g^{2j},\uno)$ and $(g^{2j},g)$ are respectively $\cF_2=9$ and $\cF_2'=3$. The $g$-action is the same as in \cite[\S4.1]{Aldazabal:2025zht}, namely $g$ acts on 6 degenerate states in $\cF_2$ as $+1$ and on 3 states as $-1$. This guarantees that  operator interpretation in the $g^2$ and $g^4$ sectors is consistent.

The group action in the $g^3$ sector is slightly different. The second line in eq. \eqref{hzglgkl} shows that the degeneracy factor and the overall constant phase in $Z(g^3,\uno)$ are respectively $\cF_3\!=\!16$ and $1$. By overall constant phase we mean the overall phase factor independent of momenta $P\in I_3^*$ and shift vector $v$. For the orbits $(g^3,g^k)$, $k\!=\!1,2$, it turns out that the degeneracy factor and overall constant phase are respectively $\cF_3'\!=\!2$ and $e^{i\pi\chi_3}\!=\!-1$ for both values of $k$ -- see \cite[eq. (3.48)]{Aldazabal:2025zht}. This 
indicates that $g$ must act on 6 degenerate states in $\cF_3$ as $e^{\frac{2\pi i}3}$, on 6 states as $e^{-\frac{2\pi i}3}$, and trivially on the remaining 4 states. This yields $g\cF_3\!=\!g^2\cF_3\!=\!(6e^{\frac{2\pi i}3}+6e^{-\frac{2\pi i}3}+4)\!=\!-2=e^{i\pi\chi_3}\cF_3'$ and $g^3\cF_3=\cF_3$, as expected. It is then straightforward to show that the $g$-action on the remaining degrees of freedom, determined by phases which depend on $P$ and $v$, is consistent. That is to say, the phase of a state in $Z(g^3,g^3)$ is equal to the third power of the phase of that state in $Z(g^3,g)$. The same result holds for the other orbit, and for the $g^6$-twisted sector.  We conclude that the HM222 $\bz_6$ model has a valid operator interpretation and thus, is an archipelago of two islands. Finally, we note that $v=0$ is a solution to the LMC for which there exist massless states in all twisted sectors which together fill 22 vector multiplets.

\subsubsection{\texorpdfstring{$\bz_{14}$}{TEXT} HM232}\label{hm232}

For all $P\in\bz_2$ classes of $I^*/I$, $2P^2\in2\bz$, and so $w=0$ is a solution which we choose. The LMC from even twisted sectors does not impose extra constraints. All the operator interpretation conditions are of the form derived in \cite{Aldazabal:2025zht} and are satisfied. We find\vs{-5}
\be\vs{-5}\begin{split}\label{hm232dat}
&t=\frac1{14}(1,1,2,3,3,4,5,5,6,7,7)\ ,\qquad\; E_1=1\ ,\qquad E_2=\frac67\ ,\qquad E_7=1\ ,\\[2pt]
&I^*/I \cong \bz_2\times\bz_{14} \cong \bz_2^2\times\bz_{7}\ ,
\qquad I_2^*/I_2\cong\bz_7^3\ ,\qquad I_7^*/I_7\cong\bz_2^{8}\ ,\\[2pt]
&\cF_1=2\ ,\qquad \cF_2=1\ ,\qquad \cF_7=16\ .
\end{split}\ee
The LMC and absence of massless states in the twisted sectors requires
\vs{-5}\be\label{lmc232}
14v^2\in2\bz\ ,\qquad\qquad 14v\in I\ ,\qquad v\notin I^*\ ,\qquad
2v\notin I^*_2\ ,\qquad 7v\notin I^*_7\ ,\vs{-5}
\ee
and there exist 29 equivalence classes of $v$ satisfying these conditions. Some examples are $\frac{1}{14}( 1 , 2 , 7 , 0 , 0 , 0 )$, $\frac{1}{14}( 0 , 5 , 1 , 2 , 2 , 2)$, and $\frac{1}{14}(3,1,1,1,1,-2)$. HM232 is therefore a populated $\bt^6/\bz_{14}$ archipelago.

\subsubsection{\texorpdfstring{$\bz_{22}$}{TEXT} HM251}\label{hm251}

This model was analyzed in \cite[\S5.2]{Aldazabal:2025zht}, where the reader is referred to for the details. Since there are vectors in 
$\bz_2$ classes of $I^*$ with half-integer length squared, the model has a non-trivial vector $w=\frac12(1,1,0,1,1,0)$. 
It was then shown that HM251 has a consistent operator interpretation and forms an archipelago of 232 islands.

In this example there exists a lattice non-isometric to $I$ that can be used to form a different $\Gamma(22,6)$ admitting the 
same $\bz_{22}$ automorphism. It can be shown that the corresponding asymmetric $\bt^6/\bz_{22}$ orbifold is consistent
and allows 169 modular invariant non-equivalent shifts that eliminate all massless states in twisted sectors.

\subsection{Order $n=3m$}\label{order3m}

Since $(3,m)=1$, the $\bt^D/\bz_{3m}$ asymmetric orbifolds have similar characteristics to those of the $\bt^D/\bz_{2m}$ models studied in \cite{Aldazabal:2025zht}. Below we first analyze the structure of the even self-dual lattice $\Gamma$, the invariant lattices $I_3$ and $I_m$, and their duals. We will then examine the two HM models of this type, namely $\bt^6/\bz_{15}$ (HM223) and $\bt^6/\bz_{21}$ (HM240). In Appendix \ref{app_Z3m}, we compute the partition function -- the final expressions are given in eqs. \eqref{Zgjglj_h3m}, \eqref{hetz3ml3}, and \eqref{hetz3mlm}. Validity of operator interpretation for the two HM models are verified in section \ref{Z3mop}.

To begin, we note that the generic discriminant group of the invariant and normal lattices is
\vs{-5}\be
I^*/I\cong N^*/N\cong \bz_3^{\ell_3'}\times\bz_m^{\ell_m'}\times\bz_{3m}^{\ell_{3m}'}
\cong\bz_3^{\ell_3}\times\bz_m^{\ell_m}\ ,\vs{-5}
\ee
where $\ell_3$ and $\ell_m$ are respectively the number of $\bz_3$ and $\bz_m$ generators. In the non-coprime sectors, $I^*_3/I_3$ has only $\bz_m$ classes and $I^*_m/I_m$ has only $\bz_3$ classes. Thus, all the $\bz_3$ classes of $N^*/N$ must have $\Theta^3$-invariant representatives that correlate with the $\bz_3$ classes of $I^*/I$ and form vectors in $I_3$. We denote the generators of the $\Theta^3$-invariant $\bz_3$ classes of $N^*$ by $f_{3,i}^N$, $i=\{1,\ldots,\ell_3\}$. Likewise, all the $\bz_m$ classes of $N^*/N$ must have $\Theta^m$-invariant representatives that correlate with the $\bz_m$ classes of $I^*/I$ and form vectors in $I_m$, and their generators are denoted by $f_{m,j}^N$, $j=\{1,\ldots,\ell_m\}$. To summarise,
\vs{-5}\be\label{f3fmgens}
\Theta^3f_{3,i}^N=f_{3,i}^N\ ,\quad1\le i\le\ell_3\ ,\qquad\qquad\Theta^mf_{m,i}^N=f_{m,i}^N\ ,\quad1\le i\le\ell_m\ .\vs{-5}
\ee
The lattice $\g$ contains vectors
\vs{-5}\be\label{Gamma3m}
\g=(N,I)+\!\!\sum_{i=1}^{\ell_3}a_i(f_{3,i}^N,f_{3,i}^I)+\sum_{j=1}^{\ell_m}b_j(f_{m,j}^N,f_{m,j}^I)\ ,
\;\; a_i=\{0,1,2\}\ ,\;\; b_j=\{0,\ldots,m-1\}\ ,\vs{-5}
\ee
where $f_{3,i}^I$ and $f_{m,j}^I$ are the generators of the $\bz_3$ and $\bz_m$  classes of $I^*\!/I$, respectively, and are correlated with those in $N^*/N$. The invariant lattices $I_3$ and $I_m$ contain vectors
\vs{-10}\be\label{I3Im}
I_3 = (N_{3},I) + \sum_{i=1}^{\ell_2} a_i (f^N_{3,i}, f^I_{3,i})\ ,\qquad\qquad
I_m = (N_{m},I) + \sum_{j=1}^{\ell_m} b_j (f^N_{m,j}, f^I_{m,j}) \ ,\vs{-15}
\ee
where $N_{3}$ and $N_{m}$ are respectively the $\Theta^3$ and $\Theta^m$-invariant sublattices of $N$.

The dual lattices $I_3^*$ and $I_m^*$ are the projections of $\g$ along $\Theta^3$ and $\Theta^m$ invariant directions. Under the action of the projection operators
\vs{-5}\be\label{proj}
\Pi_3= \frac{1}{m}(1+\Theta^3+\Theta^6+\ldots+\Theta^{3(m-1)})
\ ,\qquad\qquad\Pi_m= \frac{1}{3}(1+\Theta^m+\Theta^{2m})\ ,\vs{-5}
\ee
the generators $f^N_{3,i}$ and $f^N_{m,i}$ transform as
\vs{-5}\be\label{projV3Vm}
\Pi_mf^N_{3,i}=0\ , \qquad\qquad\Pi_3f^N_{m,j} = 0\ .\vs{-5}
\ee
We then find
\vs{-15}\be\vs{-5}\begin{split}\label{I3*}
&I_3^* =  \Pi_3\Gamma =\sum_{i=1}^{\ell_3} a_i( f^N_{3,i},f^I_{3,i})+
\sum_{j=1}^{\ell_m} b_j(0,f^I_{m,j})+(\Pi_3N,I)\ ,\\[-8pt]
&I_3^*/I_3 =   \sum_{j=1}^{\ell_m} b_j (0,f^I_{m,j}) + (\Pi_3N \!\!\!\!\mod N_{3},0)\ ,
\end{split}\ee
which contains uncorrelated $\bz_m$ classes. Similarly,
\vs{-5}\be\vs{-5}\begin{split}\label{Im*}
&I_m^* =  \Pi_m\Gamma =\sum_{i=1}^{\ell_3} a_i(0,f^I_{3,i})+
\sum_{j=1}^{\ell_m} b_j( f^N_{m,j},f^I_{m,j})+(\Pi_mN,I)\ ,\\[-8pt]
&I_m^*/I_m =   \sum_{i=1}^{\ell_3} a_i (0,f^I_{3,i}) + (\Pi_mN \!\!\!\!\mod N_{m},0)\ ,
\end{split}\ee
which has uncorrelated $\bz_3$ classes. All in all\vs{-5}
\be\label{discIs}
I^*/I\cong\bz_3^{\ell_2}\times\bz_m^{\ell_m}\ ,\qquad\qquad 
I_3^*/I_3\cong\bz_m^{\ell_m+\ell_m^N}\ ,\qquad\qquad I_m^*/I_m\cong\bz_3^{\ell_3+\ell_3^N}\ ,\vs{-5}
\ee
where $\ell_3^N$ and $\ell_m^N$ are the number of $\bz_3$ and $\bz_m$ generators along respectively the $\Theta^m$ and $\Theta^3$-invariant directions of $N$.

Let us next consider the automorphism $\Theta$ of order $3m$. Its eigenvalues have exponents $\frac{k}{3m}$ and contain $n_1$ components with $(k,3m)=1$, $n_3$ components with $(k,3m)=3$, and $n_m$ components with $(k,3m)=m$. We collect these components in a vector $t$ of the form
\vs{-5}\be\label{3meigens}
t=\frac1{3m}\Big(\big[\underbrace{1,2,4,\cdots,\tfrac{3m-1}2}_{m-1~{\rm terms}}\big]^{n_1},
\big[\underbrace{3,\cdots,3\tfrac{m-1}2}_{\frac{m-1}2~{\rm terms}}\big]^{n_3},[m]^{n_m}\Big)\ .\vs{-10}
\ee
Recall that the components $t_a$ of the above vector may be chosen such that $0\!<\! t_a\!<\!\frac12$, where $1\!\le\! a\!\le\!\frac s2$, and $s$ is the rank of the normal lattice $N$, which is 22 in the present case. The vector $t$ has only one component of type $(k,3m)\!=\!m$, i.e. $k\!=\!m$. We find
\vs{-5}\be\label{rankN}
s=(2n_1+n_3)(m-1)+2n_m\ ,\vs{-5}
\ee
where we used that the number of integers relatively prime to $3m$, $m\!>\!3$ and prime, is $2(m-1)$. The degeneracy factors in the twisted sectors, using the distribution \eqref{3meigens}, are found to be\vs{-5}
\be\label{Fs3m}
\cF_k  =\sqrt{\frac{{\rm det}^{'}(1-\Theta^k)}{|I_k^*/I_k|}}=\left\{
\begin{array}{ll}
3^{\frac12(n_m-\ell_3)}\,m^{\frac12(n_3-\ell_m)}\ , & \ k=1 \\
m^{\frac12(2n_1 + n_3 - \ell_m - \ell^N_m)}\ , & \ k=3 \\
3^{\frac12((m-1)n_1 +  n_m - \ell_3 - \ell^N_3)}\ , & \ k=m
\end{array}
\right.\ .\vs{-5}
\ee
Requiring $\cF_k\in\bz_{\ge0}$ yields
\vs{-5}\be\label{Fint}
\begin{array}{lll}
n_3-\ell_m\in2\bz_{\ge0}\ , &\quad n_m-{\ell_3}\in2\bz_{\ge0}\ , \\[5pt]
2n_1 + n_3 - \ell_m - \ell^N_m \in2\bz_{\ge0}\ , 
&\quad (m-1)n_1 +  n_m - \ell_3 - \ell^N_3 \in2\bz_{\ge0}\ , \quad \ell^N_3\in2\bz_{\ge0}\ .
\end{array}\vs{-5}\ee
The models HM223 $(\bz_{15})$ and HM240 $(\bz_{21})$ both satisfy these properties. We note that there is one asymmetric orbifold in six spacetime dimensions belonging to this family. This is the $T^4/\bz_{15}$ HM128 model and it satisfies the requirements in eq. \eqref{Fint}. 
This orbifold, however, does not give rise to a 6-dimensional island, as expected from the classification in \cite{Fraiman:2022aik}.

\subsubsection{\texorpdfstring{$\bz_{15}$}{TEXT} HM223}\label{hm223}

For this model we have 
\vs{-5}\be\vs{-5}\begin{split}\label{hm223dat}
&t=\frac1{15}(1,1,2,2,3,4,4,5,6,7,7)\ ,\qquad E_1=\frac{14}{15}\ ,\qquad E_3=1\ ,\qquad E_5=1\ ,\\[1mm]
&I^*/I\cong \bz_{15} \cong \bz_3\times\bz_5\ ,\qquad I_3^*/I_3\cong\bz_5^3\ ,\qquad I_5^*/I_5\cong\bz_3^{5}\ ,\\[1mm]
&\cF_1=1\ ,\qquad \cF_3=5\ ,\qquad \cF_5=9\ .
\end{split}\ee
The LMC \eqref{lmc_3m}, and absence of twisted sector massless states give\vs{-5}
\be\label{lmc223g}
15v^2\in2\bz\ ,\qquad\qquad 15v\in I\ ,\qquad v\notin I^*\ ,\qquad3v\notin I_3^*\ ,\qquad 5v\notin I_5^*\ .\vs{-5}
\ee
There are 18 equivalence classes for $v$, for instance $\frac1{15}( 0 , 0 , -1 , 2 , 2 , 6 )$ and $\frac{1}{15}( 2 , 2 , 2 , -1 , 1 , 5 )$. In Appendix \ref{Z3mop_223} we have verified the operator interpretation of the theory. The HM223 model is therefore an archipelago of 18 islands. For the choice $v\!=\!0$, one recovers in total 22 massless vector multiplets in the twisted sector spectrum.

\subsubsection{\texorpdfstring{$\bz_{21}$}{TEXT} HM240}\label{hm240}

The model has the following properties\vs{-5}
\be\vs{-5}\begin{split}\label{hm240dat}
&t=\frac1{21}(1,2,3,4,5,6,7,7,8,9,10)\ ,\qquad E_1=\frac{62}{63}\ ,\qquad
E_3=\frac67\ ,\qquad E_7=\frac89\ ,\\[1mm]
&I^*/I\cong \bz_3\times\bz_{21} \cong \bz_3^2\times\bz_7\ ,
\qquad I_3^*/I_3\cong\bz_7^3\ ,\qquad I_7^*/I_7\cong\bz_3^{8}\ ,\\[1mm]
&\cF_1=1\ ,\qquad \cF_3=1\ ,\qquad \cF_7=1\ .
\end{split}\ee
The LMC \eqref{lmc_3m}, and exclusion of twisted sector moduli, yield\vs{-5}
\be\label{lmc240g}
21v^2=\frac23\hs{-8}\mod2\ ,\qquad\qquad 21v\in I\ ,\qquad v\notin I^*\ ,\qquad3v\notin I_3^*\ ,\qquad 7v\notin I_7^*\ .\vs{-5}
\ee
There are 61 equivalence classes of $v$ which satisfy these conditions. For example, representatives of two families are $\frac{1}{21}(2,1,-3,-1,5,1)$ and $\frac{1}{21}(-5,-1,0,0,7,2)$. In app. \ref{Z3mop_240} we show that the operator interpretation is consistent. 
Therefore, this orbifold yields a populated archipelago.

In the HM240 the genus of $I$ has two isometry classes. Thus, there is a second invariant lattice, with identical data as in eq. \eqref{hm240dat}, which can be glued to the same 
$N$ to obtain a different $\Gamma(22,6)$ lattice admitting the same automorphism. Consistency of the new $\bt^6/\bz_{21}$ orbifold is 
shown as in app. \ref{Z3mop_240}. Besides, there exist 42 non-equivalent shifts which satisfy modular invariance and guarantee absence of massless twisted states.

\subsection{Outliers}\label{outliers}

We now consider the $\bt^6/\bz_n$ asymmetric orbifolds whose order is either $n\!=\!p^2$ with $p$ prime, or $n\!=\!2m$ with $m$ nonprime -- see sections \ref{N4} and \ref{N5}. One interesting aspect of these models is that $I^*\!/I$ may contain conjugacy classes $\bz_k$ where $k$ is neither prime nor a product of distinct prime numbers. For instance, $I^*\!/I$ may contain classes such as $\bz_4$, $\bz_9$, and $\bz_{16}$ among its components. This aspect, entangled with properties of the automorphism $\Theta$, results in a rich structure of correlated vectors in invariant lattices $I_\ell$ with $(n,\ell)\!\ne\!1$, via gluing combinations of the conjugacy classes of $I^*$ and $N_\ell^*$, where $N_\ell$ is as usual the $\Theta^\ell$-invariant part of the normal lattice.

As an example, in a $\bt^D/\bz_4$ orbifold, the invariant lattice $I_2$ may contain a glue vector of the form $(2Y_4,2f_4)$, where $Y_4$ is a $\Theta^2$-invariant $\bz_4$ generator in $N_2^*/N_2$ and $f_4$ is a $\bz_4$ generator in $I^*\!/I$. Such correlations did not exist in the previous models in sections \ref{N1}-\ref{N3}. While the diversification of correlation possibilities generates several new asymmetric orbifolds, we do not yet have a systematic construction of this type of models, and thereby call them outliers for now. Below we will construct two specific examples, one $\bt^6/\bz_8$ model and one $\bt^6/\bz_9$ model (the latter is the only HM model with a perfect square order). It would be very interesting to understand the connection between the automorphisms of sublattices of the Leech lattice and the patterns of correlations emerging in the invariant lattices of the asymmetric orbifolds built upon them. We leave this for future work. We finally note that the outlier orbifolds in six dimensions are $T^4/\bz_n$ with orbifold groups $\bz_4$ (HM99), $\bz_8$ (HM99, HM107, HM143, HM147),  $\bz_9$ (HM101), and $\bz_{12}$ (HM104, HM109, HM123, HM135, HM157). However, according to the classification of \cite{Fraiman:2022aik}, none of these models are islands.

\subsubsection{\texorpdfstring{$\bz_8$}{TEXT} HM224}\label{hm224}

This model has properties\vs{-5}
\be\vs{-5}\begin{split}\label{hm224dat}
&t=\frac1{8}(1,1,1,1,2,2,3,3,3,3,4)\ ,\qquad\qquad E_1=E_2=E_4=1\ ,\\
& I^*\!/I \cong \bz_4^2\ ,\qquad\qquad I_2^*/I_2 \cong \bz_2^2\times\bz_4^2\ ,
\qquad\qquad I_4^*/I_4\cong \bz_2^8\ ,\\
&\cF_1=4\ ,\qquad\cF_2=8\ ,\qquad\cF_4=16\ .
\end{split}\ee
We will next compute the invariant lattices $I_2$ and $I_4$, and their duals, and show that their discriminant groups are as claimed in the above equation. The full partition function of the theory is computed in app. \ref{app_Z8}, where the consistency of operator interpretation is also verified.

Denote the two $\bz_4$ generators of $I^*\!/I$ by $f^I_1$ and $f^I_2$, and the two $\bz_4$ generators of $N^*\!/N$ by $f_1^N$ and $f_2^N$. The two glue vectors are of the form $(f_1^N,f_1^I)$ and $(f_2^N,f_2^I)$, and
\vs{-10}\be\label{224Gamma22_6}
\Gamma=(N,I)+\sum_{i=1}^2 \alpha_i( f^N_{i}, f^I_{i}) \ ,\qquad \alpha_i=\{0,\ldots,3\}\ .\vs{-5}
\ee
We next denote by $N_{j}$ and $N^*_{j}$ the $\Theta^j$-invariant parts of $N$ and $N^*$, respectively. We find
\vs{-5}\be\label{hm224N2N4}
N_{2}^*/N_{2}\cong\bz_4^2\ ,\qquad\qquad
N_{4}^*/N_{4}\cong\bz_2^4\times\bz_4^2\ ,\vs{-5}
\ee
which may be obtained using the basis of generators of the normal lattice provided by \cite{HM}.

\noindent\underline{\bf{$I_2^*$ lattice}}\\
It turns out that the generators of $N^*\!/N$, i.e. $f_1^N$ and $f_2^N$, are not invariant under $\Theta^2$-action. However, $W\!=\!2f_2^N+Q$ is $\Theta^2$-invariant for some $Q\!\in \!N$. Thus, the sublattice $I_2\subset\g$ contains the correlated class
\be\label{224I2}
I_2 = (N_{2},I) + \beta (W,2f^I_{2})\ ,\qquad \beta=\{0,1\}\ .
\ee
Next, denote the two $\bz_4$ generators of $N_{2}^*/N_{2}\!\cong\!\bz_4^2$ by $Y_1$ and $Y_2$. We choose a basis where $2\,Y_2\!=\!W$ mod $N_{2}$. The dual lattice $I_2^*$ has in general $\bz_2$ and $\bz_4$ conjugacy classes -- see eq. \eqref{theta}. To determine $I_2^*$, we note that the generators of $N^*_{2}$ and $I^*$ satisfy
\vs{-5}\be\label{disc224}{
\begin{pmatrix}Y_1^2& Y_1\cdot Y_2 \\
Y_1\cdot Y_2 & Y_2^2 \end{pmatrix}=
\begin{pmatrix}\frac14 & \frac14 \\ \frac14 & \frac12 \end{pmatrix}}\ ,\qquad
\begin{pmatrix}(f_1^{I})^2& f_1^I\cdot f_2^I \\
f_1^I\cdot f_2^I & (f_2^I)^2 \end{pmatrix}=
\begin{pmatrix}\!-\frac34 & \;\;\frac74 \\ \;\;\,\frac74 & \!\!-\frac32 \end{pmatrix}\ ,\vs{-5}
\ee
where $Y_i$ and $f^I_i$ respectively represent vectors $(Y_i,0)$ and $(0,f^I_i)$. Since $(Y_1,0)\cdot(W,2f_2^I)\!\notin\!\bz$ and $(0,f_1^I)\cdot(W,2f_2^I)\!\notin\!\bz$, the vectors $(Y_1,0)$ and $(0,f_1^I)$ are not in $I_2^*$. However, $(Y_1,f_1^I)\cdot(W,2f_2^I)\!\in\!\bz$ and so the correlated vector $(Y_1,f_1^I)\!\in\! I_2^*$. We thus obtain\vs{-2}
\be\vs{-5}\begin{split}\label{hm224I2d}
&I_2^*/I_2 = \alpha_1(Y_1,f_1^I) + \alpha_2(Y_2,0) + \alpha_3 (0,f^I_2) + \beta (0,2f^I_1) \ ,\qquad
I_2^*/I_2\cong\bz_2^2\times\bz_4^2\ ,\\[5pt]
&\alpha_i=\{0,\ldots,3\}\ ,\quad i=\{1,3\}\ ,\quad \beta=\{0,1\}\ .
\end{split}\ee
Note that $(2Y_2,2f_2^I)\!\in\!I_2$ and so $(Y_2,f^I_2)$ does not generate a $\bz_4$ class, but that $(Y_2,0)$ and $(Y_2,f_2^I)$ generate respectively a $\bz_4$ and a $\bz_2$ class. Another factor of $\bz_4\times\bz_2$ is generated by $(Y_1,f^I_1)$ and $(0,2f^I_1)$.

\noindent\underline{\bf{$I_4^*$ lattice}}\\
The two $N^*\!/N$ generators $f_1^N$ and $f_2^N$ are not $\Theta^4$-invariant. However, the two vectors $\wh{W}=2f_1^N+Q_1$ and $\wh{W}'=2f_2^N+Q_2$ are invariant under the $\Theta^4$-action for some $Q_1,Q_2\in N$. In fact, $\wh{W}'=W$ mod $N$, where $W$ is the $\Theta^2$-invariant vector used to construct $I_2$ \eqref{224I2}. We obtain\vs{-2}
\be\label{224I4}
I_4 = (N_{4},I) + \beta_1 (\wh{W},2f^I_{1})+ \beta_2 (W,2f^I_{2})\ ,\qquad \beta_1,\beta_2=\{0,1\}\ .\vs{-2}
\ee
We next denote the four $\bz_2$ generators of $N_{4}^*/N_{4}\!\cong\!\bz_2^4\times\bz_4^2$ by $Z_i$, $i=1,\ldots,4$, and the two $\bz_4$ generators by $\wh{Y}_1$ and $\wh{Y}_2$. The $\bz_2$ generators have integer norms, whereas $(\wh{Y}_1)^2\!=\!\frac74$ and $(\wh{Y}_2)^2\!=\!\frac52$. We choose the $\bz_4$ generators such that $2\,\wh{Y}_1\!=\!\wh{W}$ mod $N_{4}$ and $2\,\wh{Y}_2\!=\!W$ mod $N_{4}$. Furthermore, $Z_1$ and $\wh{Y}_2$ are $\Theta^2$-invariant and hence, may be expressed in terms of $N_2^*$ generators $Y_1$ and $Y_2$. More specifically, $Z_1=2Y_1$  mod $N_4$ and $\wh{Y}_2=2Y_1-3Y_2$. The latter yields $2\,\wh{Y}_2=2\,Y_2$ mod $N_4$.

We next determine $I_4^*$, which contains only $\bz_2$ conjugacy classes.  The inner products $Z_i\cdot2\,\wh{Y}_j$, with $i=1,\ldots,4$ and $j=1,2$, are integers. This yields $Z_i\cdot \wh{W}\in\bz$ and $Z_i\cdot W\in\bz$, and so $(Z_i,0)$ are uncorrelated vectors in $I_4^*$. On the other hand, $\wh{Y}_1\cdot\wh{Y}_2=\frac34$, and therefore $(\wh{Y}_i,0)$ have to be correlated with corresponding $I^*$ generators $f_i^I$ to render integer inner products with $(\wh{W},2f^I_{1})$ and $(W,2f^I_{2})$ in $I_4$. All in all, we obtain eight $\bz_2$ generators for $I_4^*/I_4$ :\vs{-10}
\be\label{hm224I4d}
\hs{-5}I_4^*/I_4\!=\!\sum_{i=1}^2\beta_i (0,2f^I_i) + \beta_3(\wh{Y}_1,f_1^I) + 
\beta_4 (\wh{Y}_2,f^I_2)+\sum_{j=1}^4\gamma_j(Z_j,0)\ ,\; \beta_i,\gamma_j=\{0,1\}\ ,
\; I_4^*/I_4\cong\bz_2^8\ .\vs{-10}
\ee
One can verify using eq. \eqref{disc224} that $(\wh{Y}_i,f_i^I)$ have integer inner products with all vectors in $I_4$.

\noindent\underline{\bf{LMC}}\\
Starting from eq. \eqref{hzglgkl}, the LMC in co-prime sectors takes the form $8P^2+2P\cdot 8v+8v^2+16E_1\!\in\!2\bz$, $\forall P\!\in\! I^*$. Given the norms of the generators \eqref{disc224}, $4(f^I_1)^2$ is odd, $4(f^I_2)^2$ is even, and so $8P^2\!\in\!2\bz$, $\forall P\!\in\! I^*$. Since $E_1\!=\!E_2\!=\!E_4\!=\!1$, the condition reduces to $8v^2\!\in\!2\bz$, where $8v\!\in\! I$. In $g^2$ and $g^6$ sectors the condition is $4P^2\!+\!16P\cdot v\!+\!16v^2\!+\!8E_2\!\in\!2\bz$, $\forall P\!\in\! I_2^*$. Using eqs. \eqref{disc224} and \eqref{hm224I2d}, we find $4P^2\!\in\!2\bz$, $\forall P\!\in\!I_2^*$. This gives no new constraint. For the $g^4$ sector we have $2P^2\!+\!16P\cdot v\!+\!32v^2\!+\!4E_4\!\in\!2\bz$, $\forall P\!\in\!I_4^*$. Eq. \eqref{hm224I4d} gives $2P^2\!\in\!2\bz$, $\forall P\!\in\!I_4^*$, and so there is again no new condition. Requiring absence of massless states in the twisted sectors, we obtain\vs{-5}
\be\label{224lmc}
8v\in I\ ,\qquad8v^2\in2\bz\ ,\qquad v\notin I^*\ ,\qquad 2v\notin I^*_2\ ,\qquad 4v\notin I^*_4\ .\vs{-5}
\ee
There is a unique family of shift vectors which satisfy these conditions and its representative can be selected to be $\frac18(1,1,0,1,1,1)$. In app. \ref{app_Z8}, we show that operator interpretation is valid in the $g^2$- and $g^4$-twisted sectors -- cf. sections \ref{pint_224_g2} and \ref{pint_224_g4}. This concludes the construction of the $\bt^6/\bz_8$ heterotic island. Taking $v=0$, which is allowed by modular invariance, leads to additional twisted massless states belonging to 22 vector multiplets.

\subsubsection{\texorpdfstring{$\bz_9$}{TEXT} HM230}\label{hm230}

This model is characterized by \vs{-5}
\be\vs{-5}\begin{split}\label{hm230ei}
&t=\tfrac1{9}(1,1,1,2,2,2,3,3,4,4,4)\ ,\qquad\qquad\quad E_1=E_3=1\ ,\\[2pt]
& I^*\!/I \cong \bz_3\times\bz_9\ ,\qquad I_3^*/I_3 \cong \bz_3^5\ ,\qquad\qquad\cF_1=3\ ,\qquad\cF_3=9\ .
\end{split}\ee
The full partition function of the theory is computed in app. \ref{app_Z9}.

We will next compute the invariant lattice $I_3$ and its dual. Denote the generators of the $\bz_3$ and $\bz_9$ classes in $I^*\!/I$ respectively by $f_3^I$ and $f_9^I$. Likewise denote the generators in $N^*\!/N$ by $f_3^N$ and $f_9^N$. The lattice $\g$ is of the form
\vs{-5}\be\label{Gamma22_6}
\Gamma=(I,N)+ a( f^N_{3}, f^I_{3}) + b(f^N_{9}, f^I_{9})\ ,\qquad a=\{0,1,2\}\ ,\;\; b=\{0,1,\ldots,8\}\ .\vs{-5}
\ee
The generators $f_3^N$ and $f_9^N$ are not invariant under $\Theta^3$-action. However, the vector $W\!=\!3f_9^N\!+\!Q$ is $\Theta^3$-invariant for some $Q\!\in\! N$. Thus, $I_3$ contains correlated class $(W,3f_9^I)$, and\vs{-5}
\be\label{hm230I3}
I_3 = (N_{3},I) + a (W,3f^I_{9})\subset\g \ ,\qquad a=\{0,1,2\}\ ,\vs{-5}
\ee
where $N_{3}$ is the $\Theta^3$-invariant part of $N$ and has rank 4. It turns out that
\vs{-5}\be\label{hm230N3}
N_{3}^*/N_{3}\cong\bz_3^2\times\bz_9\ .\vs{-5}
\ee
We denote the generators of the two $\bz_3$ and one $\bz_9$ classes of $N_{3}^*$ respectively by $Y_3^1$, $Y_3^2$, and $Y_9$, and choose a basis in which $3Y_9=W$ mod $N_{3}$.

We next compute $I_3^*/I_3$, which has only $\bz_3$ classes. The inner products are
\vs{-5}\be\label{disc230}{
\begin{pmatrix}(Y_3^1)^2& Y_3^1\cdot Y_3^2 & Y_3^1\cdot Y_9 \\
Y_3^1\cdot Y_3^2 & (Y_3^2)^2 & Y_3^2\cdot Y_9 \\
Y_3^1\cdot Y_9 & Y_3^2\cdot Y_9 & Y_9^2  \end{pmatrix}=
\begin{pmatrix}\;\;\;\frac23 & \;\;\;\frac23 & -\frac13 \\ \;\;\;\frac13 & \;\;\;\frac43 & 
-\frac23 \\ -\frac13 & -\frac23 & \;\;\;\frac49 \end{pmatrix}}\ ,\;
\begin{pmatrix}(f_3^{I})^2& f_3^I\cdot f_9^I \\
f_3^I\cdot f_9^I & (f_9^I)^2 \end{pmatrix}=
\begin{pmatrix}-\frac43 & -\frac43 \\ -\frac43 & -\frac{4}9 \end{pmatrix}\ ,\vs{-5}
\ee
where $Y$ and $f^I$ respectively represent vectors $(Y,0)$ and $(0,f)$. We find $(Y_3^i,0)\cdot(W,3f_9^I)\in\bz$, $i=1,2$, but $(Y_9,0)\cdot(W,3f_9^I)\!\notin\!\bz$. Thus, $(Y_9,0)\!\notin\!I_3^*$ and has to be correlated. We obtain\vs{-5}
\be\label{hm230I3d}
\hs{-3}I_3^*/I_3 =   a_1 (0,f^I_{3}) + a_2 (0,3f^I_{9})+ \sum_{i=1}^{2} a_i' (Y^i_{3},0)+ c(Y_9,f_9^I)\ ,
\;\; a_i,a'_i,c=\{0,1,2\}\ ,\;\; i=\{1,2\}\ .\vs{-10}
\ee
Note that $(3Y_9,0)$ does not generate a new $\bz_3$ class since $(3Y_9,3f^I_9)\!\in\!I_3$. We thus find $I_3^*/I_3\cong\bz_{3}^5$. The LMC \eqref{lmc_9} together with absence of massless states in all twisted sectors yield\vs{-5}
\be\label{lmc230g_i}
9v\in I\ ,\qquad\qquad 9v^2\in2\bz\ ,\qquad\qquad v\notin I^*\ ,\qquad\qquad 3v\notin I_3^*\ .\vs{-5}
\ee
There exist 3 non-equivalent classes of shift vectors that satisfy these conditions. They can be taken to be $\frac19(2, -1, 1, 1, 2, 1)$, 
$\frac19(3, -1, 0, 2, 1, 1)$, and $\frac19(3, 0, 2, 2, 1, 0)$. In app. \ref{Z9op} we verify the operator interpretation in the $g^3$- and 
$g^6$-twisted sectors. We conclude that the HM230 $\bt^6/\bz_9$ orbifold is an archipelago of 3 islands.

\section{The type II story}\label{section_typeii_story}

In this section we study type II $\bt^6/\bz_n$ asymmetric orbifolds with 16 right-moving supercharges.
The orbifold action is therefore chosen to break all left-moving supersymmetries by acting appropriately on the left-moving fermions. All such actions are classified in appendix \ref{app_triple}. 
To maintain world-sheet supersymmetry, there must then be a corresponding action on the left-moving bosons, while the right-moving 
bosons and fermions are inert. In general, such orbifolds lead to 4-dimensional $\cN=4$ theories with rank reduction that were first 
analyzed in \cite{Dixon:1987yp}. In this work we specialize to constructing islands and impose that
all left-moving bosons are rotated. This is a necessary condition that only guarantees absence of massless 
vector multiplets in the untwisted sector. We will shortly determine the additional requirements to avoid 
massless states in twisted sectors.

In the following we will first discuss at length the possible lattices and automorphisms that can be employed
to construct 4-dimensional type II islands. Afterwards in subsection \ref{subsection_typeii_Z} we will derive the terms 
in the partition function which are common in all $\bt^6/\bz_n$ asymmetric orbifolds, and examine modular invariance in the
$g$-twisted sector. Model specific terms of the partition function,
which depend on the order $n$, will be computed case by case in section \ref{section_typeii_isl}, where we
will also address the consistency of the orbifold conformal field theory.

\subsection{\texorpdfstring{Lattices, orbifold action, and classification}{TEXT}}\label{subsub_typeii_latt}

For islands, it follows that the orbifold action on the $(6,6)$ lattice $\Gamma$ must be such that the
invariant lattices $I$ and $N$ have signature $(0,6)$ and $(6,0)$ respectively. Moreover, the $\bz_n$ automorphism
must be $(\Theta_L,\Theta_R)=(\Theta,\uno)$, and such that $\Theta$ has no trivial eigenvalues.
An even self-dual  lattice $\Gamma$ with these properties can be built up by taking the momenta $(P_L,P_R) \in \Gamma$ to satisfy
\vs{-3}\be\label{gammadef}
P_L, P_R \in \Lambda_\cG^*, \qquad P_L - P_R \in \Lambda_\cG\ ,\vs{-5}
\ee
where $\Lambda_\cG$ and $\Lambda_\cG^*$ are respectively the root and the weight lattice of 
a semisimple simply-laced Lie algebra $\cG$ of rank 6. Now, the purely left $\bz_n$ automorphism
must satisfy $(\Theta P_L, P_R) \in \Gamma$ for all $(P_L,P_R) $, which requires
$\Theta P_L - P_L \in \Lambda_\cG $, $\forall \ P_L \in \Lambda_\cG^*$.
This simply means that $\Theta$ sends weights into weights modulo roots. Therefore, it suffices to take $\Theta$ to be an 
order $n$ element of the Weyl group of $\cG$, denoted $W(\cG)$. The eigenvalues of $\Theta$ can be written as 
$e^{\pm 2\pi i  t_a}$, $a=1, \ldots, 3$, with $0 \le t_a \le 1$ and $nt_a \in \bz$.

The even self-dual lattice $\Gamma$ with vectors $(P_L,P_R)$ specified in eq. \eqref{gammadef} will be denoted by $\Gamma(\cG)$ 
whenever it is necessary to stress the underlying Lie algebra.
To connect with the general formalism of section \ref{section_asym}, let us discuss the invariant and normal lattices in
the present setup. Since the automorphism does not leave any left-moving directions invariant, we obtain that
$(\Theta P_L ,P_R) = (P_L, P_R)$ implies $P_L=0$ and then necessarily $P_I=(0,P_R)$ with $P_R \in \Lambda_\cG$.
We conclude that{\footnote{As usual, $L(j)$ means lattice $L$ with Gram matrix rescaled by $j$.}}
\vs{-5}
\be
I = \Lambda_\cG(-1)\, , \quad N=\Lambda_\cG \, .
\ee
The action of the orbifold generator $g$ on the momentum lattice is given by eq. \eqref{gactlatt}, namely
$g|P_N, P_I \rangle = e^{2\pi i P \cdot v} |\Theta P_N, P_I \rangle$, 
with $v$ a constant shift vector that can be chosen along the invariant right-moving directions. 

The action of the orbifold generator on world-sheet left-moving bosons and fermions is determined by the $\Theta$ eigenvalues
as specified in eqs. \eqref{gXa} and \eqref{defvf} respectively. As mentioned before, the requirement that $g$ is of order $n$ 
acting on the fermions gives the condition
\vs{-5}\be\label{orderc}
n(t_1 + t_2 + t_3) = 0 \; \text{mod}\; 2 \ . \vs{-5}
\ee
This condition is also required by modular invariance.
Additionally, demanding that all left-moving supersymmetries are broken amounts to imposing
\vs{-10}\be\label{susyc}
\pm t_1 \pm t_2 \pm t_3 \not= 0 \, \text{mod}\, 2 \ , \quad \text{ for all choices of signs} \ .
\ee
This latter condition is equivalent to requiring that $\Theta$ is not in an $SU(3)$ subgroup of $SO(6)$.

So far we have explained that the eigenvalues of the automorphism $\Theta$, encoded in the triple $t=(t_1,t_2,t_3)$,
are subject to the conditions in eqs. \eqref{orderc} and \eqref{susyc}. Additionally,  they 
are restricted by the requirement that $\Theta$ acts crystallographically on a six-dimensional lattice.
Using results in \cite{Erler:1992ki} we then find the allowed triples in Table \ref{tab_t}. With islands in mind, we have also imposed 
$t_a \not=0, \ \forall\,  a$, to ensure that $\Theta$ does not leave invariant directions.

After determining the eigenvalues we can proceed to identify the root lattice $\Lambda_\cG$, and the automorphism 
$\Theta$ belonging to the Weyl group $W(\cG)$. This is straightforward because semisimple simply-laced root lattices of dimension 6 
can only contain factors $A_i$, $1\le i\le6$, $D_j$, $j=4,5,6$, and $E_6$. Moreover, for each simple factor, the non-degenerate 
conjugacy classes of $W(\cG)$ (i.e. those without invariant subspaces) and their respective eigenvalues are known, 
see e.g. \cite[Table 11]{Lerche:1988np}. In Table \ref{tab_t} we denote by $\cA_i$, $\cD_j$ and $\cE_6$ the classes of the Coxeter 
element of the respective root lattice, whereas $\cD_j(a_m)$ and $\cE_6(a_m)$ refer to other non-degenerate classes of the Weyl group. 

\begin{table}[h!]\begin{center}
\setlength{\extrarowheight}{2.5pt}
{{\begin{tabular}{|c|c||c|c|c|}
\hline
\# & $\bz_n$ & $(t_1,t_2,t_3)$ & root lattice $\Lambda_\cG$ & automorphism $\Theta$ \\
\hline
1& *$\bz_4$ & $(\tfrac12,\tfrac12,\tfrac12)$ & $D_6, D_4+2A_1$ & $-\uno_{6\times 6}$\\[1mm]
\hline
2& $\bz_6$ & $(\tfrac13,\tfrac13,\tfrac13)$ & $A_2+A_2+A_2$ & $\cA_2\times\cA_2\times\cA_2$\\[1mm]
\hline
3& $\bz_6$ & $(\tfrac12,\tfrac12,\tfrac13)$ & $D_4+A_2$, \underline{$4A_1+A_2$} & $-\uno_{4\times 4}\times \cA_2$, 
\underline{$\cA_1^4\times\cA_2$}\\[1mm]
\hline
4& *$\bz_6$ & $(\tfrac16,\tfrac16,\tfrac23)$ & $E_6$ & $\cE_6(a_2)$\\[1mm]
\hline
5& $\bz_8$ & $(\tfrac18,\tfrac38,\tfrac34)$ & $D_6$ & $\cD_6(a_1)$\\[1mm]
\hline
6& $\bz_8$ & $(\tfrac14,\tfrac12,\tfrac12)$ & $A_3+A_1+2A_1$ & $\cA_3 \times \cA_1 \times \cA_1^2$\\
\hline
7 & *$\bz_9$ & $(\tfrac19,\tfrac29,\tfrac59)$ & $E_6$ & $\cE_6(a_1)$\\[1mm]
\hline
8& $\bz_{12}$ & $(\tfrac12,\tfrac13,\tfrac13)$ & $2A_1+A_2+A_2$ & $\cA_1^2\times \cA_2\times \cA_2$\\[1mm]
\hline
9& $\bz_{12}$ & $(\tfrac16,\tfrac12,\tfrac12)$ & $D_4+2A_1$ & $\cD_4\times \cA_1^2$\\[1mm]
\hline
10 & $\bz_{12}$ & $(\tfrac16,\tfrac16,\tfrac12)$ & $D_6$ & $\cD_6(a_2)$\\[1mm]
\hline
11 & $\bz_{12}$ & $(\tfrac1{12},\tfrac5{12},\tfrac23)$ & $E_6$ & $\cE_6$\\[1mm]
\hline
12& $\bz_{12}$ & $(\tfrac14,\tfrac14,\tfrac13)$ & $D_4+A_2$ & $\cD_4(a_1)\times \cA_2$\\[1mm]
\hline
13& $\bz_{12}$ & $(\tfrac14,\tfrac34,\tfrac13)$ & $D_4+A_2$ & $\cD_4(a_1)\times \cA_2$\\[1mm]
\hline
14& $\bz_{14}$ & $(\tfrac17,\tfrac27,\tfrac47)$ & $A_6$ & $\cA_6$\\[1mm]
\hline
15& $\bz_{15}$ & $(\tfrac15,\tfrac25,\tfrac13)$ & $A_4+A_2$ & $\cA_4 \times \cA_2$\\[1mm]
\hline
16 & $\bz_{18}$ & $(\tfrac19,\tfrac29,\tfrac49)$ & $E_6$ & $\cE_6(a_1)$\\[1mm]
\hline
17 & $\bz_{20}$ & $(\tfrac12,\tfrac15,\tfrac25)$ & $2A_1+A_4$ & $\cA_1^2\times\cA_4$\\[1mm]
\hline
18 & $\bz_{20}$ & $(\tfrac1{10},\tfrac3{10},\tfrac12)$ & $D_6$ & $\cD_6$\\[1mm]
\hline
19 & $\bz_{24}$ & $(\tfrac13,\tfrac14,\tfrac12)$ & $A_2+A_3+A_1$ & $\cA_2 \times \cA_3  \times\cA_1$\\[1mm]
\hline
20 & $\bz_{24}$ & $(\tfrac16,\tfrac14,\tfrac12)$ & $D_5+A_1$ & $\cD_5(a_1)\times\cA_1$\\[1mm]
\hline
21 & $\bz_{30}$ & $(\tfrac15,\tfrac25,\tfrac23)$ & $A_4 + A_2$ & $\cA_4 \times \cA_2$\\[1mm]
\hline
\end{tabular}
}}
\caption{Data for type II $\bt^6/\bz_n$ island candidates. The triple in column 3 encodes the eigenvalues of the automorphism $\Theta$ which breaks all left-moving supersymmetries, does not leave invariant directions, and may be realized as an element of the Weyl group of a semisimple simply-laced Lie algebra $\cG$ of rank 6. The Lie algebra root lattice and the conjugacy class of $\Theta$ are given in columns 4 and 5. The entries indicated with asterisks denote models which are explicitly constructed in section \ref{section_typeii_isl}.} \label{tab_t}\end{center}\end{table}

A few observations about Table \ref{tab_t} are in order.
For the $\bz_4$, the lattice $6A_1$ is also possible, but we will find that massless states in twisted sectors cannot be prevented.
For the \# 3 $\bz_6$ with the underlined lattice, we will see that to satisfy modular invariance without doubling the order
would require the formalism of \cite{Aldazabal:2025zht}.

In cases with $n$ even, excepting \#'s 3-5 and 11-13, 
the full orbifold generator $g$ has order $n$ but $\Theta$ has order $m\!=\!n/2$. The order of $g$ is dictated by 
the condition in eq. \eqref{orderc} to have proper action on the left-moving fermions. This means that the order of $\Theta$ is
doubled in $\text{Spin}(6)$. In these cases $g^m$ is a $\bz_2$ that leaves $\Gamma$ invariant but acts as $(-1)^{F_L}$
on left-moving fermions, where $F_L$ is left-moving fermion number.

Some of the $\bz_n$'s in Table \ref{tab_t} have already appeared in the literature. The $\bz_9$ in row 7 was first considered in \cite{Dabholkar:1998kv}, while the $\bz_n$'s in rows 3, 12, 13, and 15 were proposed in \cite{Mizoguchi:2001cp} without a 
full analysis. The remaining allowed lattices and automorphisms are new to our knowledge. 

Using the results collected in appendix \ref{app_triple}, the same procedure that yielded Table \ref{tab_t} can be applied to classify asymmetric type II $\bt^4/\bz_n$ orbifolds leading to 
six-dimensional islands. In this way we recover the Dabholkar-Harvey island with $n=5$ \cite{Dabholkar:1998kv},
as well as the cases with $n=8,10,12$ that were predicted in \cite{Fraiman:2022aik} and constructed in \cite{Baykara:2025lhl}.  

Finally, let us mention that there are other ways of assembling a $(6,6)$ even self-dual lattice with purely left automorphism
$\Theta \in W(\cG)$ by gluing $N=\Lambda_\cG$ to a $(0,6)$ lattice non-isometric to $\Lambda_\cG(-1)$ but in its same 
genus, provided this genus has more than one isometry class\footnote{We thank H\'ector Parra De Freitas for pointing out this possibility.}.
This happens when $\Lambda_\cG$ is equal to $4A_1+A_2$, $A_3+3A_1$, $2A_2+2A_1$, $2A_1+A_4$ and $A_2+A_3+A_1$.

We will shortly explain that in all $\bt^6/\bz_n$ asymmetric orbifolds constructed from the the data in Table \ref{tab_t} 
there exist shift vectors that satisfy modular invariance in the $g$-twisted sector, and eliminate all twisted massless states. 
We then refer to them as island \emph{candidates}.
To confirm that an specific orbifold is an island, one has to prove that modular invariance and consistent operator interpretation 
are fulfilled in all twisted sectors. In section \ref{section_typeii_isl}, we will provide explicit construction of the asymmetric 
orbifolds for some examples in Table \ref{tab_t} and show that they are indeed string islands. 
A proof of existence of type II islands for generic orders of the orbifold group in the Table is beyond the scope of the present work. 
We hope to make progress on this problem in the future.

\subsection{Partition function and LMC}\label{subsection_typeii_Z}

In the following we present the explicit form of some selected $ Z(g^\ell,g^k)$ common to all models.  
As we have stressed, they can all be derived from untwisted sector terms $Z(\uno, g^\ell)$ by
a chain of modular transformations.
The latter can be obtained starting with eq.\eqref{z00typeii}. Taking
into account the action of the orbifold generator, defined in eqs. \eqref{gactlatt}, \eqref{gXa} and \eqref{defvf}, yields
\be\label{z1gl}
\hs{-5}Z(\uno,g^\ell)\!=\!  
\Bigg\{\! \frac{1}{\eta^4} \!\left(\sum_{p \in V}-\sum_{p \in Sp}\right)\opm q^{\frac12 p^2}e^{2\pi i p\cdot  \ell v_f}\!\Bigg\}
\Bigg\{\frac1{\eta^{6} \bar\eta^{6}}
\prod_{\substack{a=1\\\ell t_a\notin\bz}}^{3}
\frac{2 \sin(\pi  \ell t_a)\, \eta^3}
{\vtgen{\frac12}{\frac12 - \ell t_a} }\!
\sum_{P\, \in \, I_\ell}\!\! q^{\frac12 P_L^2}\bar q^{\frac12 P_R^2}\, e^{2\pi i P \cdot \ell v}\! \Bigg\}\ ,\vs{-5}
\ee
where $I_\ell$ is the invariant lattice under $\Theta^\ell$ and $1\!\le\!\ell\!<\!n$. 
Here we have used that the action of $g$ on momentum states is given by eq. \eqref{gactlatt}.
The action of $g^2$ might have to be modified in $\bt^6/\bz_n$ orbifolds with $n$ even, as dictated by modular invariance and 
consistency of the operator interpretation. An example will be presented in subsection \ref{subsection_typeii_z4}.

Performing $S$ and $T$ transformations on $Z(\uno,g^\ell)$, we obtain
\vs{-5}\be\vs{-5}\label{zggl}\begin{split}
Z(g^\ell,g^{k\ell}) \!= &
\Bigg\{\! \frac{e^{-i\pi k\ell^2 v_f^2}}{\eta^4} \!\left(\sum_{p \in V}-\sum_{p \in Sp}\right)
\opm q^{\frac12 (p+\ell v_f)^2} \opm e^{2i\pi (p+\ell v_f)\cdot k\ell v_f}\!\Bigg\}\\[1mm]
& \hspace*{1.0cm} \times\Bigg\{\!  \frac{\cF_\ell \, e^{2\pi ik E_\ell} }{\eta^6 \bar\eta^6}\!
\prod_{\substack{a=1\\\ell t_a\notin\bz}}^{3} \frac{\eta^3}{\hvtgen{ \frac12-\widetilde{\ell t_a}}{\frac12 -k\,\widetilde{\ell t_a}}}
\sum_{P\, \in \, I_\ell^*} q^{\frac12 (P+\ell v)_L^2}\bar q^{\frac12 (P+\ell v)_R^2} \op e^{i\pi k (P+\ell v)^2} \!\Bigg\}\ ,
\end{split}\ee
where we have used the definition of $\hat\vartheta$'s in eq. \eqref{that} and of $E_\ell$ in eq. \eqref{Fs}.
Note that here and later in section \ref{section_typeii_isl} we drop the phases from the $T$ transformation of $\eta^4$ in the 
denominator because they will cancel with those from $\bar\eta^4$ in $\cz_{\bar\psi}$. We also drop the phase $e^{i \pi k p^2}$ 
from left-moving fermions that cancels against $e^{i \pi k r^2}$ from right-moving fermions. 
An important result that follows from eq. \eqref{zggl} is that to avoid massless states in $g^\ell$-twisted sectors
it must be $\ell v \notin I_\ell^*$.

The coprime sectors are those with $(\ell,n)\!=\!1$, for which $I^*_\ell=I^*_1=I^*$, and $E_\ell=E_1$. 
They only have one orbit, i.e. all $Z(g^\ell, g^j)$, $1\le j<n$, are obtained from $Z(g^\ell,\uno)$ by applying $T$-transformations. 
The non-coprime sectors have more than one orbit which, depending on the value of $n$, may be 
obtained from different paths of $S$ and $T$ transformations starting from $Z(\uno,g^\ell)$. 
We will derive all the orbits for the specific models we construct in section \ref{section_typeii_isl}.

Let us now consider modular invariance in the $g$-sector. For $g$ to have order $n$ it must be that $Z(g,g^n)=Z(g,\uno)$.
From eq. \eqref{zggl} we then obtain the constraint
\vs{-5}\be\label{lmcii}
nP^2 + 2P\cdot nv + n(v^2 + v_f^2 + 2 E_1)  \in 2\bz\, ,  \qquad\forall\, P\in I^*\ .
\ee
Here we have used that $p \cdot n v_f \in \bz$ for all $p \in V$ and $p \in Sp$, by virtue of $n t_a \in \bz$ and eq. \eqref{orderc}.
We next examine the quadratic piece in $P$. 
When $n$ is odd, $nP^2 \in 2\bz$ always holds. This is because for $P \in I^*$, 
as follows from \eqref{theta}, $nP\in I$ which implies $(nP)^2\in2\bz$ 
and $n P^2 \in \bz$. When $n$ is even, we check case by case, aided by general results for 
 $P \in \Lambda_\cG^*(-1)$ such as $3P^2 \in 2\bz$ for $\Lambda_\cG=E_6$ and $4P^2 \in 2\bz$ for $\Lambda_\cG=D_6$. 
 We find that $n P^2 \in 2 \bz$  fails only in candidate number 3 in Table \ref{tab_t} with the lattice underlined. We leave 
for future study whether in this case modular invariance could still be satisfied following the formalism in \cite{Aldazabal:2025zht}, without doubling the order. 

For the linear piece in $P$ in eq. \eqref{lmcii} to be always even it suffices to impose $nv \in I$.
The remaining constant term just constraints $v^2$. Substituting  $E_1$ in eq. \eqref{Fs},
the definition of $v_f$ in eq. \eqref{defvf} and the condition in eq. \eqref{orderc}, we finally obtain that LMC reduces to
\vs{-5}\be\label{typeiilmc}
n v \in I\, , \qquad\qquad n v^2 \in 2\bz\ .
\ee
Recall that $I=\Lambda_\cG(-1)$.
We find that in all cases in Table \ref{tab_t} there exist shifts $v$ fulfilling these LMC conditions and furthermore
$v \notin I^*$ to avoid massless states in the $g$ sector. For islands,
additional constrains might be necessary to eliminate massless states in non-coprime sectors where modular invariance and operator 
interpretation also have to be analyzed. We will carry out this analysis for each model we study in section \ref{section_typeii_isl}.

When searching for shift vectors, we impose the constraint $\ell v \notin I^\ast$ for $0<\ell<n$. This condition is sufficient to guarantee that all massless states are projected out in twisted sectors and it is simpler to impose compared to the more general condition $\ell v \notin I^\ast_\ell$. As explained in more detail in appendix \ref{appv}, the stronger restriction we use yields a lower bound on the number of inequivalent shift vectors.
An extensive search for inequivalent shift vectors for all the models listed in Table \ref{tab_t} reveals that almost all of them allow for multiple inequivalent shifts, with these ``archipelagos" of islands becoming more populated as the order of the orbifold action increases. The exceptions are $\#$ 1 $\mathbb{Z}_4$, $\#$ 2 $\mathbb{Z}_6$, $\#$ 4 $\mathbb{Z}_6$, $\#$ 5 $\mathbb{Z}_8$ and $\#$ 7 $\mathbb{Z}_9$, which turn out to be single islands. Our definition of inequivalent shifts will be explained in appendix \ref{appv}. The list of inequivalent shifts for each model is available under request.

\section{Type II islands}\label{section_typeii_isl}

In the previous section we described a class of type II asymmetric $\bt^6/\bz_n$ orbifolds in which the automorphism
and the underlying even self-dual lattice $\Gamma$ are specified through the data in Table \ref{tab_t}.
We will now study these orbifolds in more detail. In particular, we will compute the entire partition function for the cases
\# 7 $\bz_9$, \# 4 $\bz_6$ and \# 1 $\bz_4$ in Table \ref{tab_t}. In these examples we will prove the full consistency of
the orbifold conformal field theory, including validity of the operator interpretation. We will also determine the shift vectors that
satisfy modular invariance and guarantee absence of massless states in twisted sectors. It turns out that in each case
there is only one inequivalent shift vector. Therefore, each orbifold leads to a 4-dimensional type II island. 
While our analysis strongly suggests that consistency of operator interpretation holds analogously for generic values of $n$, 
in the lack of a general proof the remaining entries in Table \ref{tab_t} are left as island candidates.

\subsection{\texorpdfstring{Type II $\bt^6/\bz_9$ }{TEXT}}\label{subsection_typeii_z9}

This model was originally studied in \cite{Dabholkar:1998kv}, and was shown to possess a shift vector which satisfies the
level-matching conditions and removes all massless states in the twisted sectors. Since the operator interpretation of the world-sheet theory 
and the equivalence classes of the shift vectors were not explicitly analyzed in that reference, we revisit the model here. We find that 
the orbifold has a valid operator interpretation and that it indeed has a unique equivalence class of island shift vectors. 

In this model we will also examine the spectrum of massless and massive states. For the island we will
determine the first massive states and organize them into BPS multiplets of $\cN=4$ supersymmetry in 4 
dimensions. The results could be useful to study duality with a heterotic island. We will also show that 
for other choices of the shift that do not eliminate massless twisted sectors, supersymmetry is extended
to $\cN=8$ or $\cN=5$.

The $(6,6)$ lattice $\Gamma$ is based on $\Lambda_\cG=E_6$, and will be denoted by $\Gamma(E_6)$. 
The $\bz_9$ automorphism belongs to the conjugacy class $\cE(a_1)$ of $E_6$. 
The invariant and normal lattices are $I=E_6(-1)$ and $N=E_6$.
The model is further characterized by
\vs{-7}
\be\begin{split}\label{z9e6dat}
&t=\frac19(1,2,5)\ ,\qquad\qquad\; E_1=\frac7{27}\ ,\qquad\quad\; E_3=\frac13\ ,\\[5pt]
&I^*\!/I \cong I_3^*\!/I_3\cong \bz_3\ ,\qquad\; \cF_1=1\ ,\qquad\qquad \cF_3=3\ .
\end{split}\ee
The fact that in this example $I_3 = I$ will simplify the analysis. 

The partition function terms in the untwisted and coprime twisted sectors, i.e. $Z(g^\ell, g^k)$ with $\ell=0$ or $(\ell,n)=1$, 
are obtained from eqs. \eqref{z1gl} and \eqref{zggl}. This is also the case for $Z(g^3,g^{3k})$ and $Z(g^6,g^{6k})$.
Modular invariance requires the shift $v$ to satisfy eq. \eqref{typeiilmc}
with $n=9$. To avoid massless states in coprime sectors it must be $v \notin I^*$, whereas $3v \notin I^*$ is
needed in the $g^3, g^6$ sectors. After imposing the equivalence relation discussed in appendix \ref{appv} we find that
there is only one inequivalent shift, chosen to be
\vs{-5}\be\label{vz9ii}
v=\frac19(4,7,10,7,4,5) \ ,\vs{-5}
\ee
in the root basis. It satisfies $9v^2=2$.

We still have to analyze the non-coprime $g^\ell$-twisted sectors with $\ell=3,6$. We will focus on $\ell=3$
since $\ell=6$ is analogous. As already mentioned, the terms $Z(g^3,\uno)$ and $Z(g^3,g^3)$ are given in eq. \eqref{zggl}.
The remaining orbits are deduced performing a $T^k S$ transformation of $Z(g,g^3)$ and $Z(g^2,g^3)$.
In this way we find
\vs{-5}\be\vs{-5}\begin{split}\label{iiz9g3gm1}
 &  Z(g^{3}, g^{3k -1}) =\Bigg\{\!\frac{e^{i \pi k (3v_{f})^{2}}}{\eta^{4}} \left( \sum_{p \in V} - \sum_{p \in Sp} \right)
q^{\frac{1}{2}(p + 3v_{f})^{2}}e^{ 2 \pi i p \cdot (3kv_{f})}e^{- 2 \pi ip \cdot v_{f}} \!\Bigg\}\\[3pt]
&\hs{40}\times \Bigg\{\!\frac{i \sqrt{3}\, \cF_1\,  e^{- 3i \pi v^{2}} e^{2 \pi i k E_{3}}}{\bar{\eta}^{6} \eta^{6}} 
\prod_{a=1}^{3}\frac{\eta^{3}}{\hvtgen{\frac12-\widetilde{3t_a}}{\frac12-k \widetilde{3t_{a}} + t_a}}
\sum_{P \in I} \bar{q}^{\frac{1}{2}(P+3v)^{2}_{R}} e^{i \pi k(P+3v)^{2}} e^{- 2 \pi i P \cdot v}\!\Bigg\}\ ,
\end{split}\ee
\vs{-5}\be\vs{-3}\begin{split}\label{iiz9g3gm2}
 &  Z(g^{3}, g^{3k -2}) =\Bigg\{\!\frac{e^{i \pi k (3v_{f})^{2}}}{\eta^{4}} \left( \sum_{p \in V} - \sum_{p \in Sp} \right)
q^{\frac{1}{2}(p + 3v_{f})^{2}}e^{ 2 \pi i p \cdot (3kv_{f})}e^{- 2 \pi ip \cdot 2v_{f}} \!\Bigg\}\\[3pt]
&\hs{30}\times \Bigg\{\!\frac{-i \sqrt{3}\, \cF_1 \,  e^{3i \pi v^{2}} e^{2 \pi i k E_{3}}}{\bar{\eta}^{6} \eta^{6}} 
\prod_{a=1}^{3}\frac{\eta^{3}}{\hvtgen{\frac12-\widetilde{3t_a}}{\frac12-k \widetilde{3t_{a}} + \widetilde{2t_{a}}}}
\sum_{P \in I} \bar{q}^{\frac{1}{2}(P+3v)^{2}_{R}} e^{i \pi k(P+3v)^{2}} e^{- 2 \pi i P \cdot 2v}\!\Bigg\}\ .
\end{split}\ee
Here $\cF_1=1$. Orbits in the $g^6$-twisted sector are analogous.
The above results will be used to study the consistency of the operator interpretation. 

\noindent{\underline{\bf Operator interpretation}\\[3pt]
We concentrate on the $g^3$-twisted sector. The other non-coprime sector is similar. 
A first basic check is that the momenta that appear in the orbits in eqs. \eqref{iiz9g3gm1} and \eqref{iiz9g3gm2}, namely
$(P+3v)$ with $P \in I$, are a subset of the momenta that appear in $Z(g^3,\uno)$, i.e. $(P+3v)$, with $P \in I_3^*$. 
This is satisfied simply because in this case $I_3=I$ and $I\subset I^*$. Moreover, $I$ is the invariant part of $I^*$ under $g$.

Next, to define the action of $g$ on the $\cF_3\!=\!3$ degenerate states in $Z(g^3,\uno)$, we extract the overall constant 
factor  $i\sqrt3 \, \cF_1$ in $Z(g^3,g^{-1})$ in eq. \eqref{iiz9g3gm1}.
Since $\cF_1=1$, this suggests that $g$ must act trivially on one state and by $e^{-\frac{2\pi i}3}$ on the remaining two states. 
Notice that $g^3$ acts trivially on all states, as it should since the overall factor in $Z(g^3,g^3)$ is also $\cF_3$.
The proposed action gives the overall factor in $Z(g^3, g^{-1})$ because 
$g^{-1}\cF_3\!=1+\!e^{\frac{2\pi i}3}+\!e^{\frac{2\pi i}3}\!=\!i\sqrt3\, \cF_1$. 
The $g$-action on other degrees of freedom is then automatically satisfied, namely the phase of a state in $Z(g^3,g^3)$ times the 
third power of the phase of the state in $Z(g^3,g^{-1})$ is 1. An analogous action is defined for the $g^6$ sector. We conclude that operator 
interpretation is consistent. Therefore, the $\bt^6/\bz_9$ model on $\Gamma(E_6)$ is an island.

\subsubsection{Spectrum}

The spectrum of massless and massive states can be determined from the full partition function in eq. \eqref{pfgen}. 
These states must fill supermultiplets of ${\mathcal N}=4$ supersymmetry in 4 dimensions.
In particular, massless states must arrange either into the gravity multiplet or vector multiplets.
On the other hand, massive states are organized into BPS multiplets of spin $j$, which can be long $L_j$,  
intermediate $I_j$ ($\frac14$ BPS), or short $S_j$ ($\frac12$ BPS), in the notation of \cite{Kiritsis:1997hj} where more details can be found. 
BPS multiplets can be distinguished by the helicity supertraces $B_{2n} \equiv \text{STr}\, \lambda^{2n}$, where $\lambda$ is 
the 4-dimensional helicity. For instance, only $B_4(S_j) \not=0$, implying that only $\frac12$ BPS multiplets contribute to 1-loop 
corrections to four-derivative couplings in the effective action \cite{Gregori:1997hi}. 

From the partition function we can extract the left and right masses in a generic $g^\ell$-twisted sector. They read
\vs{-5}
\be
m_L^2= \frac12 (p+\ell v_f)^2 + \frac12 P_L^2 + N_L -\frac12 + E_\ell, \qquad 
m_R^2=\frac12 r^2 + \frac12 (P+\ell v)_R^2 + N_R -\frac12,
\label{mtw}
\ee
where $E_\ell$ is the zero point energy, cf. \eqref{Fs}. In this orbifold $E_\ell=\frac{7}{27}$ for $\ell=1,2,4,5,7,8$, 
whereas $E_\ell=\frac13$ for $\ell=3,6$. 
In general the oscillator number $N_R$ is always integer but $N_L$ can be fractional for $\ell > 0$. 
The GSO projection requires that $p$ and $r$ are either in the $V$ or the $Sp$ class of $SO(8)$.
Level-matching further imposes $m_L^2=m_R^2$.

In the untwisted sector, massless states necessarily have $p^2=r^2=1$ while everything else vanishes. From the 
right-movers all states are invariant. However, from left-movers only \mbox{$p=(\pm1,0,0,0)$} is invariant while 
other vector and spinor weights pick up non-trivial phases $e^{2i \pi p\cdot v_f}$ under the orbifold
generator. The only invariant combinations form the gravity multiplet. In the twisted sectors it follows that there
are no massless states because $r^2 \ge 1$, $P \in I_\ell^*$, and by hypothesis $\ell v \notin I_\ell^*$, imply that $m_R^2 > 0$. 

Another distinctive feature of the spectrum is that there are no massless R-R states. Thus there are no massless R-R potentials to 
source standard BPS D-branes. This is a generic property of this class of type II asymmetric
orbifolds with 16 supercharges.  It would be interesting to study non-BPS D-branes, as initiated in
\cite{Bianchi:1998vq, Gutperle:2000bf}.

Modular invariance can be satisfied with trivial shift $v=0$. With this choice in this $\bt^6/\bz_9$ asymmetric orbifold,
in the twisted sectors we find extra massless states that complete the gravity multiplet of $\cN=8$ supersymmetry in four dimensions.
Therefore, with $v=0$ there is effectively a purely toroidal compactification, which implies the non-trivial identity
$\cz_g = Z(\uno,\uno)$, where $\cz_g$ is the double sum in eq. \eqref{fullzg}.

In the example at hand, modular invariance also allows non-zero shifts with $v \notin I^*$ but $3v \in I^*$. In this case 
additional massless states only arise in the $g^3$ and $g^6$-twisted sectors. They combine with untwisted massless states to
form the gravity multiplet of $\cN=5$ supersymmetry in four dimensions
(see e.g. \cite{Blumenhagen:2016rof} for multiplets of extended supersymmetry). 

Let us now briefly discuss massive states, taking the shift $v$ in eq. \eqref{vz9ii}.
After substituting the specific $Z(g^\ell, g^k)$ in eqs. \eqref{fullzg} and \eqref{pfgen}, 
we can compute the contribution to the partition function of  the four NS-NS, NS-R, R-NS, and R-R sectors.
Up to $m_L^2\!=\!m_R^2\le1$ we obtain
\vs{-5}\be\vs{-5}\begin{split}\label{z9full}
&\cz_{\rm NS-NS}=\cz_{\rm NS-R}=16+ 32\, (q\bq)^{\frac19}+ 144\, (q\bq)^{\frac29}+
1728\, (q\bq)^{\frac13}+ 2080\, (q\bq)^{\frac49}\\
&\hs{20}+ 5760\, (q\bq)^{\frac59}+ 31104\, (q\bq)^{\frac23}+ 28800\, (q\bq)^{\frac79}+
44880\, (q\bq)^{\frac89}+ 257696\, q\bq + \cdots\ ,\\[4pt]
&\cz_{\rm R-NS}=\cz_{\rm R-R}= 0 + 32\, (q\bq)^{\frac19}
+ 192\, (q\bq)^{\frac29}+
1728\, (q\bq)^{\frac13}+ 2080\, (q\bq)^{\frac49}\\
&\hs{20}+ 6144\, (q\bq)^{\frac59}+ 31104\, (q\bq)^{\frac23}+ 28800\, (q\bq)^{\frac79}+
42432\, (q\bq)^{\frac89}+ 256928\, q\bq + \cdots
\ .
\end{split}\ee
We have checked that the same results are obtained for other $v$'s in the same equivalence class.

As mentioned before, massive states can be organized into BPS multiplets. For instance, at order $(q\bq)^{\frac19}$ states
only appear in the $g$ and $g^8$-twisted sectors and fill the multiplets $4 S_0 + 2 S_{\frac12}$.
One easily finds $B_4(4 S_0 + 2 S_{\frac12})=0$ and $B_6(4 S_0 + 2 S_{\frac12})=-\frac{45}2$.
A more efficient way to keep track of the BPS multiplets at each mass level is to compute the helicity generating partition
function and then evaluate the derivatives that give the generating function of the helicity supertraces, also denoted $B_{2n}$ \cite{Kiritsis:1997hj}. In this way we obtain
\vs{-5}\be\label{b4z9}\vs{-5}
B_{4} = \frac{\bar\eta^6}{6\, \eta^2}\sum_{\ell=0}^8 \sum_{k=0}^8 Z(g^\ell, g^k) =
3 - 9(q \bar q)^{\frac{2}{9}} -72 (q \bar q)^{\frac{5}{9}} + 459 (q \bar q)^{\frac{8}{9}} + 0 q \bar q 
+ \cdots \ ,\\[5pt]
\ee
where we kept only level-matched terms.
For the generating function $B_6$ the analytic expression is less compact. The $(q\bq)$ expansion is given by
\vs{-5}\be\label{b6z9}\vs{-5}\begin{split}
& B_6 = \frac{195}{4} - \frac{45}{2} (q \bar q)^{\frac{1}{9}} - \frac{585}{4} (q \bar q)^{\frac{2}{9}} - 
1215 (q \bar q)^{\frac{1}{3}} - \frac{585}{2} (q \bar q)^{\frac{4}{9}} -2250 (q \bar q)^{\frac{5}{9}} \\
&\hs{25}-2250(q \bar q)^{\frac{7}{9}} + \frac{29835}{4} (q \bar q)^{\frac{8}{9}} - 
\frac{24705}{2}  q \bar q + \cdots \ .
\end{split}\ee
Our results reveal that only short multiplets contribute to the $B_6$ index up to $O(q\bq)$.

\subsection{\texorpdfstring{Type II $\bt^6/\bz_6$}{TEXT}}\label{subsection_typeii_z6}

In this model the $(6,6)$ lattice is again $\Gamma(E_6)$ based on $\Lambda_\cG=E_6$ but now the $\bz_6$ automorphism $\Theta$
belongs to the conjugacy class $\cE(a_2)$ of $E_6$. This is a new $\bz_6$ model which did not appear in \cite{Mizoguchi:2001cp}.
The invariant and normal lattices are $I=E_6(-1)$ and $N=E_6$. The eigenvalues and other relevant data are
\vs{-5}\be\vs{-5}\begin{split}\label{z4d6dat}
&t=\frac16(1,1,4)\ ,\qquad\qquad\; E_1=\frac14\ ,\qquad\quad\, E_2=\frac13\ ,\qquad\! E_3=\frac14\ ,\\[3pt]
&I^*\!/I \cong I_2^*\!/I_2\cong \bz_3\ ,\qquad I_3^*\!/I_3\cong \bz_2^2\ ,\qquad \cF_1=1\ ,\qquad \cF_2=3\qquad \cF_3=2\ .
\end{split}\ee
Clearly $I_2=I$ because $\Theta^2$ does not leave invariant directions. Instead, $I_3$ is a lattice of signature $(2,6)$
that can be determined and has the discriminant group given above. Specifically
\vs{-5}
\be\label{I3Z6typeii}
I_3 = (N_3, I) + (F,F) + (\bar F, \bar F) \, .
\ee
Here $N_3$ is the $\Theta^3$-invariant part of $N$, which has rank 2 and happens to be isomorphic to $A_2(2)$. 
In the additional correlated pairs ($\bar F$) $F$ stands for an $E_6$ weight in the (anti) fundamental representation chosen 
invariant under $\Theta^3$.

The terms $Z(g^\ell, g^k)$ for $\ell=0,1,5$ are obtained from eqs. \eqref{z1gl} and \eqref{zggl}. 
The modular invariance conditions on the shift $v$ are given in eq. \eqref{typeiilmc}
with $n=6$. To avoid massless states in all twisted sectors it must be $\ell v \notin I_\ell^*$. Taking into account
the equivalence relation discussed in appendix \ref{appv}, we find only one inequivalent shift that can be chosen to be $v=\frac16(4,8,11,8,5,6)$ 
in the root basis.

We now consider the remaining $g^\ell$-twisted sectors with $\ell=2,3,4$. 
The terms $Z(g^\ell,g^{k\ell})$ were already given in \eqref{zggl}. For $\ell=2,4$ there is
an additional orbit derived applying a $T^k S$ transformation to $Z(g,g^\ell)$.
In the $g^2$-sector we obtain
\vs{-4}\be\vs{-7}\label{zg2orbsz6}
\begin{split}
Z(g^2, g^{2k-1})&=\Bigg\{\!\frac{e^{- 2 \pi i v_f^2} }{\eta^4}\!\left( \sum_{p \in V} - \sum_{p \in Sp} \right)
q^{\frac12(p+2v_f)^2}e^{-2 \pi i p \cdot v_f} e^{i \pi k (2v_f)^2} e^{2 \pi i k p \cdot 2v_f}\!\Bigg\}\\[2pt]
&\times\Bigg\{\!\frac{\cF_1e^{i \frac{2\pi}3}e^{2 \pi ikE_2}e^{2 \pi iv^2}}{\eta^6 \bar \eta^6}
\prod_{a=1}^3\frac{\eta^3}{\hvtgen{\frac12- \widetilde{2 t_a}}{\frac12-k\,\widetilde{2 t_a}+ t_a}}\\[-4pt]
&\times\sum_{b=0}^2 \sum_{P \in I}\bar q^{\frac12(P+2bf^I\!+2v)^2_R}\, e^{i \pi k (P+2bf^I\!+2v)^2}
e^{- 2 \pi i (b f^I)^2} e^{-2 \pi i(P+ 2b f^I\!+2v)\cdot v}\!\Bigg\}\ ,
\end{split}\ee
where $f^I$ is the generator of the only $\bz_3$ class in $I^*\!/I$. The result for $Z(g^4,g^{4k-1})$ is analogous.

In the $g^3$-twisted sector there are instead two additional orbits of the form
\vs{-4}\begin{align}\label{zg3orbsz6}
&Z(g^3,g^{3k-1})=\Bigg\{\!\frac{e^{- 3 \pi i v_f^2} }{\eta^4}\!\left( \sum_{p \in V} - \sum_{p \in Sp} \right)
q^{\frac12(p+3v_f)^2}e^{-2 \pi i p \cdot v_f} e^{i \pi k (3v_f)^2} e^{2 \pi i k p \cdot 3v_f}\!\Bigg\}\\[4pt]
&\times\!\Bigg\{\!\frac{\cF_1e^{i \frac{\pi}2(k-1)}e^{-3 \pi iv^2}}{\eta^6 \bar \eta^6}\,
\frac{2 \sin(- \pi t_3)\eta^3}{\vtgen{\frac12}{\frac12+ t_3}}
\prod_{a=1}^2\!\frac{\eta^3}{\vtgen{\frac12- {3 t_a}}{\frac12-k\,{3 t_a}+ t_a}}\!
\sum_{P \in I}\bar q^{\frac12(P+3v)^2_R}\, e^{i \pi k (P+3v)^2}e^{-2i \pi P \cdot v}\!\Bigg\}\ ,\nn
\end{align}
\vs{-10}\begin{align}
&Z(g^3,g^{3k-5})=\Bigg\{\!\frac{e^{- 3 \pi i v_f^2} }{\eta^4}\!\left( \sum_{p \in V} - \sum_{p \in Sp} \right)
q^{\frac12(p+3v_f)^2}e^{-2 \pi i p \cdot 5v_f} e^{i \pi k (3v_f)^2} e^{2 \pi i k p \cdot 3v_f}\!\Bigg\}\\[4pt]
&\times\!\Bigg\{\!\frac{\cF_1e^{i \frac{\pi}2(k-1)}e^{-3 \pi i(5v)^2}}{\eta^6 \bar \eta^6}\,
\frac{2 \sin(- \pi 5t_3)\eta^3}{\vtgen{\frac12}{\frac12+ 5t_3}}
\prod_{a=1}^2\!\frac{\eta^3}{\vtgen{\frac12- {3 t_a}}{\frac12-k\,{3 t_a}+ 5t_a}}\!
\sum_{P \in I}\bar q^{\frac12(P+3v)^2_R}\, e^{i \pi k (P+3v)^2}e^{-2i \pi P \cdot 5v}\!\Bigg\}\ .\nn
\end{align}
Below we will use these results to check operator interpretation in the $g^3$-twisted sector.

\noindent{\underline{\bf Operator interpretation}\\[3pt]
Let us first consider the $g^2$-twisted sector. The basic condition for operator interpretation is fulfilled because
the effective momenta that appear in e.g. $Z(g^2,g)$, cf. \eqref{zg2orbsz6}, are $P+ 2v$ with $P \in I^*$,
which are the same momenta that appear in $Z(g^2,\uno)$, cf. \eqref{zggl}. Moreover, $I^*$ is invariant under
the orbifold action.
Concerning the group action, the validity of operator interpretation can be verified using the constrains derived in 
\cite[\S4]{Aldazabal:2025zht}. The $g^4$-twisted sector is analogous.

In the $g^3$-twisted sector, the primary condition for operator interpretation is also satisfied.
For instance in $Z(g^3,g)$, cf. \eqref{zg3orbsz6}, the momenta that occur are $(P+3v)$, with $P\in I$.
They are the $g$-invariant subset of the momenta that occur in $Z(g^3,\uno)$, namely $(P+3v)$, with $P \in I_3^*$
since $I$ is the invariant part of $I_3^*$, as can be deduced from \eqref{I3Z6typeii}. 
Now let us turn to the group action. 
In the $g^3$-twisted sector, the degeneracy factor in $Z(g^3,\uno)$ is $\mathcal{F}_3\!=\!2$ 
whereas in the other two orbits, the degeneracy factors is $\cF_1\!=\!1$. To define the action of $g$ on $\cF_3$ we absorb an 
overall constant phase $-1$ in $Z(g^3,g)$ and $Z(g^3,g^2)$ into the degeneracy factor $\cF_1$. 
We then posit that $g$ acts on the states encoded in $\cF_3$ by multiplication by appropriate phases, 
concretely by $e^{\frac{2 \pi i}{3}}$ on one and by $e^{-\frac{2 \pi i}{3}}$ on the other. 
Thus $g \mathcal{F}_3\!=\!e^{\frac{2 \pi i}{3}} \!+\!e^{-\frac{2 \pi i}{3}}\!=\!-1\!=\!-\cF_1$. 
At the same time, the action of $g$ on other world-sheet degrees of freedom is determined
from the remaining phases in $Z(g^3, g^j)$. From eqs. \eqref{zggl} and \eqref{zg3orbsz6} we find
\vs{-5}
\begin{align}\label{phasesz6}
g^3|p+3v_f, P+3v\rangle &= e^{i \pi (9v_f^2+ 2p \cdot 3 v_f)}\, e^{ i \pi(2 E_3 +  (P+3v)^2)}
|p+3v_f, P+3v\rangle, \ P \in I_3^*  \, , \nn \\[1mm] 
g^{-1}|p+3v_f, P+3v\rangle &= e^{-i \pi (3v_f^2+ 2p \cdot v_f)}\, e^{i \pi/2} \, e^{- i \pi (3v^2+ 2P\cdot v)}
|p+3v_f, P+3v\rangle, \ P \in I
\, .
\end{align}
It can be verified that the phase in the first line, for common $P \in I$, multiplied by the
$3^{\text{rd}}$ power of the phase in the second line gives one. 
Similar arguments apply to the $g^2$ action.
Hence, the operator interpretation in this asymmetric orbifold is consistent. We conclude that 
the $\bt^6/\bz_6$ model realized on $E_6$ is a unique type II island.

\subsection{\texorpdfstring{Type II $\bt^6/\bz_4$}{TEXT}}\label{subsection_typeii_z4}

In this example the left-moving automorphism is the order 2 involution $\Theta=-\uno$. However, the $g$-action on left-moving
fermions is of order 4. There are three allowed $(6,6)$ lattices $\Gamma(\cG)$ depending on whether $\Lambda_\cG$ is
chosen to be $6 A_1$, $D_4 + 2 A_1$ or $D_6$. We will see that the first option does not give rise to an island because 
massless states in the $g^2$-twisted sector cannot be avoided. The relevant data are
\vs{-5}\be\label{z4dat}
 t=\frac12(1,1,1)\ ,\quad E_1=\frac38\, ,\; E_2=0\ ,\quad 
 I^*\!/I \cong \left\{\begin{array}{ll}\!  \bz_2^6 &\  6A_1 \\[1mm]
\! \bz_2^4 & \ D_4+2A_1 \\[1mm]
\! \bz_2^2 & \ D_6 \end{array}
 \right.\, ,
 \quad
 \cF_1= \left\{\begin{array}{ll}\!  1 &\  6A_1 \\[1mm]
\! 2 & \ D_4+2A_1 \\[1mm]
\! 4 & \ D_6 \end{array}
 \right.\, . 
\ee
Since $\Theta^2=\uno$, $I_2=\Gamma(\cG)$ and $\cF_2=1$ for all the three allowed $\Lambda_\cG$  lattices.

An interesting property of this orbifold is that the $g^2$-action in the untwisted sector \eqref{z1gl} has to be modified as
\vs{-5}\be\label{actg2lat}
g^2: |P_N,P_I\rangle  \to e^{2\pi i P \cdot (2v+w)} | P_N,P_I\rangle\ ,\vs{-5}
\ee
where $w$ is a constant vector with the properties
\vs{-5}
\be\label{wprops}
w\in I^*\, , \qquad P^2 = P \cdot w \, , \ \forall \, P \in I^* \, .
\ee
Such a $w$ exists for all choices of $\Lambda_\cG$ in eq. \eqref{z4dat}.
For $\Lambda_\cG\!=\!D_6$, $w$ is in the vector conjugacy class of $D^*_6/D_6$, whereas for the other two choices
$\Lambda_\cG=6A_1$ or $\Lambda_\cG\!=\!D_4+2A_1$, $w$ is in the fundamental class of each of the $A^*_1/A_1$ factors. 
Concretely, we can take
\vs{-5}
\be\label{wvecs}
w= \frac12(1,1,1,1,1,1) \ \, 6A_1 \, , \quad 
w= \frac12(0,0,0,0,1,1) \ \, D_4+2A_1 \, , \quad
w= \frac12(2,2,2,2,1,1)  \  \, D_6 \, ,
\ee
in the root basis of each $\Lambda_\cG$.

A similar modification of the $g^2$ action was introduced in \cite{Acharya:2022shu}.
To explain the reason why it is needed, let us first compute $Z(g^2,g^{-1})$ following the path of modular transformations
\vs{-5}
\be\label{Z4path}
Z(\uno,g)\xrightarrow{S}Z(g,\uno)\xrightarrow{T^{2}}Z(g,g^2)\xrightarrow{S}Z(g^2,g^{-1})\ .\vs{-5}
\ee
Using eq. \eqref{zggl}, we find\vs{-5}
\be\vs{-5}\begin{split}\label{Z4gg2}
Z(g,g^2)=\Bigg\{\!\frac{e^{-i \pi 2 v_f^2}}{\eta^4} & \left( \sum_{p \in V} - \sum_{p \in Sp} \right)
q^{\frac12(p+v_f)^2}e^{2 \pi i(p+v_f)\cdot 2v_f}\!\Bigg\}\\[2pt]
&\times\Bigg\{\!\frac{\mathcal{F}_1\, e^{4 \pi i E_1}}{\eta^6 \bar \eta^6} \prod_{a=1}^{3} \frac{\eta^3}
{\vartheta \begin{bmatrix}\frac12 -t_a\\ \frac12-2t_a\end{bmatrix}} \sum_{P \in I^*} \bar q^{\frac12 (P+v)^2_{R}} e^{i \pi 2 (P+v)^2}\!\Bigg\} \ .
\end{split}\ee
The phase inside the sum in the second line has a $P$-dependent component $e^{2\pi i P^2}$, where $P\!\in\!I^*$. 
For all choices of $\Lambda_\cG$ in eq. \eqref{z4dat}, it happens that $2 P^2$ is an integer for all $P \in I^*$, but the integer 
can be even or odd. This non-trivial phase quadratic in $P$ obstructs a direct Poisson resummation to obtain $Z(g^2,g^{-1})$.
However, a constant vector $w$ with the properties in eq. \eqref{wprops} exists for the $\Lambda_\cG$'s at play.
Hence we can write  $e^{2\pi i P^2}=e^{2\pi i P\cdot w}$ for all $P\in I^*$. 
Then, upon performing a Poisson resummation, eq. \eqref{Z4gg2} transforms into\vs{-5}
\be\vs{-5}\label{zg2gz4}\begin{split}
&Z(g^2,g^{2k-1})=\Bigg\{\!\frac{e^{i \pi 2v_f^2}}{\eta^4} \left( \sum_{p \in V} - \sum_{p \in Sp} \right)
q^{\frac12(p+2v_f)^2}e^{-2 \pi i(p+ 2v_f)\cdot v_f}e^{i \pi k (p+2v_f)^2}\!\Bigg\}\\[3pt]
&\hs{62}\times\Bigg\{\!\frac{\mathcal{F}'_1\, e^{4 \pi i E_1}\, e^{2 \pi i v^2}}{\eta^6 \bar \eta^6} 
\left(\frac{2\eta^3}{\vartheta_2} \right)^3 \sum_{P \in I} \bar q^{\frac12(P+2v + w)^2_{R}}
e^{-2 \pi i (P+2v+w)\cdot v} e^{i \pi k (P+2v+w)^2}\!\Bigg\}\ ,\\
&\cF'_2=\sqrt{|I^*/I|}\;{\cF_2}\;2^{-3}\ .
\end{split}\ee
It turns out that $\cF'_2=1$ for all choices of $\Lambda_\cG$. 

We observe that the states in the $(g^2,g)$ orbit have momenta in $I\!+\!2v\!+\!w$. Now considering eq. \eqref{zggl}, with the 
$g^2$-action as in \eqref{gactlatt}, we notice that $Z(g^2,\uno)$ has momenta in $\g\!+\!2v$. Thus, the states in 
$Z(g^2,g^{-1})$ would not be a subset of those in $Z(g^2,\uno)$, violating a basic requirement for a consistent operator interpretation.
The conclusion is that the action of $g^2$ in the untwisted sector must acquire an extra phase as defined in eq. \eqref{actg2lat}
to yield
\vs{-5}\be\vs{-5}\label{g12z4}
Z(\uno,g^2)=\Bigg\{\!\frac{1}{\eta^4}\!\left( \sum_{p \in V} - \sum_{p \in Sp} \right)
q^{\frac12p^2} e^{2 \pi i p \cdot 2 v_f}\!\Bigg\}
\Bigg\{\!\frac{1}{\eta^6 \bar \eta^6}\! \sum_{P \in \Gamma}\!\!
q^{\frac12 P_{L}^{2}} \bar q^{\frac12 P_{R}^{2}} e^{2 \pi i P \cdot (2 v+w)}\!\Bigg\}\ .
\ee
After a $T^kS$ transformation we then obtain
\vs{-5}\be\vs{-5}\label{g22kz4}\begin{split}
&Z(g^2,g^{2k}) = \Bigg\{\!\frac{e^{-i \pi k(2v_f)^2}}{\eta^4}\left( \sum_{p \in V} - \sum_{p \in Sp} \right)
q^{\frac12(p+2v_f)^2}e^{2 \pi i(p+2v_f)\cdot 2kv_f}\!\Bigg\}\\
&\hs{165}\times\Bigg\{\!\frac{1}{\eta^6 \bar \eta^6} \sum_{P \in \Gamma}
q^{\frac12(P+2v+w)^{2}_{L}} \bar q^{\frac12(P+2v+w)^{2}_{R}} e^{i \pi k (P+2v+w)^2}\!\Bigg\}\ .
\end{split}\ee
The numerical prefactor in $Z(g^2, g^{2k})$ is $\cF_2=1$ because $\Gamma$ is even self-dual.

Finally, eq. \eqref{g22kz4} shows that with the modified $g^2$-action, the states in $Z(g^2,\uno)$ have momenta in 
$\Gamma\!+\!2v\!+\!w$. Therefore, now $Z(g^2,g^{-1})$ in eq. \eqref{zg2gz4} is consistent with operator interpretation
as discussed below in greater detail.

Note that the LMC in the $g$-twisted sector is still given by eq. \eqref{typeiilmc} with $n=4$. This is because $Z(g,g^\ell)$
in eq. \eqref{zggl}, and its general consequence \eqref{lmcii}, are not modified, while $4P^2 \in 2\bz$ for all choices of
$\Lambda_\cG$. 
Besides the LMC, we also demand that the shifts $v$ ensure absence of massless states in twisted sectors.
In the $g$-twisted sector the constraint is $v \notin I^*$, as deduced from $Z(g,1)$ in eq. \eqref{zggl}. In
the $g^2$-twisted sector, as follows from $Z(g^2,\uno)$ in eq. \eqref{g22kz4},
the requirement is instead $2v + w \notin I^*$, which simplifies to $2 v \notin I^*$.
For $\Lambda_\cG=6A_1$ we find that there exist no shift vectors satisfying the latter requisite and LMC.
For $\Lambda_\cG$ equal to $D_4 + 2 A_1$, or $D_6$, there exists a unique equivalence class of shift vectors that 
satisfy the LMC and remove all massless states in all twisted sectors. The representatives can be chosen to be
\vs{-5}
\be\label{vz4typeii}
v=\frac14(2,3,2,2,2,1) \ \ D_4 + 2A_1\, , \qquad v=\frac14(2,3,4,5,3,3) \ \ D_6 \, .
\ee
in the root lattice of each $\Lambda_\cG$.

\noindent{\underline{\bf Operator interpretation}}\\[3pt]
The above results allow to verify the validity of operator interpretation in the $g^2$-twisted sector. 
We first remark that the momenta appearing in $Z(g^2,g^{-1})$ are $P+2v+w$ with $P\in I$.
These are precisely the momenta that survive upon inserting $g^{-1}$ in the trace of the Hilbert space in 
$Z(g^2,\uno)$ where states have $P+2v+w$ with $P\in \Gamma$.

We next examine the phases. Since $\mathcal{F}_2=1$ and there are two T-orbits we have to consider a $\mathbb{Z}_2$ 
action on the fixed locus. We impose trivial $g$-action $g \cF_2=\cF_2$.
From eqs. \eqref{zg2gz4} and \eqref{g22kz4} we then obtain
\vs{-5}\begin{align}\label{phasesz4}
g^{-1}|p+2v_f, P+2v+w\rangle &= e^{-2i \pi (v_f^2+ p \cdot v_f)}\, e^{2 i \pi (2 E_1 + v^2- (P+2v+w)\cdot v)}
|p+2v_f, P+2v+w\rangle, \ P \in I \, , \nn \\[1mm] 
g^2|p+2v_f, P+2v+w\rangle &= e^{4i \pi (v_f^2+ p \cdot v_f)}\, e^{ i \pi (P+2v+w)^2}
|p+2v_f, P+2v+w\rangle, \ P \in \Gamma \ .
\end{align}
It can be checked that the square of the phase in the first line multiplied by the phase in the second line gives one, 
provided that $w^2$ is odd. This is always the case for the $w$'s given in eq. \eqref{wvecs}. Thus the operator interpretation 
is consistent.

To conclude we make some comments about the massless spectrum.
For $\Lambda_\cG$ equal to $D_4 + 2A_1$ or $D_6$, choosing the shift vectors in eq. \eqref{vz4typeii}, we find as expected that the
massless states arise only in the untwisted sector and fill the gravity multiplet of $\cN=4$ supersymmetry. 
We also obtain that setting $v=0$, which is a solution to the LMC, there will be additional massless states in the twisted sectors
that enhance supersymmetry to $\cN=8$ for $\Lambda_\cG=D_6$, to $\cN=6$ for $\Lambda_\cG=D_4 + 2 A_1$, and to
$\cN=5$ for $\Lambda_\cG=6A_1$.

\section{Final comments}\label{section_disc}

We have built 4-dimensional,  $\cN=4$ supersymmetric string vacua without massless vector multiplets as $\bt^6/\bz_n$ asymmetric 
orbifolds of type II and heterotic strings. Such vacua, referred to as islands, are isolated in
the sense that they have no continuous moduli other than the dilaton belonging to the universal gravity multiplet.
String islands have been realized by way of asymmetric orbifolds before, see e.g.  \cite{Dabholkar:1998kv, Mizoguchi:2001cp, Baykara:2025lhl}. 
This work, and the precedent \cite{Aldazabal:2025zht}, stand apart because the partition function is derived rigorously and the
consistency of the orbifold conformal field theory is proven beyond the trivial case of $n$ prime.
In particular, we have established the validity of the operator interpretation by carrying out a careful analysis of
the Hilbert spaces in twisted sectors. 
We have also remarked that in some cases there appear {\em archipelagos}, i.e. groups of islands emerging
from the same $\bt^6/\bz_n$ orbifold but differing in the massive spectrum.

A comprehensive construction and validation of $\bt^D/\bz_n$ heterotic asymmetric orbifolds having
$n=2m$, with $m$ prime greater than 2, was already accomplished in \cite{Aldazabal:2025zht}. Here we have delivered
general results for $n=3m$, with $m$ prime greater than 3, and we have also discussed other orders in
detail. In addition, we have extended the analysis to type II setups. We have concentrated on 4-dimensional vacua
but 6-dimensional cases can be considered as well.

Although we only presented some specific islands to illustrate the approach, our methods can be applied to work out many other examples.
In fact, in the heterotic case, in Table \ref{HMisl},  we have provided the list of all possibilities based on using Leech lattice data 
to design the $\Gamma(22,6)$ lattice and to select the $\bz_n$ automorphism. In the type II case we also classified 
all possible $\bz_n$ automorphisms acting on $\Gamma(6,6)$ lattices built from ADE root lattices.

We have focused on even-self dual $\Gamma(22,6)$ lattices built from data associated with automorphisms of the Leech lattice
because this ensures that the normal lattice has no roots, a necessary condition for rank reduction and, in particular, for 
heterotic islands \cite{Aldazabal:2025zht}. However, it is conceivable that islands could also be found using a different construction
of the $\Gamma(22,6)$ lattice and the automorphism, provided they satisfy two requirements: (1) the invariant lattice and normal lattices 
have signatures $(0,6)$ and $(22,0)$ respectively, and (2) the normal lattice has no roots. A classification of all possible automorphisms
of order two \cite{Nikulin80int} and three \cite{Brandhorst19} of $\Gamma(22,6)$ shows that the first 
requirement can be met, unlike in the Leech scheme, but whether the second is satisfied remains to be determined. 
This interesting problem is being addressed in upcoming work \cite{HPDF}.

The formalism of asymmetric orbifolds can be exploited to survey other components of the moduli space of 4-dimensional,  
$\cN=4$ supersymmetric string vacua, which are characterized by the rank of the gauge group carried by massless vector multiplets.
For type II vacua, it has been shown that only even ranks between 0 and 6 are allowed by unitarity of the
underlying superconformal field theory, which also constraints the possible gauge groups \cite{Dixon:1987yp}. 
For heterotic vacua, the same ranks arise and additionally ranks $22,14,10$ and $8$ are known to appear
\cite{Chaudhuri:1995fk, Chaudhuri:1995bf, Chaudhuri:1995ee, Gregori:1997hi, Mikhailov:1998si, deBoer:2001wca, Persson:2015jka, 
Fraiman:2022aik}. It would be interesting to look into the question of why some even ranks seem to be forbidden.

As a byproduct of our scan of 4-dimensional $\cN=4$ type II islands, we found 4-dimensional vacua with $\cN=5$ and $\cN=6$ supersymmetry.
This occurred in cases when the shift vector was absent or allowed massless states in twisted sectors, as in an example found
in \cite{Dabholkar:1998kv}. Type II asymmetric orbifolds leading to 4-dimensional vacua with $\cN=5,6$ supersymmetry
were originally considered in \cite{Ferrara:1989nm}, see also \cite{Bianchi:2008cj, Bianchi:2010aw, Blumenhagen:2016rof, Bianchi:2022tbr, Bianchi:2023hbo}. A feasible next step is to search systematically for such type II asymmetric orbifolds.

We were further motivated to consider 4-dimensional $\cN=4$ supersymmetric type II and heterotic islands in
order to study their possible strong coupling duals, which are 
expected to be islands as well, as argued in \cite{Dabholkar:1998kv}. 
While our results could be useful to address this open problem, we leave a closer investigation for future 
work\footnote{
We have been informed that this problem is considered in a forthcoming publication in which 4-dimensional string islands
are also discussed\cite{HPDF2}.
}.
Some aspects of S-duality in models with reduced rank have been studied in \cite{Persson:2015jka,Persson:2017lkn}.

\section*{Acknowledgments}
We thank H. Parra De Freitas and S. Theisen for helpful discussions. We are grateful to H. Parra De Freitas for reading the manuscript and sharing insightful comments.
G.A. acknowledges support by a Simons Targeted Grant to Instituto Balseiro. His work was partially supported by the CONICET PIP 2025, 11220250101185. A.F. thanks the Edwin Schr\"odinger Institute, within the program ``The Unreasonable Effectiveness of Toric Geometry: Bridging Mathematics, Computation, and String Theory", as well as the Instituto Balseiro, within the J.M. Maldacena Visiting Professors Program, for hospitality and support during completion of this work. I.G.Z. was supported by a UKRI Future Leaders Fellowship (project reference UKRI2349).

Mathematica, Magma, and SageMath were used to compute non-equivalent shift vectors and the spectra. ChatGPT, Claude, Copilot, and Gemini were also used to help with the computation of non-equivalent shift vectors.

\appendix

\section{\texorpdfstring{$\bt^6/\bz_8$}{TEXT}}\label{app_Z8}
In this appendix we compute the partition function of the $\bz_8$ model HM224 of section \ref{hm224}. The exponents of the eigenvalues of $\Theta$ are collected in vector $t$ \vs{-5}
\be\label{8eigens}
t=\frac1{8}\Big([1,3]^{n_1},[2]^{n_2},[4]^{n_4}\Big)\ ,\qquad n_1=4\ ,\;n_2=2\ ,\;n_4=1\ .\vs{-5}
\ee
The partition function in the coprime sectors and the $(g^\ell,\uno)$ orbits in the even twisted sectors is given by eq. \eqref{hzglgkl}. Since the $g^2$ and $g^6$ sectors have two orbits each, whereas the $g^4$ sector has four orbits, we will analyze them individually.

\subsection{\texorpdfstring{$g^2$ and $g^6$}{TEXT} sectors}\label{app_z8_g2g6}
To obtain the $(g^\ell,g^{-1})$ orbit, with $\ell\!=\!2,6$, we start from the untwisted sector and take the following path of $S$ and $T$ modular transformations
\vs{-5}\be\label{pellm1}
Z(\uno,g)\xrightarrow{S}Z(g,\uno)\xrightarrow{T^{\ell}}Z(g,g^{\ell})
\xrightarrow{S}Z(g^\ell,g^{-1})\xrightarrow{T^k}Z(g^\ell,g^{k\ell-1})\ ,
\qquad\ell=2,6\ .\vs{-5}
\ee
The first line of eq. \eqref{hzglgkl} gives\vs{-5}
\be\label{Zgglj_z8}
Z(g,g^{\ell})=\frac{\cF_1}{\eta^{22} \bar\eta^{6}}\, 
\prod_{a=1}^{11}\frac{ e^{i\pi(\frac12 - t_a+\ell t_a^2)}\, \eta^3}
{\vtgen{\frac12- t_a}{\,\frac12-\ell t_a} }\sum_{Q\, \in \, I^*} 
\bar q^{\frac12 (Q+ v)_R^2}\, e^{i\pi \ell (Q+ v)^2}\ ,\vs{-10}
\ee
where $e^{i\pi\mu}\!=\!1$ as seen in eq. \eqref{mu}. We next compute the $T^kS$ transformation of this partition function.

Let us first examine the contribution from right-moving compact directions, which contains the lattice sum, and is denoted by
\vs{-5}\be\label{Zgglj_z8_lat}
Z_{\rm lat}(g,g^{\ell}):=\frac1{\bar\eta^{6}}\!\sum_{Q\, \in \, I^*} 
\bar q^{\frac12 (Q+ v)_R^2}\, e^{i\pi \ell (Q+ v)^2}\ .\vs{-10}
\ee
Recall from section \ref{hm224} that $I^*\!/I\cong \bz_4^2$ with generators $f_1^I$ and $f_2^I$. Any vector $Q\!\in\! I^*$ may be written as
\vs{-5}\be\label{Istar224}
Q=P+(2a+a')f_1^I+(2b+b')f_2^I\ ,\qquad\qquad a,a',b,b'\in\{0,1\}\ ,
\qquad\qquad P\in I\ .\vs{-5}
\ee
The norm of this vector is
\vs{-5}\be\vs{-5}\begin{split}\label{Qnorm224}
Q^2&=P^2+2P\cdot\big((2a+a')f_1^I+(2b+b')f_2^I\big)+\big((2a+a')f_1^I+(2b+b')f_2^I\big)^2\\
&=a'^2(f_1^I)^2+2a'b'\,f_1^I\cdot f_2^I+b'^2(f_2^I)^2\;\;{\rm mod}\;\bz=
-\frac34a'^2-\frac32b'^2-\frac12a'b'\;\;{\rm mod}\;\bz\ ,
\end{split}\ee
where in the second line we used eq. \eqref{disc224}. Inserting eqs. \eqref{Istar224} and \eqref{Qnorm224} in eq. \eqref{Zgglj_z8_lat}, we obtain
\vs{-5}\be\label{Zgglj_z8_lat_i}
Z_{\rm lat}(g,g^{2j})=\frac{e^{-2\pi ijv^2}}{\bar\eta^{2d}}\!\!\sum_{a',b'=0}^1\!F(a'f_1^I+b'f_2^I)\;
e^{-2\pi ij\big(\tfrac {3}4a'^2+\tfrac {b'}2(a'+3b')\big)}\ ,\vs{-10}
\ee
where $j=1,3$, and
\vs{-10}\be\vs{-5}\begin{split}\label{Zgglj_z8_lat_ii}
F(a'f_1^I+b'f_2^I)&:=\sum_{\!\!a,b=0}^1\sum_{P\, \in \, I}
\bar q^{\frac12 (P+(2a+a')f_1^I+(2b+b')f_2^I+v)_R^2} e^{2\pi i(P+(2a+a')f_1^I+(2b+b')f_2^I+v)\cdot 2jv}\\
&\;=\sum_{P'\in I'}\bar q^{\frac12 (P'+a'f_1^I+b'f_2^I+v)_R^2} e^{2\pi i (P'+a'f_1^I+b'f_2^I+v)\cdot 2jv}\ .
\end{split}\ee
The vectors $P'=P+2af_1^I+2bf_2^I$ have norm $P'^2\in\bz$ and form the lattice $I'\!\subset\!I^*$ which is odd self-dual. We next perform a Poisson resummation on the sum $F$ in eq. \eqref{Zgglj_z8_lat_i}, and find
\vs{-5}\be\vs{-5}\begin{split}\label{Zgljg_z8_lat}
Z_{\rm lat}(g^{2j},g^{-1})&=\frac{e^{-2\pi ijv^2}}{\bar{\eta}^{2d}}\hs{-5}
\sum_{a,b=0}^1\sum_{P\in I}
\bar q^{\frac12 (P+2jaf_1^I+2jbf_2^I+ 2jv)_R^2} e^{-2\pi i (P+2jaf_1^I+2jbf_2^I)\cdot v}\\[-6pt]
&\hs{90}\times\sum_{a',b'=0}^1\!\!e^{2\pi i j\big(\tfrac32aa'+\tfrac12(ab'+a'b)\big)}
e^{-2\pi ij\big(\tfrac {3}4a'^2+\tfrac{b'}2(a'+3b')\big)}\ ,
\end{split}\ee
where in the first line we used $(j,2)=1$ to rescale the generators of $I'$, and in the second line we used eq. \eqref{disc224}. The sum in the second line of the above expression is evaluated to be
\vs{-5}\be\vs{-5}\begin{split}\label{Zgljg_z8_lat_i}
&1-e^{i\pi ja}+e^{i\frac\pi2j}e^{i\pi jb}+e^{i\frac\pi2j}e^{i\pi j(a+b)}=
1-e^{i\pi a}+e^{i\frac\pi2j}e^{i\pi b}+e^{i\frac\pi2j}e^{i\pi(a+b)}=:2e^{i\pi\theta(a,b)}\ ,\\
&\theta(0,0)=-\theta(0,1)=\tfrac j2\ ,\qquad
\theta(1,0)=-\theta(1,1)=0\ .
\end{split}\ee

The lattice sum in the $(g^{2j},g^{-1})$ orbit \eqref{Zgljg_z8_lat} is over $I'$ which contains vectors $I+2af_1^I+2bf_2^I$. The vectors $2af_1^I $ and $2bf_2^I$ are indeed elements in $I_2^*$, as observed in eq. \eqref{hm224I2d}. Therefore, $I'\subset I_2^*$ and no modification by $w$-type vectors to the $g^{2}$ and $g^6$ action in the untwisted sector is required. We note that $I_2^*$ contains $f_2^I$, but that $I'$ only contains $2f_2^I$. Given eqs. \eqref{Zgljg_z8_lat} and \eqref{Zgljg_z8_lat_i}, we conclude that $g$ acts by $-2i$ on $I+2f_2^I$, and by zero on $I+f_2^I$ and $I+2f_1^I$ in $Z(g^{2},\uno)$. One may obtain the same result \eqref{Zgljg_z8_lat} using the method introduced in 
\cite[app. D]{Aldazabal:2025zht} via defining the function $L$ to perform the Poisson resummation.

To obtain the partition function $Z(g^{2j},g^{2kj-1})$, we also need to compute the $T^kS$ transformation of the contribution from the left-moving compact directions in $Z(g,g^{2j})$ in eq. \eqref{Zgglj_z8}, which contains the oscillators modes. All in all, we find
\vs{-5}\be\vs{-5}\begin{split}\label{hetz8g2}
&Z(g^{2j},g^{2kj-1})\!=\!\cF'_2\,\frac{e^{2\pi ik E_2}e^{i\pi(\Delta_1(2j,1,8)+\Delta_2(2j,1,8))}
e^{2\pi ijv^2}}{\eta^{22} \bar\eta^{6}}\!\!\!\prod\limits_{\substack{a=1\\2 t_a\in\bz}}^{11}\!
\frac{-2\sin(\pi t_a) \eta^3}{\vtgen{\frac12}{\frac12+ t_a}}\!\prod_{\substack{a=1\\2 t_a\notin\bz}}^{11}\!
\frac{\eta^3}{\hvtgen{\frac12- \widetilde{2j t_a}}{\frac12-k\,\widetilde{2j t_a}+ {t_a}}}\\[-5pt]
&\times\sum_{a,b=0}^{1}\sum_{P \in I}\bar q^{\frac12 (P+2j a f^I_1+2j b f^I_2+2j v)_R^2}
e^{i\pi k(P+2j a f^I_1+2j b f^I_2+2j v)^2}e^{-2\pi i(P+2j a f^I_1+2j b f^I_2+2j v)\cdot v}
e^{i\pi\theta(a,b)}\ ,\nn
\end{split}\ee
\be
\hs{-300}\cF'_2=2\,\frac{\cF_1}{2^{n_4}}=4\ ,\qquad j=1,3\ .
\ee
Above we have defined
\vs{-10}\begin{align}
&e^{i\pi\Delta_1(\ell,j,n)}:=\prod\limits_{\substack{a=1\\\ell t_a\in\bz}}^{11}\!\!
e^{i\pi n\fm(t_a-jt_a^2)}e^{-i\pi\ell t_a(1+jt_a)}\ ,\label{hetz3mdel1}\\
&e^{i\pi\Delta_2(\ell,j,n)}:=\!\prod_{\substack{a=1\\\ell t_a\notin\bz}}^{11}\!
e^{\frac{i\pi}2}e^{i\pi n\fm(t_a-jt_a^2)}e^{i\pi(\lfloor \ell t_a\rfloor-\widetilde{jt_a})}\,
e^{i\pi(\widetilde{\ell t_a}-\lfloor \ell t_a\rfloor)jt_a}\ ,\vs{-10}\label{hetz3mdel2}
\end{align}
where $\fm$ is a non-negative integer determined by the path of modular transformations, and $\fm\!=\!0$ in the present case. We will soon encounter paths with non-zero $\fm$ -- see eq. \eqref{z8m}.

\subsubsection{Operator interpretation}\label{pint_224_g2}
For states appearing in both $Z(g^{2j},g^{2j})$ and $Z(g^{2j},g^{-1})$, the phase in the former times the ${2j}^{\rm th}$ power of the phase in the latter must be one. As for the degeneracy factors, $g$-action on $\cF_{2}=8$ in $Z(g^{2j},\uno)$ must yield $\cF'_{2}=4$ in $Z(g^{2j},g)$.

The states with $P'\!=\!P+2j\,af_1^I+2j\,bf_2^I$, with $P\!\in\! I$, appear in both orbits. The phase in $Z(g^{2j},g^{2j})$ is given by eq. \eqref{hzglgkl}\vs{-5}
\be\label{ph22}
e^{i \pi 2E_2}e^{i\pi (P+2jaf_1^I+2jbf_2^I+2j v)^2}=e^{-i\pi 3a^2}
e^{2\pi i(P+2jaf_1^I+2jbf_2^I)\cdot 2jv}e^{i\pi4j^2v^2}\ ,\vs{-5}
\ee
where we used eq. \eqref{disc224} together with $j=1,3$ and $E_2=1$. The phase in $Z(g^{2j},g^{-1})$ is given by eq. \eqref{hetz8g2} with $k=0$\vs{-5}
\be\label{ph22_i}
e^{2\pi ik E_2}e^{i\pi(\Delta_1(2j,1,8)+\Delta_2(2j,1,8))}
e^{-i\pi 2jv^2}e^{-2\pi i(P+2j a f^I_1+2j b f^I_2)\cdot v}e^{i\pi\theta(a,b)}\ ,\vs{-5}
\ee
where $e^{i\pi(\Delta_1(2j,1,8)+\Delta_2(2j,1,8))}=e^{i\frac{\pi}2j}$. Using eq. \eqref{Zgljg_z8_lat_i}, we find that the phase \eqref{ph22} times the $2j^{\rm th}$ power of the phase \eqref{ph22_i} is indeed 1
\vs{-5}\be\label{ph22_ii}
e^{-i\pi 3a^2}e^{i\pi j^2}e^{i\pi2j\theta(a,b)}=1\ ,\;\;\forall\, a,b\ .
\ee

For the degeneracy factors, we note that $\cF_2-\cF'_2=8-4=4$. Hence, $g$ acts on 6 states in $\cF_2$ as $+1$, and on the remaining 2 states as $-1$, and yields $\cF'_2=4$. We conclude that operator interpretation in the $g^2$ and $g^6$ sectors is consistent.

\subsection{\texorpdfstring{$g^4$}{TEXT} sector}\label{app_z8_g4}
The path of modular transformations for the $(g^4,g^{-j})$ orbits is
\vs{-5}\be\label{pathg4}
Z(\uno,g^j)\xrightarrow{S}Z(g^j,\uno)
\xrightarrow{T^l}Z(g^j,g^{l j})\equiv Z(g^j,g^4)
\xrightarrow{S}Z(g^4,g^{-j})\xrightarrow{T^k}Z(g^4,g^{4k-j})\ ,\vs{-5}
\ee
where
\be\label{z8m}
l=\frac{4}{(j,4)}\ ,\qquad l j=\frac{4j}{(j,4)}=8\fm+4\ ,\qquad j=\{1,2,3\}\ .
\ee
The partition function $Z(g^j,g^{lj})$ is obtained from eqs. \eqref{hzglgkl} and \eqref{mu} to be of the form\vs{-5}
\be\label{Zg4g4l_z8}
Z(g^j,g^{l j})=\frac{\cF_j}{\eta^{22} \bar\eta^{6}}\,\prod_{a=1}^{11}
\frac{ e^{i\pi\big(\frac12 -j  t_a+\lfloor j t_a\rfloor+l(j t_a)^2\big)}\, \eta^3}
{\vtgen{\frac12-j t_a}{\,\frac12-l j t_a} }\!\sum_{Q\, \in \, I^*_j} \!
q^{\frac12 (Q+j v)_L^2}\,\bar q^{\frac12 (Q+j v)_R^2}\, e^{i\pi l (Q+j v)^2}\ .\vs{-5}
\ee
We shall next compute $Z(g^4,g^{-j})$ by applying $T^kS$ transformations to the above partition function. Since the lattice sum is over $I^*$ for $j\!=\!1,3$, and over $I_2^*$ for $j\!=\!2$, we will treat the contribution from the right-moving part separately for these two cases.

\subsubsection{\texorpdfstring{$Z_{\rm lat}(g^4,g^{-j})$, $j=1,3$}{TEXT}}\label{s_hetz8g4latt_i}
For $j=1,3$, eq. \eqref{z8m} gives $l=4$ and $\fm=0,1$. Also, $ljv\!=\!4v$ mod $I$ since $8v\in I$. Using eq. \eqref{Zg4g4l_z8}, we define
\be\label{Zgjglj_z8_lat}
Z_{\rm lat}(g^j,g^{4})=
\frac{1}{\bar\eta^{6}}\! \sum_{Q\, \in \, I^*}\!
\bar q^{\frac12 (Q+ jv)_R^2}\, e^{2\pi i (2Q^2+ Q\cdot\,4v+2j^2v^2)}\ .
\ee
Is there a vector $v_p\in I^*$ such that $2Q^2=Q\cdot v_p$ mod $\bz$ for all $Q\in I^*$ ? The answer is yes. Using $Q$ and its norm in eqs. \eqref{Istar224} and \eqref{Qnorm224}, together with eq. \eqref{disc224}, one can show that the only such possibility is for $v_p=2f_2^I$ where we have\vs{-5}
\be\label{Qvp}
2Q^2-Q\cdot2f_2^I=-\frac32a'^2+\frac12a'=\frac12a'(a'+1)\;{\rm mod}\;\bz\in\bz\ ,\quad a'=\{0,1\}\ .\vs{-5}
\ee
In fact, $v_p$ is the analogue of $w$ in $\bz_{2m}$ models with $m$ odd prime \cite{Aldazabal:2025zht}. Eq. \eqref{Zgjglj_z8_lat} then reads\vs{-5}
\be\label{Zgjglj_z8_lat_i}
Z_{\rm lat}(g^j,g^{4})=\frac{e^{i\pi 4v^2}}{\bar\eta^{2d}}\hs{-3}\sum_{Q\, \in \, I^*}\!
\bar q^{\frac12 (Q+ jv)_R^2}\, e^{2\pi i (Q+jv)\cdot(4v+v_p)}e^{-2\pi i (4v+v_p)\cdot jv}\ ,\vs{-5}
\ee
Upon Poisson resummation, we obtain
\vs{-5}\be\label{Zg4gj_z8_lat}
Z_{\rm lat}(g^{4},g^{-j})=\sqrt{|I^*/I|}\;\frac{e^{i\pi 4v^2}}{\bar\eta^{2d}}\hs{-2}\sum_{P\, \in \, I}
\bar q^{\frac12 (P+4v+v_p)_R^2}\, e^{-2\pi i(P+4v+v_p)\cdot jv}\ .\vs{-5}
\ee

Two comments are in order. First, the lattice sum in the $(g^4,g^{-j})$ sectors with, $j=1,3$, is $I$ shifted by $4v+v_p$. As observed from eqs. \eqref{hm224I4d} and \eqref{hzglgkl}, $v_p=2f_2^I\!\in\! I_4^*$, and $I_4^*$ shifted by $4v$ is the lattice sum in $Z(g^4,\uno)$. Therefore, these orbits do not impose a modification to the $g^4$-action in the untwisted sector. Second, we observe that the $\bz_2$ class in $I_4^*$ generated by $2f_1$ does not survive the $g^j$-action in the $g^4$ sector for $j=1,3$.

\subsubsection{\texorpdfstring{$Z_{\rm lat}(g^4,g^{-2})$}{TEXT}}\label{s_hetz8g4latt_ii}
The path \eqref{pathg4} in this case has parameters $j\!=\!2$, $l\!=\!2$, $\fm\!=\!0$, and the right-moving contribution of eq. \eqref{Zg4g4l_z8} is
\vs{-10}\be\label{Zg2g4_z8_lat}
Z_{\rm lat}(g^{2},g^{4})=\frac{1}{\bar\eta^{2d}}\! \sum_{Q\, \in \, I^*_2}\!
q^{\frac12 (Q+2v)_L^2}\, \bar q^{\frac12 (Q+2v)_R^2}\,
e^{2\pi i (Q^2+Q\cdot4v+4v^2)}\ .\vs{-5}
\ee
Using eq. \eqref{hm224I2d}, and defining $\alpha_i=2c_i+c_i'$, where $c_i,c_i'=\{0,1\}$, and $i=\{1,2,3\}$, we find\vs{-5}
\be\label{I2d_norm}
Q^2=\frac12\big(c_1'^2+c_2'^2+c_3'^2+c_1'(c_2'+c_3')\big)\,{\rm mod}\,\bz=\begin{cases}
      \text{0 mod $\bz$}  & \text{$c_1'$ and $c_2'+c_3'$ even}\\
      \text{$\frac12$ mod $\bz$} & \text{otherwise}
    \end{cases}\ .\vs{-5}
\ee
Consider the vectors $P'\in I_2^*$ with norm $P'^2\!\in\!\bz$ (i.e. $e^{2\pi iP'^2}\!=\!1$). They define a lattice which we denote by $I_2'$.  Using eq. \eqref{I2d_norm}, one can show that the vectors $P'$ are of the form\vs{-5}
\be\label{I2p}
P+\big(2c_1Y_1+(2c_2+\epsilon)Y_2\,,\,2\beta f_1^I+(2c_3+\epsilon) f_2^I\big)\ ,
\qquad P\in I_2\ , \qquad \beta,\epsilon=\{0,1\}\ .\vs{-5}
\ee
The vectors $Q\in I_2^*/I_2'$ are generated by $(Y_1,f_1^I)$ and $(0,f_2^I)$ and give the phase $e^{2\pi iQ^2}=-1$ for all $Q$. Eq. \eqref{Zg2g4_z8_lat} may then be written as
\vs{-5}\be\label{Zg2g4_z8_lat_i}
Z_{\rm lat}(g^{2},g^{4})=\frac{e^{-4\pi iv^2}}{\bar\eta^{2d}}\hs{-8}\sum_{c_1',c_3'=0}^1\hs{-6}F(I')\,
e^{i\pi\big(\delta_{c_1',0}\delta_{c_3',1}+\delta_{c_1',1}\big)}\ ,\vs{-15}
\ee
where
\vs{-5}\be\label{Zg2g4_z8_lat_ii}
F(I')=\!\sum_{P'\in I_2'}q^{\frac12 (P'+c_1'Y_1)_L^2}
\bar q^{\frac12 (P'+c_1'f_1^I+c_3'f_2^I+2v)_R^2}
e^{2\pi i (P'+(c_1'Y_1\,,\,c_1'f_1^I+c_3'f_1^2)+2v)\cdot 4v}\ .\vs{-7}
\ee

The dual lattice $I_2'^*$ contains vectors\vs{-5}
\be\label{I2pd}
P+\big(2aY_1\,,\,2a f_1^I+2bf_2^I\big)\ ,
\qquad P\in I_2\ , \qquad a,b=\{0,1\}\vs{-5}
\ee
and ${\rm vol}_{I'_2}^{-1}\!=\!\sqrt{|I_2'/I_2'^*|}\!=\!2$. Note that $(2Y_2,2f_2^I)\in I_2$ and so $(2Y_2,0)\in I_2'^*$. Poisson resumming eq. \eqref{Zg2g4_z8_lat_i} over $I_2'$ gives\vs{-5}
\begin{align}
&Z_{\rm lat}(g^{4},g^{-2})\!=\!\frac1{{\rm vol}_{I_2'}}\frac{e^{-4\pi iv^2}}{\bar{\eta}^{2d}}
\!\!\!\!\sum_{c_1',c_3'=0}^1\sum_{P'\in I_2'^*}\!
q^{\frac12 (P+4v)_L'^2}\bar q^{\frac12 (P'+4v)_R^2} 
e^{-2\pi i P'\cdot(c_1'Y_1\,,\,c_1'f_1^I+c_3'f_2^I+2v)}
e^{i\pi\big(\delta_{c_1',0}\delta_{c_3',1}+\delta_{c_1',1}\big)}\nn\\[-2pt]
&=\frac1{{\rm vol}_{I_2'}}\frac{e^{4\pi iv^2}}{\bar{\eta}^{2d}}
\hs{-2}\sum_{a,b=0}^1\sum_{P\in I_2}
q^{\frac12 (P+2aY_1+4v)_L'^2}\bar q^{\frac12 (P+2a f_1^I+2bf_2^I+4v)_R^2}
e^{-2\pi i (P+(2aY_1\,,\,2a f_1^I+2bf_2^I)+4v)\cdot2v}\nn\\[-4pt]
&\times\sum_{c_1',c_3'=0}^1e^{i\pi\big(c_1'b+c_3'a\big)}
e^{i\pi\big(\delta_{c_1',0}\delta_{c_3',1}+\delta_{c_1',1}\big)}\ .\label{Zg4g2_z8_lat}
\end{align}
The sum in the last  line is $1-e^{i\pi a}-e^{i\pi b}-e^{i\pi(a+b)}=2\,e^{i\pi\delta_{a,0}\,\delta_{b,0}}$. The above result may be obtained using the $L$ function method in \cite{Aldazabal:2025zht}.

Two comments are in order. First, the vectors in $I_2'^*$ \eqref{I2pd} may be written in terms of the vectors in $I_4^*$ \eqref{hm224I4d} as $P+a(Z_1,2f_1^I)+b(0,2f_2^I)$, $P\in I_2$. Thus, the lattice sum in $Z(g^4,g^{-2})$ is a subset of the one in $Z(g^4,\uno)$. It was shown below eq. \eqref{Zg4gj_z8_lat} that this is also the case for $(g^4,g^{-1})$ and $(g^4,g^{-3})$ sectors. Hence, we conclude that the $g^4$-action in the untwisted sector need not be modified.\footnote{In the $\bz_{2m}$ models of \cite{Aldazabal:2025zht} ($m$ odd prime), the lattice sum in $Z(g^{2j},g^{-1})$ is $I$ shifted by some vectors including $jw$ ($2w\!\in\! I$). However, $w$ is not an element in $I_2^*$. Hence, the action of $g^{2j}$ in the untwisted sector on $P\!\in\! I_{2}$ must be defined as $e^{2\pi iP\cdot(2jv+jw)}$ so then, upon Poisson resummation, the shift $jw$ is added to $I_2^*$ in $Z(g^{2j},\uno)$.} Second, the vectors $(Z_i,0)$, $j=\{2,3,4\}$, $(\wh{Y}_1,f_1^I)$, and $(\wh{Y}_2,f_2^I)$ in $I_4^*$ are not $g^2$-invariant. Furthermore, $(2\wh{Y}_1,0)$, $(2\wh{Y}_2,0)$, and $(0,2f_1^I)$ are invariant under $g^2$ action, but not under $g$- and $g^3$-action. Finally, $(0,2f_2^I)$ is $g^j$-invariant for $j=\{1,2,3\}$.

\subsubsection{\texorpdfstring{$Z(g^{4},g^{4k-j})$}{TEXT}}
Taking into account the $T^kS$ modular transformation of the left-moving part of eq. \eqref{Zg4g4l_z8}, we find the partition functions
\vs{-5}\be\vs{-5}\begin{split}\label{hetz8g4}
&Z(g^{4},g^{4k-j})=\cF'_4\,\frac{e^{2\pi ik E_4}e^{i\pi(\Delta_1(4,j,8)+\Delta_2(4,j,8))}
e^{i\pi 4v^2}}{\eta^{22} \bar\eta^{6}}\!\!\prod\limits_{\substack{a=1\\4 t_a\in\bz}}^{11}\!
\frac{-2\sin(\pi jt_a) \eta^3}{\vtgen{\frac12}{\frac12+ jt_a}}\prod_{\substack{a=1\\4 t_a\notin\bz}}^{11}\!
\frac{\eta^3}{\hvtgen{\frac12- \widetilde{2j t_a}}{\frac12-k\,\widetilde{4 t_a}+ {jt_a}}}\\[-7pt]
&\hs{30}\times\sum_{P \in I}\bar q^{\frac12 (P+2 f^I_2+4v)_R^2}
e^{i\pi k(P+2 f^I_2+4 v)^2}e^{-2\pi i(P+2 f^I_2+4v)\cdot jv}\ ,\nn
\end{split}\ee
\be
\hs{-205}\cF'_4=\sqrt{|I^*/I|}\;{\cF_1}\;2^{-(\frac{n_2}2+n_4)}=4\ ,\qquad j=\{1,3\}\ ,
\ee
and 
\vs{-5}\be\vs{-5}\begin{split}\label{hetz8g4_i}
Z(g^{4},g^{4k-2})&=\cF''_4\,\frac{e^{2\pi ik E_4}e^{i\pi(\Delta_1(4,2,8)+\Delta_2(4,2,8))}
e^{i\pi 4v^2}}{\eta^{22} \bar\eta^{6}}\!\!\prod\limits_{\substack{a=1\\4 t_a\in\bz}}^{11}\!
\frac{-2\sin(\pi 2t_a) \eta^3}{\vtgen{\frac12}{\frac12+ 2t_a}}\prod_{\substack{a=1\\4 t_a\notin\bz}}^{11}\!
\frac{\eta^3}{\hvtgen{\frac12- \widetilde{4 t_a}}{\frac12-k\,\widetilde{4 t_a}+ {2t_a}}}\\[-7pt]
&\times\sum_{a,b=0}^1\sum_{P\in I_2}
q^{\frac12 (P+2aY_1+4v)_L'^2}\bar q^{\frac12 (P+2a f_1^I+2bf_2^I+4v)_R^2}
e^{i\pi k(P+(2aY_1\,,\,2a f_1^I+2bf_2^I)+4v)^2}\\[-7pt]
&\hs{150}\times e^{-2\pi i (P+(2aY_1\,,\,2a f_1^I+2bf_2^I)+4v)\cdot2v}\,
e^{i\pi\delta_{a,0}\delta_{b,0}}\ ,\nn
\end{split}\ee
\be
\hs{-280}\cF''_4=2\,\sqrt{|I_2'/I_2'^*|}\;{\cF_2}\;2^{-(\frac{n_2}2+n_4)}=8\ ,
\ee
where $\Delta_1(4,j,8)$ and $\Delta_2(4,j,8)$ are given by eqs. \eqref{hetz3mdel1} and \eqref{hetz3mdel2}.

\subsubsection{Operator interpretation}\label{pint_224_g4}
The phases and degeneracy factors in the $g^4$-twisted sector orbits are given respectively by eqs. \eqref{hzglgkl}, \eqref{hetz8g4}, and \eqref{hetz8g4_i}, and are summarized below \vs{-5}
\begin{align}
&(g^4,g^4):e^{i \pi 2E_4}e^{i\pi (P'+4 v)^2}\ ,\quad P'\in I_4^*\ ,\quad \cF_4=16\ ,\label{224g4ph}\\[3pt]
&(g^4,g^{-j}):e^{i\pi(\Delta_1(4,j,8)+\Delta_2(4,j,8))}e^{-i\pi 4v^2}
e^{-2\pi i(P+2 f^I_2)\cdot jv}\ ,\quad P\in I\ ,\;j=\{1,3\}\ ,\quad \cF'_4=4\ ,\nn\\[3pt]
&(g^4,g^{-2}):e^{i\pi(\Delta_1(4,2,8)+\Delta_2(4,2,8))}e^{-i\pi 4v^2}
e^{-2\pi i (P+(2aY_1\,,\,2a f_1^I+2bf_2^I))\cdot2v}e^{i\pi\delta_{a,0}\delta_{b,0}}\ ,\; P\in I_2\ ,\; \cF''_4=8\ ,\nn
\end{align}
where $e^{i\pi(\Delta_1(4,j,8)+\Delta_2(4,j,8))}=e^{i\frac{\pi}2j}$, $j=\{1,3\}$, $e^{i\pi(\Delta_1(4,2,8)+\Delta_2(4,2,8))}=1$, $E_4=1$, and $8v^2\in2\bz$. 

States with momenta $P+2f_2^I$, $P\!\in\! I$, appear in all orbits in the $g^4$ sector. For the $(g^4,g^{-2})$ sector, this corresponds to $a=0$, $b=1$, and so $e^{i\pi\delta_{a,0}\delta_{b,0}}=1$. Consistency of the group action requires that the phase in $Z(g^{4},g^{4})$ times the fourth power of the phase in $Z(g^{4},g^{-1})$ must be one, and the phase in $Z(g^{4},g^{4})$ times the square of the phase in $Z(g^{4},g^{-2})$ must also be one. We find\vs{-10}
\be\vs{-5}\begin{split}\label{224g4ph_i}
&e^{i\pi (P+2f_2^I+4 v)^2}(e^{-i\pi 4v^2}e^{-2\pi i (P+2f_2^I)\cdot v} )^4=1\ ,\,\,\qquad\forall \,P\in I\ ,\\[3pt]
&e^{i\pi (P+2f_2^I+4 v)^2}(e^{-i\pi 4v^2}e^{-2\pi i (P+2f_2^I)\cdot2v} )^2=1\ ,\qquad\forall \,P\in I\ ,
\end{split}\ee
i.e., both conditions are satisfied. Note that the phase $e^{i\pi(\Delta_1(4,j,8)+\Delta_2(4,j,8))}$ in the $(g^4,g^{-j})$ sectors ($j=1,3$) are not included above : they dress $\cF'_4$ to define the $g$-action on the degeneracy factors. More specifically, $g$, $g^2$, and $g^3$ must act on  $\cF_4\!=\!16$ in $Z(g^4,\uno)$ and yield $i\cF_4'\!=\!4i$ in $Z(g^4,g)$, $\cF_4''\!=\!8$ in $Z(g^4,g^2)$, and $-i\cF_4'\!=\!-4i$ in $Z(g^4,g^3)$. To achieve this, $g$ must act on 4 out of 16 degenerate states in $\cF_4$ as $i$, on 6 states as $-1$, and on the remaining 6 states as $+1$. This action gives $g\cF_4\!=\!4i-6+6\!=\!4i$, $g^2\cF_4\!=\!-4+6+6\!=\!8$, $g^3\cF_4\!=\!-4i-6+6\!=\!-4i$, and $g^4\cF_4\!=\!4+6+6\!=\!16$, as required. Thus, the $g$-action on the states $P+2f_2^I$  is consistent for all $P\!\in\! I$.

States with momenta $P+(2aY_1\,,\,2a f_1^I+2bf_2^I)$, $P\!\in\! I_2$, excluding $a\!=\!0$, $b\!=\!1$, and $P\!\in\! I$, appear only in the $(g^4,\uno)$ and $(g^4,g^2)$ orbits. Consistency of the group action requires that the phase in $Z(g^{4},g^{4})$ times the square of the phase in $Z(g^{4},g^{-2})$ must be one. We find that\vs{-5}
\be\vs{-10}\begin{split}\label{224g4ph_ii}
e^{i\pi (P+(2aY_1\,,\,2a f_1^I+2bf_2^I)+4 v)^2}\big(e^{-i\pi 4v^2}
e^{-2\pi i (P+(2aY_1\,,\,2a f_1^I+2bf_2^I))\cdot2v}e^{i\pi\delta_{a,0}\delta_{b,0}} \big)^2=1\ ,\qquad\forall P\in I\ ,
\end{split}\ee
is indeed satisfied, where we used eq. \eqref{disc224}. As for the $g^2$-action on the degeneracy factors, we note that $\cF_4-\cF''_4\!=\!8$ is even. Hence, $g^2$ must act on 12 states in $\cF_4$ by $+1$, and on the remaining 4 states by $-1$. This yields $g^2\cF_4\!=\!\cF''_4\!=\!8$ and $g^4\cF_4\!=\!16$, as expected. All in all, we conclude that operator interpretation in the $g^4$-twisted sector is consistent.

\section{\texorpdfstring{$\bt^6/\bz_9$}{TEXT}}\label{app_Z9}
In this appendix we compute the partition function and examine the consistency of operator interpretation for the HM230 $\bz_9$ model in section \ref{hm230}. The eigenvalue distribution is of the form
\vs{-5}\be\label{9eigens}
t=\frac1{9}\Big([{1,2,4}]^{n_1},[3]^{n_3}\Big)\ ,\qquad n_1=3\ , \quad n_3=2\ .
\ee
The partition function in the coprime sectors and the orbits of $(g^3,\uno)$ and $(g^6,\uno)$ are given by eq. \eqref{hzglgkl}. To compute $Z(g^\ell,g^{k\ell-j})$, where $\ell=3,6$ and $j,k=1,2$, we consider the path
\vs{-5}\be\label{gellomj}
Z(\uno,g^j)\xrightarrow{S}Z(g^j,\uno)\xrightarrow{T^{3\zeta}}Z(g^j,g^{3\zeta j})\equiv
Z(g^j,g^{\ell})\xrightarrow{S}Z(g^\ell,g^{-j})\xrightarrow{T^k}Z(g^\ell,g^{k\ell-j})\ ,\vs{-5}
\ee
where\vs{-5}
\be\label{Nmjl_9}
3\zeta j=9\mathfrak m+\ell\ , \qquad \zeta,j=1,2\ ,\qquad \mathfrak m\in\bz_{\ge0}\ .
\ee
For $j=1$, $\mathfrak m=0$ and $\ell=3\zeta$. For $j=2$ and $\ell=3$, $\mathfrak m=1$ and $\zeta=2$, whereas for $j=2$ and $\ell=6$, $\mathfrak m=0$ and $\zeta=1$. Using the first line of eq. \eqref{hzglgkl} we find\vs{-5}
\be\label{Zgjglj_h}
Z(g^j,g^{3\zeta j})=\frac{\cF_1}{\eta^{22} \bar\eta^{6}}\, \prod_{a=1}^{11}
\frac{ e^{i\pi\big(\frac12 -j  t_a+\lfloor jt_a\rfloor+3\zeta(j t_a)^2\big)}\, \eta^3}
{\vtgen{\frac12-j t_a}{\,\frac12-3\zeta j t_a} }\sum_{Q\,\in\,I^*} 
\bar q^{\frac12 (Q+j v)_R^2} e^{i\pi 3\zeta (Q+j v)^2}\ ,\vs{-5}
\ee
where we used the fact that $(j,9)=1$ and so $\cF_j=\cF_1$. We next compute the $S$ modular transformation of the contribution of the right-moving compact directions
\vs{-5}\be\label{gjglj230_i}
Z_{\rm lat}(g^j,g^\ell)=\frac1{\bar{\eta}^{6}}\!\sum_{Q\, \in \, I^*}
\bar q^{\frac12 (Q+j v)_R^2} e^{i\pi 3\zeta (Q+j v)^2}\ .\vs{-5}
\ee

We recall that the HM230 model has $I^*\!/I\!=\!\bz_3\!\times\!\bz_9$, generated by $f_3^I$ and $f_9^I$ which satisfy $3(f_3^{I})^2\in2\bz$ and $9(f_9^{I})^2\in2\bz$ -- cf. eq. \eqref{disc230}. Any vector $Q\in I^*$ may be written as\vs{-5}
\be\label{Istar230}
Q=P+af_3^I+3bf_9^I+cf_9^I\ ,\qquad\qquad a,b,c\in\{0,1,2\}\ ,\qquad\qquad P\in I\ ,\vs{-5}
\ee
The right-moving part of eq. \eqref{Zgjglj_h} may then be written as\vs{-5}
\be\label{gjglj230}
Z_{\rm lat}(g^j,g^\ell)=
\frac1{\bar{\eta}^{6}}\!\!\sum_{a,b,c=0}^2\sum_{P\, \in \, I}
\bar q^{\frac12 (P+af_3^I+3bf_9^I+cf_9^I+j v)_R^2} e^{i\pi 3\zeta (P+af_3^I+3bf_9^I+cf_9^I+j v)^2}\ .\vs{-5}
\ee
In the exponent of the phase on the \textsc{rhs}, the coefficient of $2\pi i$ is\vs{-5}
\be\vs{-5}\begin{split}\label{gm230exp}
\frac {3\zeta}2 (P+af_3^I+3bf_9^I+cf_9^I+j v)^2&=\frac {3\zeta}2 (P+af_3^I+3bf_9^I+cf_9^I)^2\\[-5pt]
&+ 3\zeta(P+af_3^I+3bf_9^I+cf_9^I+jv)\cdot jv-\frac {3\zeta}2j^2v^2\ ,
\end{split}\ee
where we used eq. \eqref{Nmjl_9}. Using eq. \eqref{disc230} and that $P\in I$, we find\vs{-5}
\be
\frac{3\zeta}2(P+af_3^I+3bf_9^I+cf_9^I)^2=\frac{3\zeta}2(af_3^I+3bf_9^I+cf_9^I)^2\;{\rm mod}\;\bz=
-\frac{2\zeta}3c^2\;{\rm mod}\;\bz\ .\vs{-5}
\ee
Using the LMC \eqref{lmc230g_i}, the \textsc{rhs} of eq. \eqref{gm230exp} is found to be\vs{-5}
\be\label{gm230exp_ii}
-\frac{2\zeta}{3}c^2+(P+af_3^I+3bf_9^I+cf_9^I+jv)\cdot \ell v-\frac \ell2jv^2\;{\rm mod}\;\bz\ .\vs{-5}
\ee
Eq. \eqref{gjglj230} then reads
\be\label{gjglj230_ii}
Z_{\rm lat}(g^j,g^\ell)=\frac{e^{-i\pi\ell jv^2}}{\bar{\eta}^{6}}\sum_{c=0}^2
F(cf_9^I)\,e^{-2\pi i\frac{2\zeta}{3}c^2}\ ,\vs{-5}
\ee
where\vs{-10}
\be\vs{-10}\begin{split}\label{gjglj230_iii}
F(cf_9^I)&:=\sum_{\!\!a,b=0}^2\sum_{P\, \in \, I}\,
\bar q^{\frac12 (P+af_3^I+3bf_9^I+cf_9^I+j v)_R^2}\, e^{2\pi i (P+af_3^I+3bf_9^I+cf_9^I+j v)\cdot\ell v}\\
&\;=\sum_{P'\in I'}\bar q^{\frac12 (P'+cf_9^I+jv)_R^2}\, e^{2\pi i (P'+cf_9^I+jv)\cdot\ell v}\ .
\end{split}\ee
The lattice $I'\!\subset\! I^*$ contains vectors $P+af_3^I+3bf_9^I$, where $P\!\in\! I$ and $a,b=0,1,2$. The dual lattice $I'^{*}$ contains vectors $P+3bf_9^I$ with $P\!\in\! I$ and $b=0,1,2$. Hence, $I^*\!/I'^*=\bz_3$ with generator $f_3^I$, and we obtain $I\subset I'^*\subset I'\subset I^*$. Preforming a Poisson resummation on $F(cf_9^I)$, eq. \eqref{gjglj230_ii} transforms to
\vs{-5}\be\vs{-5}\begin{split}\label{glgmj230}
Z_{\rm lat}(g^\ell,g^{-j})=\sqrt3\,\frac{e^{-i\pi\ell jv^2}}{\bar{\eta}^{6}}\!\!\sum_{P'\in I'^*}\!
\bar q^{\frac12 (P'+\ell v)_R^2} e^{-2\pi i P'\cdot jv}\!
\sum_{c=0}^2e^{-2\pi i P'\cdot cf_9^I}e^{-2\pi i\frac{2\zeta}{3}c^2}\ .
\end{split}\ee
Recalling eq. \eqref{hm230I3d}, we observe that $I'^*\subset I_3^*$. Note that $I_3^*$ contains two $\bz_3$ classes along the directions of $I$ generated by $f_3^I$ and $3f_9^I$, whereas $I'^*$ contains only one $\bz_3$ class with generator $3f_9^I$. Thus, only one conjugacy class along $I$-directions survives the $g$-action on $(g^\ell,\uno)$.

The vectors $P'\in I'^*$ are of the form $P'=3bf_9^I$ mod $I$, where $b\!=\!0,1,2$. Since $(\zeta,3)\!=\!1$, we may rescale the basis and write $P'\!=\!3\zeta bf_9^I$ mod $I$. Eq. \eqref{glgmj230} then reads\vs{-5}
\be\vs{-5}\begin{split}\label{glgmj230_i}
Z_{\rm lat}(g^\ell,g^{-j})=\sqrt3\;\frac{e^{-i\pi\ell jv^2}}{\bar{\eta}^{6}}\sum_{b=0}^2\sum_{P\in I}
\bar q^{\frac12 (P+3\zeta bf_9^I+\ell v)_R^2} e^{-2\pi i (P+3\zeta bf_9^I)\cdot jv}
\sum_{c=0}^2e^{2\pi i \frac{4\zeta}3bc}e^{-i\pi\frac{4\zeta}{3}c^2}\ ,
\end{split}\ee
where we used $(f_9^I)^2=-\frac{4}9$. The sum over $c$ may then be written as\vs{-5}
\be\label{sumc}
\sum_{c=0}^2e^{2\pi i \frac{4\zeta}3bc}e^{-i\pi\frac{4\zeta}{3}c^2}=
e^{i\pi\frac{4\zeta}{3}b^2}\sum_{c=0}^2e^{-i\pi\frac{4\zeta}{3}(c-b)^2}=
e^{i\pi\frac{4\zeta}{3}b^2}\sum_{c=0}^2e^{-i\pi\frac{4\zeta}{3}c^2}=:
C_3(\zeta)e^{i\pi\frac{4\zeta}{3}b^2}\ ,\vs{-5}
\ee
where in the third equality we defined $c-b\to c$. We obtain\vs{-5}
\be\label{C9q}
C_3(\zeta)=\begin{cases}\;\;\;i\sqrt3 & \zeta=1\\-i\sqrt3 & \zeta=2\end{cases}=:
e^{i\pi\theta_3^{\zeta}}\sqrt3\ ,\qquad\theta_3^{\zeta}=
\begin{cases}\;\;\;\frac12 & \zeta=1\\-\frac12 & \zeta=2\end{cases}\ .\vs{-5}
\ee
Inserting eq. \eqref{sumc} and \eqref{C9q} in eq. \eqref{glgmj230_i} we obtain\vs{-5}
\be\vs{-10}\begin{split}\label{glgmj230_ii}
Z_{\rm lat}(g^\ell,g^{-j})=3\;\frac{e^{i\pi\theta_3^{\zeta}}
e^{-i\pi\ell jv^2}}{\bar{\eta}^{6}}\sum_{b=0}^2\,\sum_{P\in I}
\bar q^{\frac12 (P+3\zeta bf_9^I+\ell v)_R^2}\, e^{-2\pi i (P+3\zeta bf_9^I)\cdot jv}\,
e^{i\pi\frac{4\zeta}{3}b^2}\ .\vs{-5}
\end{split}\ee
The above result may be obtained using the method of $L$ functions in \cite{Aldazabal:2025zht}.

We next compute the transformation of the contribution of the left-moving compact directions in $Z(g^j,g^\ell)$ in eq. \eqref{Zgjglj_h}, and obtain the partition function
\vs{-5}\be\vs{-5}\begin{split}\label{hetz9l3}
&Z(g^{\ell},g^{k\ell-j})=\cF'_3\,\frac{e^{2\pi ik E_\ell}e^{i\pi(\Delta_1(\ell,j,9)+\Delta_2(\ell,j,9))}
e^{i\pi\theta_3^{\zeta}}e^{i\pi\ell jv^2}}{\eta^{22} \bar\eta^{6}}\\
&\qquad\qquad\quad\times\prod\limits_{\substack{a=1\\3 t_a\in\bz}}^{11}
\frac{-2\sin(\pi jt_a) \eta^3}{\vtgen{\frac12}{\frac12+j t_a}}\prod_{\substack{a=1\\3 t_a\notin\bz}}^{11}
\frac{\eta^3}{\hvtgen{\frac12- \widetilde{\ell t_a}}{\frac12-k\,\widetilde{\ell t_a}+ \widetilde{jt_a}}}\\[-2pt]
&\qquad\qquad\quad\times\sum_{b=0}^{2}\sum_{P \in I}\bar q^{\frac12 (P+3\zeta b f^I_9+\ell v)_R^2}
e^{i\pi k(P+3\zeta b f^I_9+\ell v)^2}
e^{-2\pi i(P+3\zeta b f_9^I+\ell v)\cdot jv}e^{i\pi\frac{4\zeta}3 b^2}\ ,\\[-2pt]
\end{split}\ee
where
\vs{-5}\be
\cF'_3=\frac{3^{\frac{\ell_3+\ell_9}2}}{3^{\frac{n_3}2}}\cF_1=3^{\frac12(n_1-\ell_9)}=3\ ,
\quad \ell=3,6\ ,\quad \zeta,j=1,2\ ,\quad 3\zeta j=9\fm+\ell\ ,\vs{-5}
\ee
and $\Delta_1(\ell,j,9)$ and $\Delta_2(\ell,j,9)$ are defined in eqs. \eqref{hetz3mdel1} and \eqref{hetz3mdel2}.

\subsection{Operator interpretation}\label{Z9op}
Let us now examine the operator interpretation in the $g^3$ and $g^6$ sectors. The phase in the $Z(g^{3\zeta},g^{3\zeta})$, with $\zeta=1,2$, is obtained from eq. \eqref{hzglgkl}\vs{-5}
\be\label{931ph}
e^{i \pi 2E_3}e^{i\pi (P'+3\zeta v)^2}\ ,\qquad P'\in I_3^*\ .\vs{-5}
\ee
Vectors $P'$ that are also elements of the lattice $I'^*$ in $Z(g^{3\zeta},g^{-1})$ are of the form $P'\!=\!P+3\zeta bf_9^I$. The above phase is then\vs{-5}
\be\label{931ph_i}
e^{i \pi 2E_3}e^{i\pi (3\zeta bf_9^I)^2}e^{i\pi (3\zeta v)^2}
e^{2\pi i(P+3\zeta bf_9^I)\cdot3\zeta v}\ ,\qquad P\in I\ .\vs{-5}
\ee
The phase in the $(g^{3\zeta},g^{-1})$ sector is obtained from eq. \eqref{hetz9l3} to be\vs{-5}
\be\label{93gph}
e^{i\pi(\Delta_1(3\zeta,1,9)+\Delta_2(3\zeta,1,9))}
e^{i\pi\theta_3^{\zeta}}e^{-i\pi3\zeta v^2}e^{-2\pi i (P+3\zeta bf_9^I)\cdot v}e^{-i \pi 3\zeta (b f_9^I)^2}\ .\vs{-5}
\ee
For $j=1$, $3\zeta =\ell$ (i.e. $\mathfrak m=0$) and, using eqs. \eqref{hetz3mdel1} and \eqref{hetz3mdel2}, we find\vs{-5}
\be\label{9del}
e^{i\pi\Delta_1(3\zeta,1,9)}\!=\!\!\prod\limits_{\substack{a=1\\3 t_a\in\bz}}^{11}\!\!
e^{-i\pi 3\zeta t_a(1+t_a)}\ ,\qquad e^{i\pi\Delta_2(3\zeta,1,9)}=
\!\prod_{\substack{a=1\\3 t_a\notin\bz}}^{11}\!
e^{\frac{i\pi}2}e^{i\pi(\lfloor 3\zeta t_a\rfloor-{t_a})}\,
e^{i\pi(\widetilde{3\zeta t_a}-\lfloor 3\zeta t_a\rfloor)t_a}\ .\vs{-5}
\ee
The phase in eq. \eqref{93gph} to power of $3\zeta$ times the phase in \eqref{931ph_i} is found to be\vs{-5}
\be\vs{-5}\begin{split}\label{9ph_i}
&\hs{60}e^{i\pi3\zeta\theta_3^{\zeta}}e^{-i\pi(3\zeta v)^2}\!e^{-2\pi i(P+3\zeta b f_9^I)\cdot 3\zeta v}
e^{-i\pi (3\zeta b f^I_9)^2}\!e^{-i\pi 2E_3}e^{i\pi 3\zeta(\frac{11-n_3}2+\delta(\zeta))}\times\\
&e^{i \pi 2E_3}e^{i\pi (3\zeta b f_9^I)^2}e^{i\pi(3\zeta v)^2}
e^{2\pi i(P+3\zeta b f_9^I)\cdot 3\zeta v}=e^{i\pi3\zeta(\theta_3^{\zeta}+11-n_3+\delta(\zeta))}\ ,
\end{split}\ee
where\vs{-10}
\be\label{delta}
e^{i\pi\delta(\zeta)}:={\rm exp}\bigg(i\pi\sum\limits_{\substack{a=1\\3 t_a\notin\bz}}^{11}
\lfloor 3\zeta t_a\rfloor\bigg)\ ,\qquad \delta(\zeta)=0,1\;{\rm mod}\;2\ .
\ee
The requirement that eq. \eqref{9ph_i} is equal to 1 gives\vs{-5}
\be\label{g9ph}
\zeta\Big(\theta_m^{3\zeta}+\frac{11-n_3}2+\delta(\zeta)\Big)\in2\bz\ .\vs{-5}
\ee
This condition is indeed satisfied for the HM230 model.

For the $g$-action on the degeneracy factors, since $\cF_3-\cF_3'=9-3=6$, $g$ must have a cyclic $\bz_3$ action on 2 copies of $3$-plet states. In other words, $g$ acts on 2 degenerate states in $\cF_3$ as $e^{\frac{2\pi i}3}$, on 2 states as $e^{\frac{4\pi i}3}$, and on the remaining 5 states as $+1$. We thus conclude that the operator interpretation in the HM230 $\bz_9$ model is consistent.

\section{\texorpdfstring{$\bt^D/\bz_{3m}$}{TEXT}}\label{app_Z3m}
In this appendix we consider $\bt^D/\bz_n$ heterotic theories with $n\!=\!3m$, where $m$ is prime and $m\!>\!3$. The invariant lattices $I_3$ and $I_m$, and their duals, are computed in section \ref{order3m}. Here, we first compute the partition function and analyse the orbits in the non-coprime $g^\ell$-twisted sectors, $(\ell,3m)\!\ne\!1$. Then in section \ref{Z3mop} we verify consistency of operator interpretation for the two heterotic models $\bt^6/\bz_{15}$ (HM223) and $\bt^6/\bz_{21}$ (HM240) of sections \ref{hm223} and \ref{hm240}.

The partition function in the coprime twisted sectors and in the $(g^\ell,\uno)$ orbits with $(\ell,3m)=1$ is given by eq. \eqref{hzglgkl}. For the $(g^\ell,g^{-j})$ orbits, $0<j<(\ell,3m)$, we consider the path
\vs{-5}\be\label{gellomj_3m}
Z(\uno,g^j)\xrightarrow{S}Z(g^j,\uno)\xrightarrow{T^{l}}Z(g^j,g^{lj})\equiv
Z(g^j,g^{\ell})\xrightarrow{S}Z(g^\ell,g^{-j})\xrightarrow{T^k}Z(g^\ell,g^{k\ell-j})\ ,\vs{-5}
\ee
where
\vs{-5}\be\label{Nmjl_3m}
lj=3m\,\mathfrak m+\ell\ , \qquad(j,3m)=1\ ,\qquad
\mathfrak m\in\bz_{\ge0}\ ,\qquad g^n=\uno\ .\vs{-5}
\ee
For $j\!=\!1$, one may choose $\fm\!=\!0$ and so $l\!=\!\ell$. It will be convenient to choose values of $j$ and 
$\mathfrak m$ such that $l\!=\!\ell$ in all orbits, i.e. $\ell(j-1)\!=\!3m\,\mathfrak m$. Once the desired value of $j$ is chosen in a 
given orbit, other sectors $(g^\ell,g^{k\ell-j})$ are obtained from $T$ transformations. We thus find
 \vs{-5}\be\vs{-5}\begin{split}\label{path}
&Z(\uno,g^j)\xrightarrow{S}Z(g^j,\uno)\xrightarrow{T^{\ell}}Z(g^j,g^{\ell j})\equiv
Z(g^j,g^{\ell})\xrightarrow{S}Z(g^\ell,g^{-j})\xrightarrow{T^k}Z(g^\ell,g^{k\ell-j})\ ,\\[5pt]
&\ell j=3m\,\mathfrak m+\ell\ , \qquad(j,3m)=1\ ,
\qquad\mathfrak m\in\bz_{\ge0}\ .
\end{split}\ee
Since $(\ell,3m)\ne1$, we have\vs{-5}
\be\label{elljs}
\begin{cases}
j=\frac{m\,\mathfrak m}{q}+1 &\ell=3q\ ,\;\;\; q=\{1,\ldots,m-1\}\\
j=\frac{3\mathfrak m}{q'}+1 &\ell=mq'\ ,\;q'=\{1,2\}
\end{cases}\ .\vs{-5}
\ee

Using the first line of eq. \eqref{hzglgkl}, together with eq. \eqref{mu}, we obtain\vs{-5}
\be\label{Zgjglj_h3m}
Z(g^j,g^{\ell j})=\frac{\cF_1}{\eta^{16+2d} \bar\eta^{2d}}\, 
\prod_{a=1}^{8+d}
 \frac{ e^{i\pi\big(\frac12 -j  t_a+\lfloor jt_a\rfloor+\ell(j t_a)^2\big)}\, \eta^3}
{\vtgen{\frac12-j t_a}{\,\frac12-\ell j t_a} }\sum_{P\, \in \, I^*} 
\bar q^{\frac12 (P+j v)_R^2} e^{i\pi \ell (P+j v)^2}\ ,\vs{-5}
\ee
where $D=2d$, $\cF_j\!=\!\cF_1$ as $(j,3m)\!=\!1$. To compute the $T^kS$ transformation of this partition function, we start with the contribution from the right-moving compact directions
\vs{-5}\be\label{Zlatjl}
Z_{\rm lat}(g^{j},g^{\ell})=\frac1{\bar\eta^{2d}}\sum_{P\, \in \, I^*} 
\bar q^{\frac12 (P+j v)_R^2} e^{i\pi \ell (P+j v)^2}\ .\vs{-5}
\ee
Any vector in $P\in I^*$ may be written as\vs{-10}
\be\label{Istar3m}
\hs{-5}P=X+Y\hs{-10}\mod I\ ,\; X:=\sum_{i=1}^{\ell_3}a_if_{3,i}^I\ ,\;
Y:=\sum_{\iota=1}^{\ell_m}b_\iota f_{m,\iota}^I\ ,\; a_i=\{0,1,2\}\ ,b_\iota=\{0,\ldots,m-1\}\ .\vs{-5}
\ee
The sum over $P\in I^*$ in eq. \eqref{Zgjglj_h3m} may then be written as a double sum over $X$ and $Y$. We may then choose to perform a Poisson resummation on one of the sums to obtain the desired expression. We will consider the $\ell=3q$ and $\ell=mq'$ cases separately.

\subsection{\texorpdfstring{$g^{3q}$ sectors, $0<q<m$}{TEXT}}

The Poisson resummation in eq. \eqref{Zlatjl} is performed over the infinite sum over $X$ and $I$. We obtain\vs{-10}
\be\label{3m_l3_ph6}
Z_{\rm lat}(g^{3q},g^{-j})=3^{\frac{\ell_3}2}\frac{e^{i\pi\theta_m^{3q}}e^{i\pi3q j^2v^2}}{\bar\eta^{2d}}
\sum_{b_\iota=0}^{m-1}\,\sum_{P \in I}\bar q^{\frac12 (P+3q b_\iota f^\iota_m+3q v)_R^2}
e^{-2\pi i(P+3q b_\iota f_m^\iota+3q v)\cdot jv}e^{-i\pi3q (b_\iota f^\iota_m)^2}\ .\vs{-5}
\ee
where\vs{-5}
\be\label{Cml}
C_m(\ell):=\hspace{-18pt}\sum_{\substack{Y\in Z_m\\{\rm classes\,of\,} I^*/I}}\hs{-15}e^{i\pi\ell Y^2}
\!=e^{i\pi\theta_m^\ell}\sqrt{m^{\ell_m}}\ ,\qquad{\rm vol}_3=\sqrt{\frac{m^{\ell_m}}{3^{\ell_3}}}\ ,\qquad
\frac{C_m(3q)}{{\rm vol}_3}=\vs{-5}\sqrt{3^{\ell_3}}e^{i\pi\theta_m^{3q}}\ ,\vs{-5}
\ee
and\vs{-5}
\be\label{theta3q}
2\theta_m^{3q}\,=\,2\phi_m\;{\rm mod}\;2\ ,\quad 4\theta_m^{3q}\in2\bz\ ,\quad
\quad 4\phi_m\in2\bz\ ,\quad(3q,m)=1\ ,\quad q=\{1,\ldots,m-1\}\ .
\ee

\subsection{\texorpdfstring{$g^{mq'}$ sectors, $0<q'<3$}{TEXT}}

The Poisson resummation in eq. \eqref{Zlatjl} is now performed over the infinite sum over $Y$ and $I$, and gives\vs{-10}
\be\label{3m_lm_ph}
Z_{\rm latt}(g^{mq'}\!\!,g^{-j})\!=\!m^{\frac{\ell_m}2}\frac{e^{i\pi\theta_3^{mq'}}\!\!e^{i\pi mq'\! j^2v^2}}{\bar\eta^{2d}}
\!\!\sum_{b_i=0}^{2}\sum_{P \in I}\!\bar q^{\frac12 (P+mq' b_i f^i_3+mq' v)_R^2}
e^{-2\pi i(P+mq' b_i f_3^i+mq' v)\cdot jv}e^{-i\pi mq' (b_i f^i_3)^2}\ ,\vs{-5}
\ee
where\vs{-5}
\be\label{C3l}
C_3(\ell):=\hspace{-18pt}\sum_{\substack{X\in Z_3\\{\rm classes\,of\,} I^*/I}}\hs{-15}e^{i\pi\ell X^2}
\!=e^{i\pi\theta_3^\ell}\sqrt{3^{\ell_3}}\ ,\qquad{\rm vol}_m=\sqrt{\frac{3^{\ell_3}}{m^{\ell_m}}}\ ,\qquad
\frac{C_3(mq')}{{\rm vol}_m}=\vs{-5}\sqrt{m^{\ell_m}}e^{i\pi\theta_3^{mq'}}\ ,\vs{-5}
\ee
and\vs{-5}
\be\label{theta5qp}
2\theta_3^{mq'}=\,2\phi_3\;{\rm mod}\;2\ ,\quad 4\theta_3^{mq'}\in2\bz\ ,\quad
\quad 4\phi_3\in2\bz\ ,\quad(3,mq')=1\ ,\quad q'=\{1,2\}\ .
\ee

\subsection{\texorpdfstring{$Z(g^\ell,g^{\ell k-j})$}{TEXT}}\label{app_N3m_all}
We compute the contribution from the left-moving compact directions in $Z(g^j,g^{j\ell})$ in eq. \eqref{Zgjglj_h3m} and combine it with the right-moving contribution obtained above to find
\vs{-5}\be\begin{split}\label{hetz3ml3}
&Z(g^{3q},g^{3qk-j})=\cF'_3\frac{e^{2\pi ik E_3}e^{i\pi(\Delta_1(3q,j,3m)+\Delta_2(3q,j,3m))}
e^{i\pi\theta_m^{3q}}e^{i\pi3q j^2v^2}}{\eta^{16+2d} \bar\eta^{2d}}\\
&\qquad\qquad\times\prod\limits_{\substack{a=1\\3 t_a\in\bz}}^{8+d}
\frac{-2\sin(\pi jt_a) \eta^3}{\vtgen{\frac12}{\frac12+j t_a}}\prod_{\substack{a=1\\3 t_a\notin\bz}}^{8+d}
\frac{\eta^3}{\hvtgen{\frac12- \widetilde{3q t_a}}{\frac12-k\,\widetilde{3q t_a}+ \widetilde{jt_a}}}\\[-5pt]
&\qquad\qquad\times\sum_{b_\iota=0}^{m-1}\sum_{P \in I}
\bar q^{\frac12 (P+3q b_\iota f^\iota_m+3q v)_R^2}
e^{i\pi k(P+3q b_\iota f^\iota_m+3q v)^2}
e^{-2\pi i(P+3q b_\iota f_m^\iota+3q v)\cdot jv}e^{-i\pi3q (b_\iota f^\iota_m)^2}\ ,\\
&\cF'_3:=\frac{\cF_1\sqrt{3^{\ell_3}}}{(3q,3m)^{\frac{n_m}2}}=m^{\frac12(n_3-\ell_m)}\ ,
\qquad\qquad q,k\in\{1,\ldots,m-1\}\ ,
\end{split}\ee
\be\begin{split}\label{hetz3mlm}
&Z(g^{mq'}\!,g^{mq'\!k'\!-j})=\cF'_m\frac{e^{2\pi ik' E_m}e^{i\pi(\Delta_1(mq'\!,j,3m)+\Delta_2(mq'\!,j,3m))}
e^{i\pi\theta_3^{mq'}}\!\!e^{i\pi mq'\! j^2v^2}}{\eta^{16+2d} \bar\eta^{2d}}\\
&\hs{20}\times\prod\limits_{\substack{a=1\\m t_a\in\bz}}^{8+d}
\frac{-2\sin(\pi jt_a) \eta^3}{\vtgen{\frac12}{\frac12+j t_a}}\prod_{\substack{a=1\\m t_a\notin\bz}}^{8+d}
\frac{\eta^3}{\hvtgen{\frac12- \widetilde{mq' t_a}}{\frac12-k'\,\widetilde{mq' t_a}+ \widetilde{jt_a}}}\\[-5pt]
&\hs{20}\times\sum_{b_i=0}^{2}\sum_{P \in I}\bar q^{\frac12 (P+mq' b_i f^i_3+mq' v)_R^2}
e^{i\pi k'\!(P+mq' b_i f^i_3+mq' v)^2}
e^{-2\pi i(P+mq' b_i f_3^i+mq' v)\cdot jv}e^{-i\pi mq' (b_i f^i_3)^2}\ ,\\
&\cF'_m:=\frac{\cF_1\sqrt{m^{\ell_m}}}{(mq'\!,3m)^{\frac{n_3}2}}=3^{\frac12(n_m-\ell_3)}\ ,\qquad\qquad q',k'\in\{1,2\}\ ,
\end{split}\ee
where $\Delta_1(\ell,j,3m)$ and $\Delta_2(\ell,j,3m)$ are given in eqs. \eqref{hetz3mdel1} and \eqref{hetz3mdel2}.

\subsection{Operator interpretation}\label{Z3mop}
The results obtained above show that the lattice sums appearing in the $(g^{3q},g^j)$ and $(g^{mq'}\!\!,g^j)$ orbits are subsets of respectively $I_3^*$ and $I_m^*$ -- c.f. eqs. \eqref{I3*} and \eqref{Im*}. We will next have to examine operator interpretation for the action of $g$ on the phase and degeneracy factors. To do so, we will focus on the two island candidates HM223 and HM240 with $D=6$.

\subsubsection{\texorpdfstring{$\bz_{15}$}{TEXT} HM223}\label{Z3mop_223}
The phase of a state in the $(g^{3q},g^{3q})$ sector, using eq. \eqref{hzglgkl}, is $e^{i \pi 2E_3}e^{i\pi (P'+3q v)^2}$, where $P'\in I_{3}^*$. Since $P'$ must also appear in $Z(g^{3q},g^{-1})$, it must be of the form $P'\!=\!P+3q b_\iota f_m^\iota$, $P\!\in\! I$ -- see eq. \eqref{hetz3ml3}. Hence, the phase is of the form\vs{-5}
\be\label{ph33}
e^{i \pi 2E_3}e^{i\pi (3q b_\iota f_m^\iota)^2}e^{i\pi(3q v)^2}
e^{2\pi i(P+3q b_\iota f_m^\iota)\cdot 3qv}\ .\vs{-5}
\ee
The phase of this state in $Z(g^{3q},g^{-1})$, using eq. \eqref{hetz3ml3}, is\vs{-5}
\be\label{ph31}
e^{i\pi(\Delta_1(3q,1,15)+\Delta_2(3q,1,15))}e^{i\pi\theta_m^{3q}}
e^{i\pi3q v^2}e^{-2\pi i(P+3q b_\iota f_m^\iota+3q v)\cdot v}e^{-i\pi3q (b_\iota f^\iota_m)^2}\ .\vs{-15}
\ee

The degeneracy factors is $\cF_3\!=\!5$ in $(g^{3q},\uno)$ orbit and $\cF_3'\!=\!1$ in $(g^{3q},g)$ and $(g^{3q},g^2)$ orbits. We combine the overall constant phase in $Z(g^{3q},g^{3qk-j})$ with $\cF_3'$ to define the $g$-action on degenerate states. This phase is independent of $P$ and $v$, and in the present model is given by $e^{i\pi\Delta}:=e^{i\pi\theta_5^{3q}}e^{i\pi(\Delta_1(3q,j,15)+\Delta_2(3q,j,15))}$. It turns out that $e^{i\pi\Delta}\!=\!-1$ in $Z(g^{3q},g)$ and $Z(g^{3q},g^2)$ for all $0\!<\!q\!<\!5$. This suggests that $g$ must act on 2 out of 5 states in $Z(g^{3q},\uno)$ as $e^{\frac{2\pi i}3}$, on 2 other states as $e^{\frac{4\pi i}3}$, and on the one remaining state as $+1$. The result is $g\cF_3\!=\!2e^{\frac{2\pi i}3}+2e^{\frac{4\pi i}3}+1=\!-1=e^{i\pi\Delta}\cF_3'$, as required. Similarly, $g^2\cF_3\!=\!2e^{\frac{4\pi i}3}+2e^{\frac{2\pi i}3}+1=e^{i\pi\Delta}\cF_3'$.

The condition on phases now requires that the phase of a state in $Z(g^{3q},g^{3q})$ \eqref{ph33} times the ${3q}^{\rm th}$ power of the phase of that state in $e^{-i\pi\Delta}Z(g^{3q},g^{-1})$ \ref{ph31} must be 1. We have\vs{-5}
\be\label{ph31_ii}
e^{-i\pi(3q v)^2}\!e^{-2\pi i(P+3q b_\iota f_m^\iota)\cdot 3qv}
e^{-i\pi (3qb_\iota f^\iota_m)^2}\;e^{i \pi 2E_3}e^{i\pi (3q b_\iota f_m^\iota)^2}
e^{i\pi(3q v)^2}e^{2\pi i(P+3q b_\iota f_m^\iota)\cdot 3qv}=1\ ,\vs{-5}
\ee
which is satisfied without imposing further constraints. 

The situation in the $g^{mq'}$-twisted sectors is the same. The constant overall phase obtained from eq. \eqref{hetz3mlm} is $e^{i\pi\Delta'}:=e^{i\pi\theta_3^{5q'}}e^{i\pi(\Delta_1(5q',j,15)+\Delta_2(5q',j,15))}$. We combine this phase with $\cF_5'=1$ to derive the $g$-action on the degeneracy factors. It turns out that $e^{i\pi\Delta'}=-1$ in $Z(g^{5q'},g^i)$ for all $q'=1,2$ and $0\!<\!i\!<\!5$. This suggests an action of $g$ on the $\cF_5=9$ degenerate states in $Z(g^{5q'},\uno)$ of the following form : $g\cF_5=2e^{\frac{2\pi i}5}+2e^{\frac{4\pi i}5}+2e^{\frac{6\pi i}5}+2e^{\frac{8\pi i}5}+1=-1=e^{i\pi\Delta'}\cF_5'$. The same result holds for the $g^i$-action with $i=2,3,4$. The condition on the phases, namely that the phase of a state in $Z(g^{5q'},g^{5q'})$ times the ${5q'}^{\rm th}$ power of the phase of that state in $e^{-i\pi\Delta'}Z(g^{5q'},g^{-1})$ must be 1, then gives\vs{-5}
\be\label{phm1_ii}
e^{-i\pi(5q' v)^2}\!e^{-2\pi i(P+5q b_i f_3^i)\cdot 5q'v}
e^{-i\pi (5q'b_i f^i_3)^2}e^{i \pi 2E_5}e^{i\pi (5q' b_i f_3^i)^2}e^{i\pi(5q' v)^2}
e^{2\pi i(P+5q' b_i f_3^i)\cdot 5q'v}=1\ ,\vs{-5}
\ee
which is automatically satisfied. We thus conclude that the operator interpretation in all the twisted sectors of the HM223 model is valid.

\subsubsection{\texorpdfstring{$\bz_{21}$}{TEXT} HM240}\label{Z3mop_240}
The degeneracy factors in this model are $\cF_3=\cF_3'=1$ and $\cF_7=\cF_7'=1$. The $g$-action on the ground state is therefore trivial. The condition on the phases is derived as in the case of the HM223 model and is satisfied for all sectors. Thus, the HM240 model has a valid operator interpretation.

\section{Non-equivalent shift vectors}\label{appv}

Even if  there might be several shift vectors satisfying the  consistency requirements, a subset of them can actually be equivalent in the sense that they lead to the same partition function. Here we provide some criteria to keep just a representative of these equivalent configurations.

This problem was already addressed in \cite{Aldazabal:2025zht} by looking into symmetries of the partition function. 
In fact, it is enough to concentrate on the untwisted sector since twisted sectors are obtained by modular transformations.
To simplify the analysis we only discuss heterotic orbifolds, since the contribution of left-moving fermions in type II
can be easily taken into account. The key observation is -- see eq. \eqref{hz1gk} -- that $Z(\uno, g^\ell)$ involves  a sum over $P \in I_\ell$ and 
depends on $v$ through $e^{2 i \pi P \cdot (\ell v)}$. 

We will consider three kinds of equivalences: 
\begin{enumerate}
\item Translations : $v_1 \equiv v_2 = v_1 + Q \quad Q \in I^*$.
\item Point automorphisms : $v_1 \equiv v_2 = \theta v_1$.
\item Product with a coprime to the order : $v_1 \equiv v_2 = z
v_1 \quad (z,n)=1$.
\end{enumerate}

Altogether, the equivalence relation is given by
\be
v_1 \equiv v_2 \quad \text{iff} \quad \forall
\ell \ \, \exists \{z,\theta,Q\} \,||\,  v_2 = z \theta v_1 + Q\ , \quad(z,n)=1\ ,\quad \theta \in \Aut(I_\ell) \ , \quad Q\in I^*_\ell \, .
\ee
In practice we used  $Q\in I^*$ instead of
 $Q\in I^*_\ell$, which is equivalent, as shown in subsection
 \ref{nes.translations}. For reasons that we will explain below, we
  also considered $\theta \in \Aut(I)$ as opposed to $I_\ell$. This
  change leads to a lower bound on the number of non-equivalent
  shifts. The actual relation applied is
\be
v \equiv z\theta v +
Q, \quad (z,n)=1\ , \quad \theta \in \Aut(I)\ ,  \quad Q\in I^*.
\ee

Of course, there may be some hidden symmetry where two shifts,
not being equivalent as above, lead to the same partition function. The
numbers of non-equivalent shifts in this work do not take into account
this kind of symmetries.

\subsection{Translations} \label{nes.translations}
Let us consider two shift vectors $v_1$ and $v_2=v_1+Q$, and $P_\ell\in I_\ell$, namely,  $\Theta^\ell P_\ell = P_\ell$, where $\Theta \in \Aut(\Gamma)$. When $(\ell,n)=1$,
then $I_\ell=I$ and $e^{2 i \pi \ell P_\ell \cdot v_2}=e^{2 i \pi \ell P_\ell \cdot
  (v_1+Q)}=e^{2 i \pi \ell P_\ell \cdot v_1}$ if $Q\in I^*$. 

Also, in the case when $(\ell,n)\neq 1$, the phases are the same, as we now show.
As $I^* \subset \spanop(I)$, then $\Theta Q = Q$. Moreover $\Theta \in
\Aut(\Gamma)$ implies  
\be \label{QPl}
Q \cdot P_\ell = (\Theta^kQ) \cdot (\Theta^k P_\ell) = Q \cdot (\Theta^k P_\ell).
\ee
And we can define, for  any momentum $P_\ell \in I_\ell$, an invariant vector in $I$. Namely,
\begin{equation}
P_{\text{inv}} \equiv \sum_{k=0}^{\ell-1} \Theta^k P_\ell.
\end{equation}
Since $P_\ell \in \Gamma$ and $\Theta \in \Aut(\Gamma)$, each term $\Theta^k P_\ell \in \Gamma$ and, since $\Gamma$ is closed under addition, $P_{\text{inv}} \in \Gamma$.
Thus, $P_{\text{inv}}$ is a vector in $\Gamma$ that is left invariant
by the action of the automorphism $\Theta$ so it  belongs to the
invariant sublattice $I = \{ P \in \Gamma \mid \Theta P = P \}$. Then
$Q \cdot P_{\text{inv}} \in \bz$ and
\be
Q \cdot P_{\text{inv}} = Q \cdot \left( \sum_{k=0}^{\ell-1} \Theta^k P_\ell
\right) = \sum_{k=0}^{\ell-1} Q \cdot (\Theta^k P_\ell) = \ell (Q \cdot P_\ell)\ ,
\quad \forall P_\ell \in I_\ell,
\ee
where we have used \eqref{QPl}. Therefore, $\ell Q\in I_\ell^*$. In conclusion, for $Q\in I^*$, $v_1\equiv v_2 = v_1+Q$.

\subsection{Point automorphisms}
From \eqref{hz1gk} it can be seen that if $v_2=\theta v_1$, with
$\theta \in \Aut(I_\ell)$, the sum over $P\in I_\ell$ gets rearranged,
giving the same value for both $v_1$ and $v_2$, and then $v_2\equiv
v_1$.

There are only a few distinct $I_\ell$, because $I_{\ell_1} = I_{\ell_2}$ when $(\ell_1,n)=(\ell_2,n)$, as can be seen from 
the definition of $I_\ell$ and that $\Theta$ is of order $n$. In particular, for $(\ell,n)=1$, $I_\ell=I$.
In all the cases considered in this work, the lattice $I$ is definite, while $I_\ell$, $(\ell,n)\neq 1$, is not. Since we could only compute
the point automorphism group of definite lattices, using Magma \cite{magma} or SageMath \cite{sagemath}, we simply 
checked equivalence among shift vectors related by $\Aut(I)$. Thus, the number of non-equivalent shift vectors given 
in the text is a lower bound of the actual number. In some of the models, we could extend all of $\Aut(I)$ to $\Aut(I_\ell)$ for all 
$\ell$, in these cases the number of non-equivalent shifts given is the absolute number, not just a lower bound.

The $w$ vector in \eqref{gactlatt2} may change under $\Aut(I)$, but
the new one will also satisfy the condition $e^{2\pi iP^2}=e^{2\pi i
P\cdot w} \quad \forall P\in I^*$, and so will give the same partition
function.

\subsection{Product with coprime to the order}

In this section $z$ is a number such that $(z,n)=1$. Let us consider two
shifts, $v_1$ and $v_2=zv_1$, and call $Z_1(a,b)=Z(g^a,g^b)$ with
$v$ replaced by $v_1$, similarly for $Z_2(a,b)$ but with
$v_2=zv_1$. For example, eq. \eqref{hz1gk} would be $Z_1(0,\ell)$ if we
replaced $v$ with $v_1$. Then
\be
Z_2(0,\ell) = Z_1(0,z\ell)
\ee
and the terms in the sum implementing the projection on the untwisted
sector $Z_2(0,\ell)_{0<\ell<n}$ are a rearrangement of the
$Z_1(0,\ell)_{0<\ell<n}$. In a similar, but more laborious way, it can
be shown that all the set $\{Z_2(\ell_1,\ell_2)\}$ is a rearrangement
of the set $\{Z_1(\ell_1,\ell_2)\}$.

\subsection{Conclusion}
The conclusion is that in all our asymmetric $\bz_n$ orbifolds (both
heterotic and type II) the equivalence relation $v \equiv z\theta v +
Q$, where $(z,n)=1$, $\theta \in \Aut(I)$, and $Q\in I^*$, allows us to
set a lower bound of non-equivalent models.
This is the same relation applied in \cite{Aldazabal:2025zht}.

\section{Type II orbifold actions}\label{app_triple}

In this appendix, we classify all crystallographic $\bz_n$ actions in $SO(6)$ that are not in $SU(3)$ and lift to $\text{Spin}(6)$ 
with the same order $n$. 
They can be used to break all left-moving supersymmetries in the asymmetric type II $\bt^6/\bz_n$ settting considered in the text

We use the proper crystallographic actions classified in \cite{Erler:1992ki} as building blocks for $\bz_n$ actions in six dimensions. 
Following the conventions introduced in section \ref{subsection_Zn}, we represent the $SO(6)$ elements by triples $(t_1,t_2,t_3)$ 
with $nt_a \!\in\! \mathbb{Z}$ and $0 \le t_a \le 1$. Recall that the corresponding $\mathbb{Z}_n$ eigenvalues are $e^{\pm 2 \pi i t_a}$. 
We retain triples satisfying eq. \eqref{susyc}, which ensures that the rotation is not in $SU(3)$. On the other hand, 
eq. \eqref{orderc} guarantees that the lift to $\text{Spin}(6)$ has order $n$.

\begin{table}[t!] \begin{center} \small \setlength{\extrarowheight}{0pt} \renewcommand{\arraystretch}{1.25} \setlength{\tabcolsep}{6pt} \begin{tabular}{|c|>{\advance\baselineskip by 5pt}p{\dimexpr\textwidth-1cm-4\tabcolsep-3\arrayrulewidth\relax}|} \hline $\mathbb{Z}_n$ & $(t_1,t_2,t_3)$ \\ \hline $\mathbb{Z}_2$ & $(1,0,0)$ \\[3pt] \hline $\mathbb{Z}_3$ & $\left(\frac23,0,0\right)$ \\[3pt] \hline $\mathbb{Z}_4$ & $\left(\frac12,\frac12,\frac12\right)$, $\left(\frac12,0,0\right)$, $\left(\frac14,\frac34,0\right)$ \\[3pt] \hline $\mathbb{Z}_5$ & $\left(\frac15,\frac35,0\right)$ \\[3pt] \hline $\mathbb{Z}_6$ & $\left(\frac13,\frac13,\frac13\right)$, $\left(\frac13,\frac23,0\right)$, $\left(\frac13,0,0\right)$, $\left(\frac16,\frac56,0\right)$, $\left(\frac12,\frac12,\frac13\right)$, $\left(\frac16,\frac16,\frac23\right)$, $\left(\frac16,\frac12,0\right)$ \\[3pt] \hline $\mathbb{Z}_8$ & $\left(\frac14,\frac14,\frac14\right)$, $\left(\frac14,\frac14,\frac34\right)$, $\left(\frac14,0,0\right)$, $\left(\frac34,0,0\right)$, $\left(\frac18,\frac38,0\right)$, $\left(\frac18,\frac58,0\right)$, $\left(\frac18,\frac38,\frac34\right)$, $\left(\frac14,\frac12,\frac12\right)$, $\left(\frac14,\frac12,0\right)$ \\[3pt] \hline $\mathbb{Z}_9$ & $\left(\frac19,\frac29,\frac59\right)$ \\[3pt] \hline $\mathbb{Z}_{10}$ & $\left(\frac15,\frac25,0\right)$, $\left(\frac{1}{10},\frac{3}{10},0\right)$, $\left(\frac{1}{10},\frac{7}{10},0\right)$ \\[3pt] \hline $\mathbb{Z}_{12}$ & $\left(\frac16,\frac16,\frac16\right)$, $\left(\frac16,\frac16,\frac56\right)$, $\left(\frac16,0,0\right)$, $\left(\frac56,0,0\right)$, $\left(\frac12,\frac13,0\right)$, $\left(\frac12,\frac13,\frac13\right)$, $\left(\frac16,\frac13,\frac13\right)$, $\left(\frac16,\frac13,\frac23\right)$, $\left(\frac16,\frac12,\frac12\right)$, $\left(\frac16,\frac16,\frac12\right)$, $\left(\frac16,\frac13,0\right)$, $\left(\frac16,\frac23,0\right)$, $\left(\frac{1}{12},\frac{5}{12},0\right)$, $\left(\frac{1}{12},\frac{7}{12},0\right)$, $\left(\frac{1}{12},\frac{5}{12},\frac23\right)$, $\left(\frac{1}{12},\frac{5}{12},\frac16\right)$, $\left(\frac{1}{12},\frac{5}{12},\frac56\right)$, $\left(\frac13,\frac14,\frac14\right)$, $\left(\frac13,\frac14,\frac34\right)$, $\left(\frac16,\frac14,\frac14\right)$, $\left(\frac16,\frac14,\frac34\right)$ \\[3pt] \hline $\mathbb{Z}_{14}$ & $\left(\frac17,\frac27,\frac47\right)$ \\[3pt] \hline $\mathbb{Z}_{15}$ & $\left(\frac15,\frac25,\frac13\right)$ \\[3pt] \hline $\mathbb{Z}_{18}$ & $\left(\frac19,\frac29,\frac49\right)$ \\[3pt] \hline $\mathbb{Z}_{20}$ & $\left(\frac{1}{10},\frac{3}{10},\frac12\right)$, $\left(\frac12,\frac15,\frac25\right)$ \\[3pt] \hline $\mathbb{Z}_{24}$ & $\left(\frac{1}{12},\frac{5}{12},\frac14\right)$, $\left(\frac{1}{12},\frac{5}{12},\frac34\right)$, $\left(\frac13,\frac14,0\right)$, $\left(\frac13,\frac34,0\right)$, $\left(\frac13,\frac14,\frac12\right)$, $\left(\frac13,\frac14,\frac16\right)$, $\left(\frac13,\frac14,\frac56\right)$, $\left(\frac13,\frac13,\frac14\right)$, $\left(\frac13,\frac13,\frac34\right)$, $\left(\frac16,\frac14,0\right)$, $\left(\frac16,\frac34,0\right)$, $\left(\frac16,\frac14,\frac12\right)$, $\left(\frac16,\frac16,\frac14\right)$, $\left(\frac16,\frac16,\frac34\right)$, $\left(\frac18,\frac38,\frac16\right)$, $\left(\frac18,\frac38,\frac56\right)$, $\left(\frac18,\frac38,\frac13\right)$, $\left(\frac18,\frac38,\frac23\right)$ \\[3pt] \hline $\mathbb{Z}_{28}$ & $\left(\frac{1}{14},\frac{3}{14},\frac{5}{14}\right)$, $\left(\frac{1}{14},\frac{3}{14},\frac{9}{14}\right)$ \\[3pt] \hline $\mathbb{Z}_{30}$ & $\left(\frac15,\frac25,\frac23\right)$, $\left(\frac{1}{10},\frac{3}{10},\frac13\right)$, $\left(\frac{1}{10},\frac{3}{10},\frac23\right)$ \\[3pt] \hline $\mathbb{Z}_{36}$ & $\left(\frac{1}{18},\frac{5}{18},\frac{7}{18}\right)$, $\left(\frac{1}{18},\frac{5}{18},\frac{11}{18}\right)$ \\[3pt] \hline $\mathbb{Z}_{40}$ & $\left(\frac15,\frac25,\frac14\right)$, $\left(\frac15,\frac25,\frac34\right)$ \\[3pt] \hline $\mathbb{Z}_{60}$ & $\left(\frac{1}{10},\frac{3}{10},\frac16\right)$, $\left(\frac{1}{10},\frac{3}{10},\frac56\right)$, $\left(\frac15,\frac25,\frac16\right)$, $\left(\frac15,\frac25,\frac56\right)$ \\[3pt] \hline \end{tabular} 
\end{center} 
\caption{Crystallographic $\bz_n$ actions in $SO(6)$ that are not in $SU(3)$ and lift to $\text{Spin}(6)$ with the same order $n$. 
The eigenvalues are $e^{\pm 2 i \pi t_a}$, where triples $(t_1,t_2,t_3)$ are listed in the second column.} 
\label{tab_triple} \end{table}

To classify inequivalent actions, we must also account for identifications between triples. We embed the $SO(6)$ elements into $SO(8)$ as $(0,t_1,t_2,t_3)$, defining the action on left-moving fermions in terms of $v_f$  as in \eqref{defvf}. For all intents and purposes, the partition function defined with $v_f$ is equivalent to the partition function obtained with $\tilde v_f$ where
\vs{-5}\begin{align}
\tilde v_f = \omega(v_f) + \lambda \, , \qquad \omega \in W(D_4),
\quad \lambda \in \Lambda_{\mathcal{G}}(D_4) \ .\vs{-5}
\end{align}
The Weyl group of $SO(8)$ acts by permutations of the components and even numbers of sign flips. We only retain a representative for each equivalence class and collect the results in Table \ref{tab_triple}.

The actions listed in Table \ref{tab_triple} may as well be applied symmetrically to both left and right-movers, in which case they would cause supersymmetry to be completely broken. Such actions have been classified up to order 6 in \cite{Font:2002pq}. In this wider sense, our classification of type II asymmetric orbifold actions with 16 supercharges encompasses a classification of type II symmetric orbifold actions with zero supercharges. Finally, as already alluded to in the main text, Table \ref{tab_triple} can be used to reproduce models discussed in \cite{Fraiman:2022aik, Baykara:2025lhl}.

\pagebreak
\small\baselineskip=.87\baselineskip
\let\bbb\bibitem\def\bibitem{\itemsep1.5pt\bbb}

\bibliographystyle{utphys}
\bibliography{isl}
\end{document}